\documentclass[twocolumn]{aastex631}

\received{May 2021}
\accepted{June 2021}

\shorttitle{The complete Swift light curve of OJ 287}
\shortauthors{Komossa et al.}
\graphicspath{{./}{figures/}}
\begin{document}

\title{MOMO IV: The complete Swift X-ray and UV/optical light curve and  characteristic variability of the blazar OJ 287 during the last two decades}

%% you desire. In v6.31 you can include a footnote in the title.

%% The \author command is the same as before except it now takes an optional
%% argument which is the 16 digit ORCID. The syntax is:
%% \author[xxxx-xxxx-xxxx-xxxx]{Author Name}
%%
%% This will hyperlink the author name to the author's ORCID page. 
%%
%\correspondingauthor{August Muench}
%\email{greg.schwarz@aas.org, gus.muench@aas.org}

%\author[0000-0002-0786-7307]{G. Mueller}
%\affiliation{American Astronomical Society \\
%1667 K Street NW, Suite 800 \\

\author{S. Komossa}
\affiliation{Max-Planck-Institut f\"ur Radioastronomie, 
Auf dem H{\"u}gel 69, 
53121 Bonn, Germany}

\author{D. Grupe}
\affiliation{Department of Physics, Earth Science, and Space System Engineering, Morehead State University, 235 Martindale Dr, Morehead, KY 40351, USA}

\author{L.C. Gallo}
\affiliation{Department of Astronomy and Physics, Saint Mary’s University, 923 Robie Street, Halifax, NS, B3H 3C3, Canada}

\author{A. Gonzalez}
\affiliation{Department of Astronomy and Physics, Saint Mary’s University, 923 Robie Street, Halifax, NS, B3H 3C3, Canada}

\author{S. Yao}
\affiliation{Max-Planck-Institut f\"ur Radioastronomie, Auf dem H{\"u}gel 69, 53121 Bonn, Germany}

\author{A.R. Hollett}
\affiliation{Department of Astronomy and Physics, Saint Mary’s University, 923 Robie Street, Halifax, NS, B3H 3C3, Canada}

\author{M.L. Parker}
\affiliation{European Space Agency (ESA), European Space Astronomy Centre (ESAC), E-28691 Villanueva de la Canada, Madrid, Spain}
\affiliation{Institute of Astronomy, University of Cambridge, Madingley Road, Cambridge CB3 0HA, UK} 

\author{S. Ciprini}
\affiliation{Istituto Nazionale di Fisica Nucleare (INFN) Sezione di Roma Tor Vergata, Via della Ricerca Scientifica 1, 00133, Roma, Italy}
\affiliation{ASI Space Science Data Center (SSDC), Via del Politecnico, 00133, Roma, Italy}



%% abstract limited to 250 words. 
\begin{abstract}

We are carrying out a dense monitoring of the blazar OJ\,287 with Swift
since late 2015 as part of our 
project MOMO (Multiwavelength Observations and Modeling of OJ\,287).  This is the densest existing monitoring of OJ\,287 involving X-ray and UV data.
%We are presenting the results in a sequence of publications. %Previous work focussed on:  the luminous 2016/17 synchrotron outburst (paper I, II), the 2020 outburst with its exceptional Swift, XMM-Newton and NuSTAR spectral components (paper II), two decades of XMM-Newton X-ray spectroscopy, and a multiwavelength study quasi-simultaneous with EHT in 2018 (paper III). % 
In this latest publication of a sequence, 
we characterize the multiwavelength variability of OJ\,287  based on 
% nearly two decades of Swift observations also including public archival data since 2005; 
$>$4000 Swift single-wave-band data sets 
including archival data since 2005. 
A structure function analysis reveals a characteristic timescale of $\sim$5 days in the optical--UV at epochs of low-level activity, and larger during outbursts.
The discrete correlation function 
shows zero lag between optical and UV,
with $\tau=0\pm1$ days at the epoch of densest cadence.
During outbursts (in 2016/17 and 2020) the X-rays follow the UV with near-zero lags. However, during quiescence, the delay is 7--18 days with X-rays leading or lagging,   
interpreted as due to a different X-ray component dominated by inverse Compton emission.
Scaling relations are used to derive the  characteristic length scales of broad-line region and torus in OJ\,287. 
A remarkable, symmetric UV--optical deep fade is identified 
in late 2017, lasting for 2 months. 
We rule out occultation from the passage of a dusty cloud and a model where the secondary black hole deflects the jet between the primary and observer. We speculate about a temporary dispersion 
or jet swing event in the core or in a bright quasi-stationary jet feature.
The deep fade reveals an additional, spatially distinct X-ray component.
The epoch 2020.9--2021.1 was searched for precursor flare activity predicted by the binary black hole model of OJ\,287. 

\end{abstract}



\section{Introduction} \label{sec:intro}

Blazars harbor powerful, long-lived jets of relativistic particles that are launched in the immediate vicinity of the supermassive black holes (SMBHs) at their centers.  The accretion disk -- jet interface represents one of the most extreme astrophysical environments where magnetic fields, high gas density, and (special and general) relativistic astrophysics all play a crucial role and shape the multiwavelength electromagnetic emission of these systems \citep{Blandford2019}.

The spectral energy distribution (SED) of blazars shows two broad emission humps \citep[][]{Marscher2009, Ghisellini2015}:
%% alternate SED papers: Abdo2010, and Ghisellini2017
one at low energies peaking between the submillimeter and EUV, sometimes extending into the X-ray regime, and explained as synchrotron radiation of a population of accelerating jet electrons, and a second maximum in the hard X-ray/$\gamma$-ray regime, usually explained as inverse Compton (IC) radiation from photons that scatter off the jet electrons. The photons are located either inside the jet  (synchrotron-self-Compton radiation; SSC) or they are emitted by an external region like the broad-line region (BLR) or torus (external Comptonization; EC).
Additionally, or alternatively, hadronic processes (ultra-relativistic protons) may contribute at high energies \citep[e.g.,][]{Boettcher2019}.  

OJ 287 is a nearby bright blazar at redshift $z$=0.306, remarkable for its multiwavelength properties and its bright semiperiodic outbursts
\citep{Sillanpaa1988}.  
\citet{Takalo1994} called OJ 287 the ``Rosetta stone of blazars''.  
OJ 287 was classified as a BL Lac object, based on the faintness of its optical emission lines from the BLR, only occasionally detected in continuum low states \citep{SitkoJunkkarinen1985, Nilsson2010}.
The SED-type of OJ 287 was classified as an 
LSP (low synchrotron peak frequency; $<10^{14}$ Hz; \citet{Abdo2010}) based on the new scheme, 
equivalent to an LBL \citep[low-frequency-peaked blazar;][] {Padovani1995, Sambruna1996} in the traditional scheme. 

OJ 287 was first detected in the radio regime in the mid 1960s with the Vermilion River Observatory radio survey (VRO) at 610.5 MHz \citep{Dickel1967} and the Ohio radio survey (OJ) at 1415 MHz \citep{Dixon1968}. OJ 287 derived its name from that survey, where O stands for Ohio.  
Its optical variability was detected early, and \citet{Kinman1971} concluded that OJ 287 is one of the most variable sources. 
 
It received particular attention because of indications of periodic variability. OJ 287 is among the blazars with the largest number of reported periodicities,
with periods ranging from tens of minutes to tens of years, reported in one/some epochs, but absent in others
\citep[e.g.][]{Visvanathan1973, Carrasco1985, Valtaoja1985, Kinzel1988, Pihajoki2013a, 
Sandrinelli2016,
Sillanpaa1988, 
Kidger2000, Valtonen2006, 
Goyal2018, 
Dey2018}. 

OJ 287 is bright enough to be detected on photographic plate surveys that date back to the 1880s \citep[e.g.][]{Hudec2013}. Its optical light curve is characterized by sharp and violent outbursts as bright as 12th magnitude.
The bright optical double peaks repeat every
$\sim$12 yr \citep[$\sim$9 yr in the system's rest frame;][] {Sillanpaa1988}. The remarkable light curve triggered
 unprecedented optical monitoring campaigns \citep[e.g.][and references therein]{Pursimo2000, Sillanpaa1996, Valtonen2006, Villforth2010, Valtonen2016, Fan2009, Dey2018, Wehrle2019}. 

The majority of the early models traced back the semiperiodicity in the optical light curve to the presence of a pair of SMBHs.
Different variants of binary SMBH models were studied initially
\citep[e.g.,][]{Lehto1996,  Sillanpaa1988, Katz1997, Valtaoja2000, Villata1998, 
Liu2002, Qian2015, Britzen2018, Dey2018}. 
Detailed modeling of the historic and recent light curves has strongly favored a binary model that explains the double peaks as 
the times when the secondary SMBH impacts the disk around the primary twice during its $\sim$12 yr
orbit \citep[``impact flares" hereafter;][]{Lehto1996, Valtonen2019}. 

The orbital modeling of the binary system successfully reproduces the
overall long-term light curve of OJ 287 until 2019 \citep[][and references therein]{Valtonen2016, Dey2018, Laine2020}, with impact flares observed most recently in 2015 and 2019 (with Spitzer, because OJ 287 was unobservable with Swift and with ground-based optical observatories due to its solar proximity).
This model requires a compact binary with a semi-major axis of 9300 au and eccentricity 0.66, a primary SMBH mass of $1.8\times10^{10}$ M$_{\odot}$ and spin of 0.38, 
and a secondary SMBH mass of $1.5\times10^8$ M$_{\odot}$. 
Because of the strong general-relativistic (GR) precession of the secondary's orbit of $\Delta \Phi=38.6\deg$ per orbit, the impact flares are not always separated by 12 yr. Their separation varies strongly with time in a predictable manner. 
In addition to the impact flares, the model predicts ``after-flares''. These arise when the impact disturbance reaches the inner accretion disk \citep{Sundelius1997, Valtonen2009} and new jet activity is triggered.
The bright X-ray--UV--optical outburst in 2020 could represent the latest after-flare \citep{Komossa2020}. 

OJ 287 harbors a structured, relativistic jet 
that is well aligned with our line of sight
\citep[e.g.][]{Jorstad2005, Hodgson2017, Britzen2018}
and is highly polarized \citep{Cohen2018, Myserlis2018, Goddi2021}. 
Short-time variability in the jet position angle was interpreted as either the sign of a turbulent injection process and/or a clumpy accretion disk \citep{Agudo2012}, or as wobble induced by the binary \citep{Dey2021}. 

OJ 287 is a bright high-energy source, first detected in X-rays with Einstein \citep{Madejski1988}, in $\gamma$-rays with CGRO/EGRET \citep{Shrader1996}, and at $E>100$ GeV with VERITAS \citep{OBrien2017}. It was then observed with most of the
major X-ray missions \citep{Sambruna1994, Comastri1995, Idesawa1997, Massaro2003, Ciprini2007, Massaro2008, Seta2009, Marscher2011, Komossa2020} and with the Fermi $\gamma$-ray observatory \citep{Abdo2009}. OJ 287 is not always detected in $\gamma$-rays, but it exhibits some epochs of bright flaring \citep{Abdo2009, Agudo2011, Hodgson2017}. 
Its X-ray spectrum, based on XMM-Newton between 2005 and 2020 \citep{Komossa2021a}, is highly variable, making it one of the most spectrally variable blazars known in the soft X-ray band. Bright outbursts are driven by supersoft synchrotron flares \citep{Komossa2020}. 

Early observations (PI data and archival studies) with the Neil Gehrels Swift observatory (Swift hereafter) in 2005--2015 \citep[e.g.][] {Massaro2008, StrohFalcone2013, 
Williamson2014, Siejkowski2017, Valtonen2016} were followed by our dedicated project MOMO \citep[Multiwavelength Observations and Modeling of OJ 287;][]{Komossa2017, Komossa2020, Komossa2021a, Komossa2021b, Komossa2021c}. 
In the course of the MOMO program, we carry out a dense monitoring at $>$13 frequencies from radio to X-rays, especially with the Effelsberg 100m radio telescope and with Swift since 2015 December. 
The majority of Swift observations of OJ 287 in recent years were obtained by us. 
Results are presented in a sequence of publications and so far include:
(1) Our detection of two major nonthermal X-ray--UV--optical outbursts with Swift in 2016/17, and 2020
\citep{Komossa2017, Komossa2020}. (2) The detection of variable radio polarization in 2016 \citep{Myserlis2018}. 
(3) The rapid follow-up of the 2020 outburst with Swift, XMM-Newton
and NuSTAR establishing the spectral components up to $\sim$ 70 keV including a giant soft X-ray excess of synchrotron origin and an unexpectedly steep spectrum in the NuSTAR band \citep{Komossa2020}.  (4) XMM-Newton and Swift spectroscopy during Event Horizon telescope (EHT) campaigns catching OJ 287 at an intermediate flux level with synchrotron and IC spectral components in 2018, and a comprehensive analysis of all XMM-Newton spectra during the last two decades \citep{Komossa2021a}. 
The community was alerted about the  outbursts/low states of OJ 287 we detected with Swift (and at Effelsberg) in a sequence of {\sl{Astronomer's Telegrams}} between 2015 and 2020 (ATel \#8411, \#9629, \#10043, \#12086, \#13658, \#13702, \#13785, and \#14052). 

Here, we present the full Swift light curve of OJ 287 until 2021 March 1. The data obtained in the MOMO program represent by far the densest monitoring of OJ 287 involving X-ray and UV bands with important implications for the emission timescales and emission mechanisms of this nearby bright blazar. 
This paper is structured as follows: In Sect. 2 we present the analysis and spectral fits of the Swift data. The characteristic variability properties of all wave bands are established in Sect. 3 (fractional rms variability), Sect. 4 (structure function, SF) and Sect. 5 (discrete correlation functions, DCFs). In Sect. 6 we derive some characteristic length scales in OJ 287, discuss implications of the variability analyses, and discuss the most outstanding features in the long-term light curve and especially the symmetric UV--optical deep fade in 2017. A summary and conclusions are provided in Sect. 7.

Timescales are given in the observer's frame when reporting measurement results, except when noted otherwise. 
We use a cosmology with 
$H_{\rm 0}$=70 km\,s$^{-1}$\,Mpc$^{-1}$, $\Omega_{\rm M}$=0.3 and $\Omega_{\rm \Lambda}$=0.7 throughout this paper. At the distance of OJ 287, this corresponds to a scale of 4.5 kpc/arcsec (Wright 2006). 
% At z=0.306, This gives a scale of 4.515 kpc/". The luminosity distance DL is 1588.6 Mpc or 5.181 Gly. 

\begin{table}
\scriptsize
	\centering
	\caption{Log of Swift observations. $\Delta t$ is the exposure time of each single data set.}
	\label{tab:obs-log}
	\begin{tabular}{lclc} 
		\hline
		Instrument & Band & Date & $\Delta t$ (ks)\\
		\hline
		XRT & 0.3-10 keV & 2005--05--20 -- 2021--03--01 & 0.2-4 \\
		UVOT & V,B,U,W1,M2,W2 & 2005--05--20 -- 2021--03--01
		& 0.2--4 \\
		\hline
%\begin{tablenotes}
%{Notes: $^{1}$ }.
%\end{tablenotes}
	\end{tabular}
\end{table}

\begin{figure}
\includegraphics[clip, trim=1.8cm 2.1cm 2.2cm 0.3cm, angle=-90, width=\columnwidth]{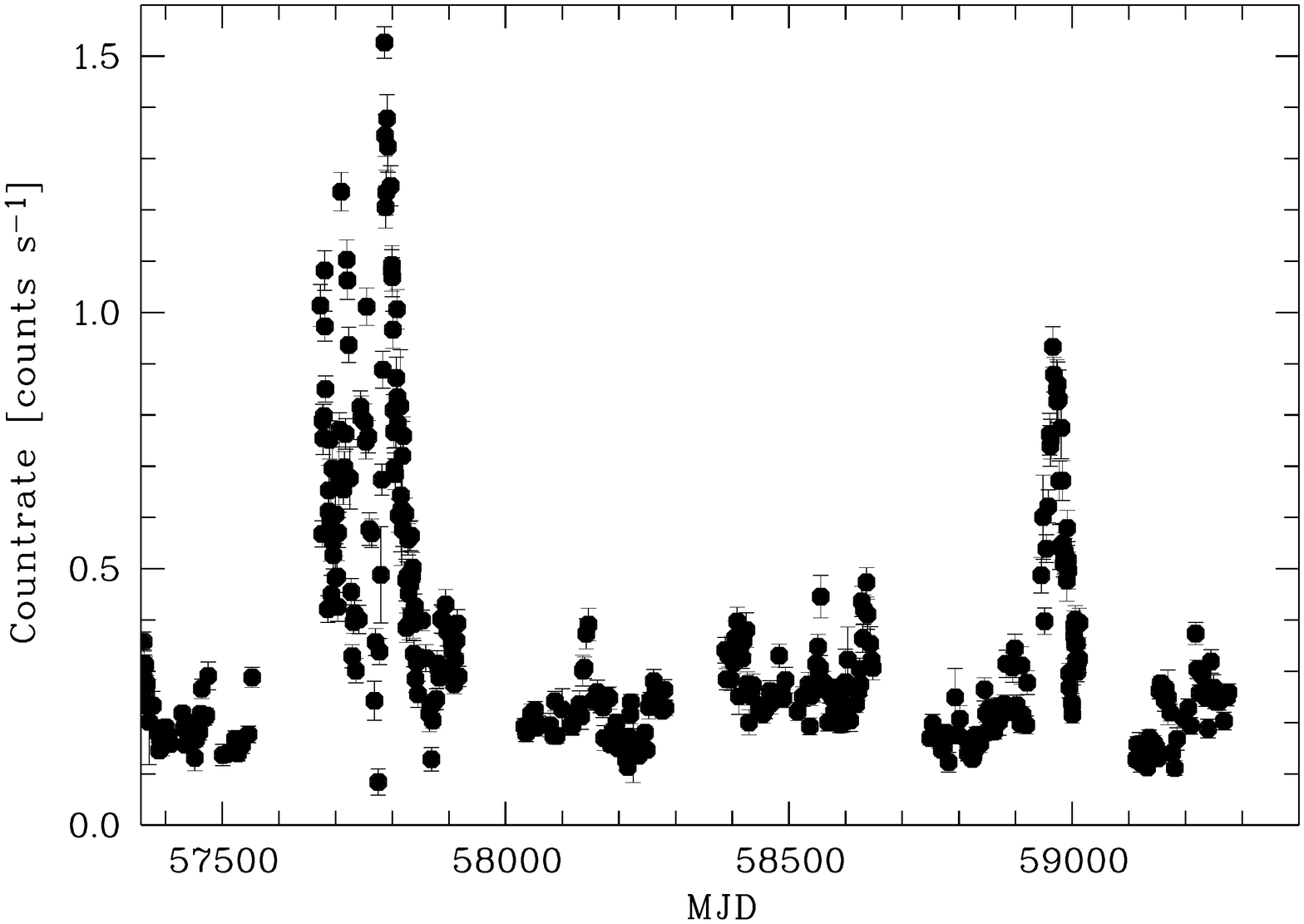}
	\caption{MOMO (0.3-10 keV) X-ray light curve of OJ 287 obtained with Swift (time interval: 2015 December -- 2021 March 1) including the two outbursts in 2016/17 and 2020. Error bars are always plotted but are sometimes smaller than the symbol size.  
     }
     \label{fig:Swift_Xlight-MOMO}
\includegraphics[width=\columnwidth]{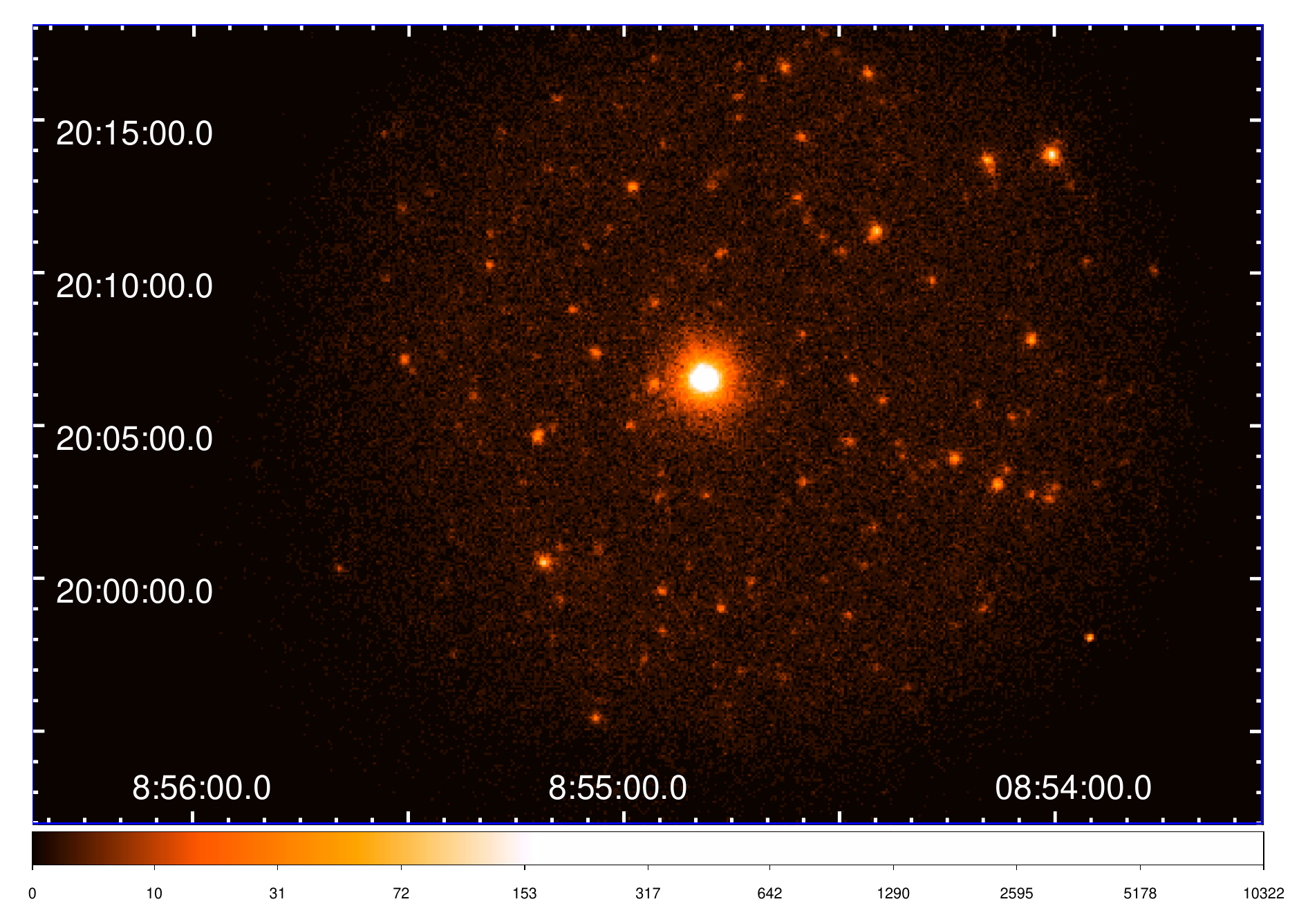}
    \caption{Deep composite X-ray image in the (0.3--10) keV band centered on OJ 287, created by combining Swift single-epoch (PC mode) images,
    % until 2021 Jan 30, 
    amounting to a total on-source exposure time of 670 ks. The axes are in right ascention (R.A.) and declination (decl). }
    \label{fig:Xima-Swift-main}
\end{figure}

\section{Swift observations and data analysis}
 \label{sec:Swift}

\begin{figure*}
\includegraphics[trim=1.8cm 5.6cm 1.3cm 2.6cm, width=18.3cm]{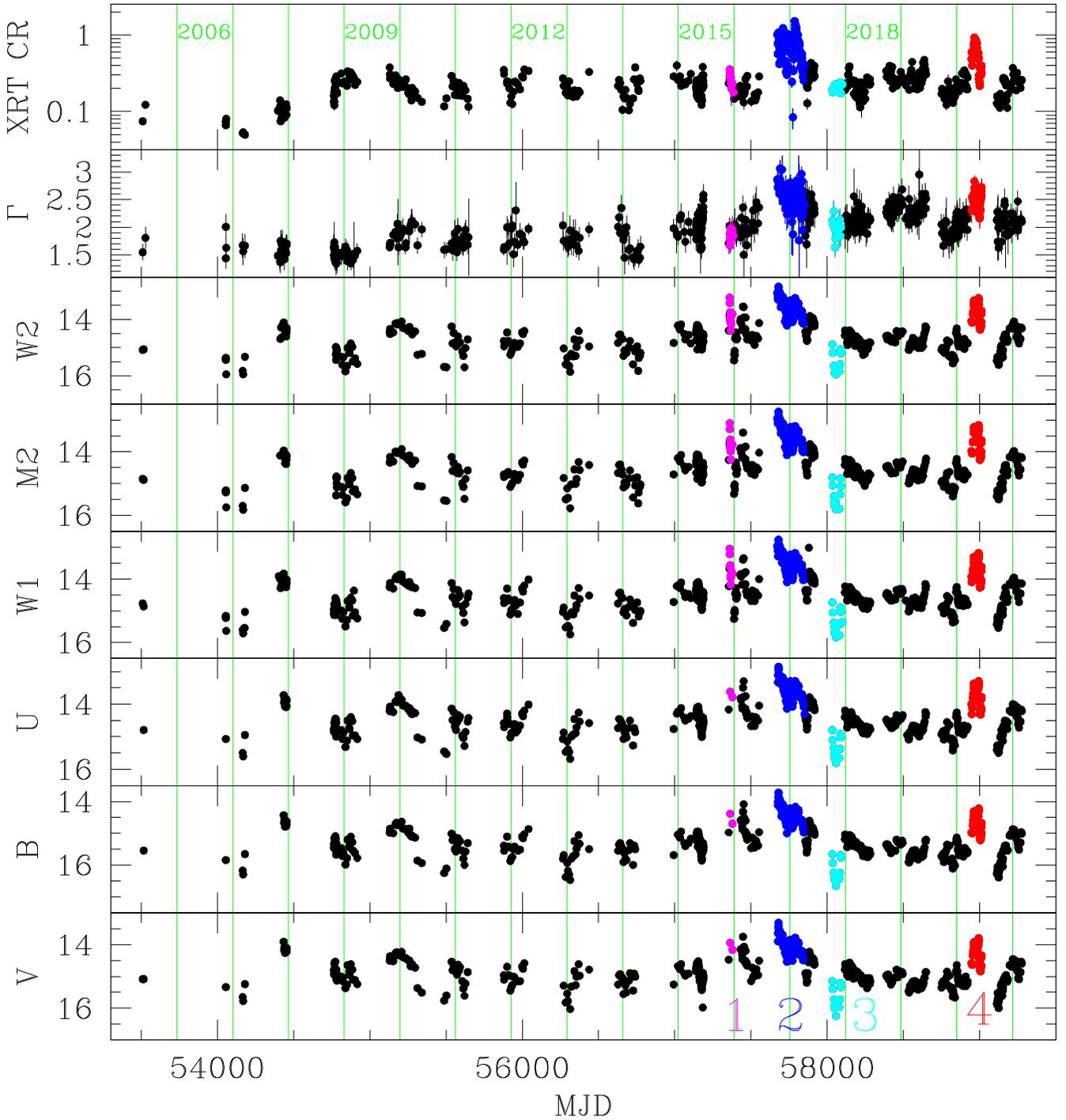}
    \caption{Swift XRT and UVOT light curve of OJ 287 since 2005. Our dense monitoring started in 2015 December. 
    The (0.3-10) keV XRT count rate  is given in cts/s. The UV magnitudes, uncorrected for extinction, are reported in the VEGA system. 
     $\Gamma_{\rm x}$ is the X-ray power-law photon index. Four epochs are marked in color: (1) the 2015 ``impact flare'', (2) the 2016/17 outburst, (3) the 2017 UV--optical deep fade, and (4) the 2020 April-June outburst. The green vertical bars mark January 1st of each year between 2006 and 2021. 
     The last data point is from 2021 March 1. 
     Error bars are always plotted but are often smaller than the symbol size.
     Due to the dense coverage, data points of several epochs can no longer be resolved individually in this long-term light curve.  In Appendix A, higher-resolution annual light curves that resolve all epochs are presented (Fig. \ref{fig:lc-Swift-2015}--\ref{fig:lc-Swift-2019}). 
     }
    \label{fig:lc-Swift-CR}
\end{figure*}

\begin{figure*}
\centering
\includegraphics[clip, trim=1.8cm 5.6cm 1.3cm 2.6cm, width=15.1cm]{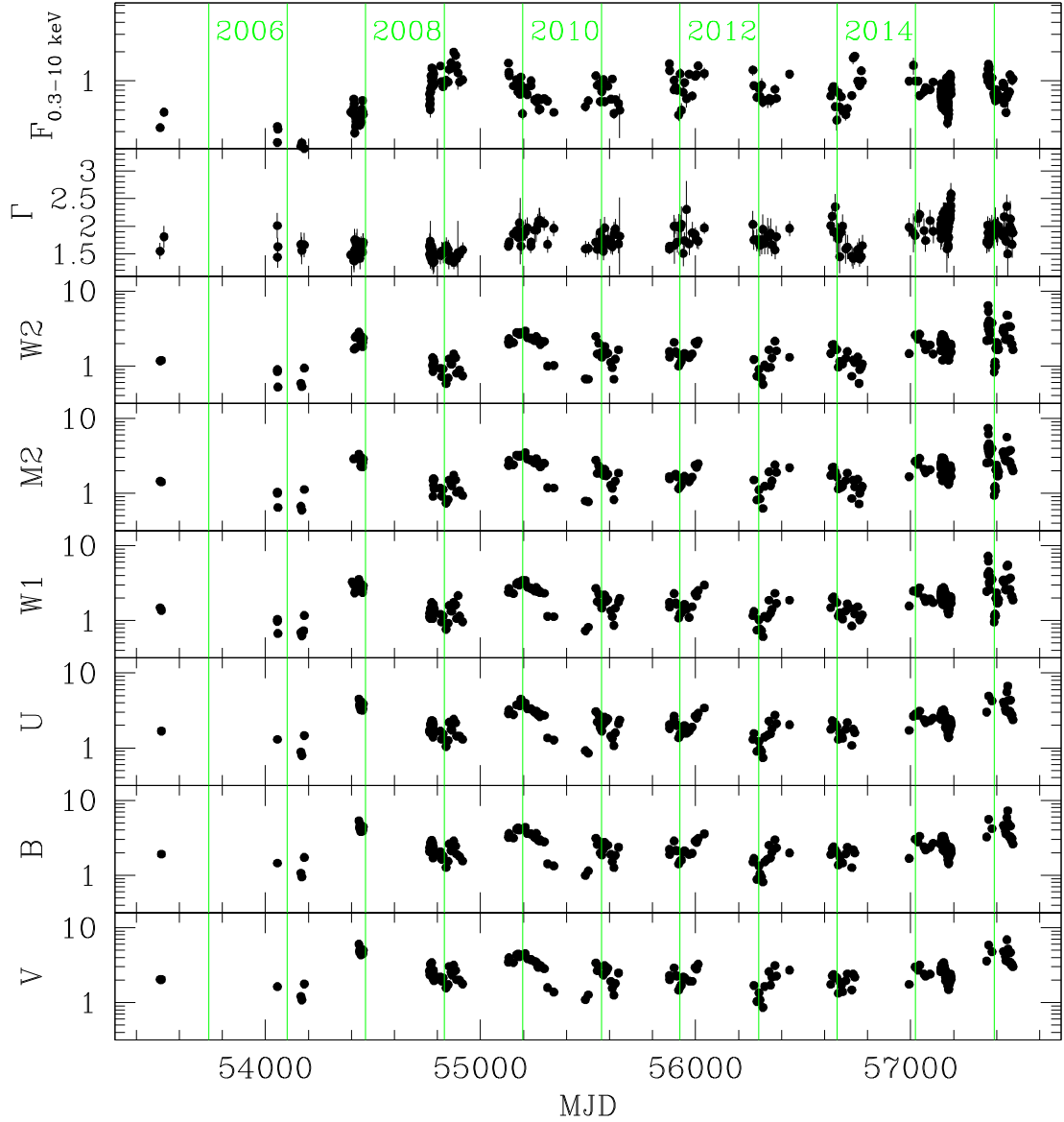}
\caption{Swift XRT and UVOT light curve of OJ 287 between 2005 and 2015 in flux units.
The X-ray flux (in the observed 0.3-10 keV band and corrected for Galactic absorption) and the optical-UV fluxes (corrected for Galactic extinction) are reported in units of 10$^{-11}$ erg/s/cm$^2$. $\Gamma_{\rm x}$ is the X-ray power-law photon index. The green vertical bars mark January 1 of each year between 2006 and 2016. Error bars are always plotted but are often smaller than the symbol size.   
}
    \label{fig:lc-Swift-fluxes2005}
\end{figure*}

\begin{figure*}
\centering
\includegraphics[clip, trim=1.8cm 5.6cm 1.3cm 2.6cm, width=15.1cm]{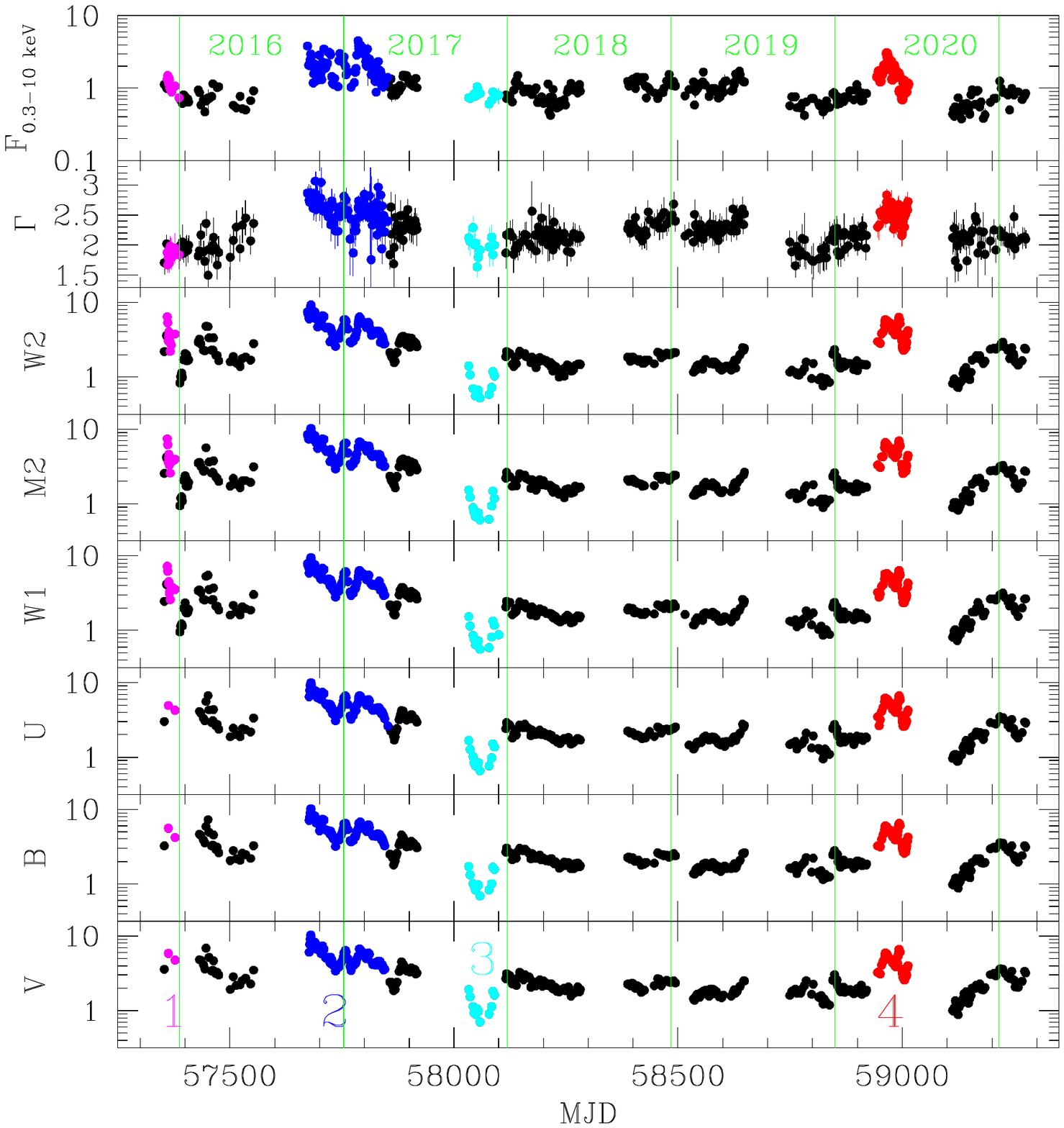}
\caption{Swift XRT and UVOT light curve of OJ 287 between 2016 and 2021 in flux units during our dense monitoring in the course of the MOMO program.
The X-ray flux (in the observed 0.3-10 keV band and corrected for Galactic absorption) and the optical-UV fluxes (corrected for Galactic extinction) are reported in units of 10$^{-11}$ erg/s/cm$^2$. $\Gamma_{\rm x}$ is the X-ray power-law photon index.
Four epochs are marked in color: (1) the 2015 ``impact flare'', (2) the 2016-17 outburst, (3) the 2017 optical-UV deep fade, and (4) the 2020 April outburst. The green vertical bars mark January 1 of each year between 2016 and 2021. Error bars are always plotted but are often smaller than the symbol size.   
}   
    \label{fig:lc-Swift-fluxes2016}
\end{figure*}

\subsection{Project database and monitoring cadence}

We have been densely monitoring  OJ 287 with Swift \citep{Gehrels2004}
since 2015 December \citep[][our Tab. \ref{tab:obs-log}, Fig. \ref{fig:Swift_Xlight-MOMO}]{Komossa2017, Komossa2020}. The monitoring continues in 2021.  Public archival data since 2005 were added to our analysis. 
One of the goals of MOMO is the characterization of the variability of OJ 287 and the measurement of cross-band time delays. Sporadically covered light curves on timescales of weeks are not sufficient for this purpose. On the other hand, continuous light curves of deep pointings with space missions like XMM-Newton often last much less than a day (10--50 ks in most observations of OJ 287; \citet{Komossa2021a}). 
To bridge this gap, we are obtaining high-cadence light curves with Swift.  
Our coverage of OJ 287 with Swift is denser during outbursts (cadence 1--3 days), and less during more quiescent epochs (cadence 3--7 days) with longer gaps when OJ 287 remained constant for several subsequent observations. 
Occasional gaps in the cadence are due to scheduling of a higher-priority target (mostly GRBs). Other gaps in the light curve arise when OJ 287 is in Swift moon constraint ($\sim$3--4 days each) and in Swift Sun constraint ($\sim$3 months each year). During one epoch in 2015 lasting $\sim$6 weeks, OJ 287 was observed at higher cadence, twice daily (PI: R. Edelson). 

Exposure times are in the range 0.2-4 ks in X-rays - typically 2 ks when OJ 287 was faint, typically 1 ks when it was bright
(Tab. \ref{tab:obs-log}).
Exposure times for the UV-optical telescope (UVOT) are in the same range as the XRT observations. Under normal circumstances, the UVOT filters V:B:U:W1:M2:W2 are observed with a ratio of 1:1:1:2:3:4 of the total exposure time, respectively \citep[e.g.][]{Grupe2010}. This may vary when the observation is interrupted by a burst or a high-priority target-of-opportunity (ToO) observation. Most of the time, all six filters of UVOT were employed in order to measure SEDs and cross-band time delays.

\subsection{Swift XRT data analysis} 

During most of the observations, the Swift X-ray telescope \citep[XRT;][]{Burrows2005} was operating in {\sl{photon counting}} (PC) mode
\citep{Hill2004}. Data above 
$\sim$1 ct s$^{-1}$ were obtained in {\sl{windowed timing}} (WT) mode \citep{Hill2004}. In that mode, only the central 4$\times$4 arcminutes of the field of view are read out, in order to avoid the effect of photon pileup. 
The XRT data analysis was performed with the {\sc{xrtdas}} package developed at the ASI Science Data Center (SSDC) and included in the {\sc{heasoft}} package (version No. 6.28). X-ray count rates were determined using the XRT product tool at the Swift data center in Leicester \citep{Evans2007}.
For spectral analysis, and in order to create the long-term light curves (Figs \ref{fig:lc-Swift-CR}, \ref{fig:lc-Swift-fluxes2005}, \ref{fig:lc-Swift-fluxes2016}), event files were created in the energy range (0.3--10) keV based on events with grades 0--12 (PC mode) and 0--2 (WT mode). 

OJ 287 is off axis in most data sets, as is typical for Swift monitoring observations. However, the point-spread function (PSF) does not strongly depend on the location within the inner several arcminutes of the field of view \citep{Moretti2005}. 
In order to carry out further spectral and temporal analysis, source photons were extracted within a circular region with a radius of 20 detector pixels, where one pixel is equivalent to 2.36$\arcsec$. 
This source extraction size does include the X-ray jet of OJ 287. The jet was detected with the Chandra observatory and has an extent of 20\arcsec.  However, the integrated Chandra ACIS-S jet emission of $\sim$0.03 cts/s with an average $\Gamma_{\rm x} = 1.61$ \citep{Marscher2011} only corresponds to a Swift XRT count rate  of 0.009 cts/s. Its contribution to the integrated emission is therefore negligible even in X-ray low states.
Background photons were collected in a nearby circular region of radius 236\arcsec. 

Swift data above a count rate  of $\sim$0.7 cts/s are affected by pile-up. 
To correct for it,  we first created a region file where the inner circular 
area of the PSF 
%%(typically 5--7 arcsec)  
was excluded from the analysis. 
The loss in counts is then corrected by creating a new ancillary response file 
based on this annular region that is used in XSPEC to correct the flux measurement.

X-ray spectra of source and background in the band (0.3-10) keV were then generated 
% and binned; but unbinned for W-STAT
and the software package XSPEC \citep[version 12.10.1f;][]{Arnaud1996} was used for spectral analysis. 
The X-ray field of OJ 287 based on the Swift (PC mode) observations between 2005 and 2021 
is shown in Fig. \ref{fig:Xima-Swift-main}. The total on-source exposure time amounts to 670 ks. In Appendix B we 
provide properties 
of the brightest serendipitous X-ray sources in the field of view.  All sources in the field are much fainter than OJ 287. 
 
The Swift X-ray source position agrees well with the coordinates of OJ 287 
(R.A.: 08$^{\rm h}$54$^{\rm m}$48.87$^{\rm s}$, decl: +20$^{\circ}$06$\arcmin$30.6$\arcsec$) within the XRT measurement errors. 
During the 2020 outburst (MJD 58954), R.A.=08$^{\rm h}$54$^{\rm m}$49.02$^{\rm s}$ and decl=+20$^{\circ}$06$\arcmin$29.8$\arcsec$ with an error radius of 3.5$\arcsec$ (90\% confidence). The enhanced source position (that makes use of UVOT field astrometry; \citet{Evans2007}) is R.A.=08$^{\rm h}$54$^{\rm m}$48.89$^{\rm s}$, and decl=+20$^{\circ}$06$\arcmin$30.9$\arcsec$) with an error radius of 1.8$\arcsec$ (90\% confidence). 

\subsection{X-ray spectral fits}

Spectra were fit with single power laws of photon index $\Gamma_{\rm X}$ 
(defined as $N(E) \propto E^{- \Gamma}$), and taking into account the background spectrum extracted as described above.
% Gamma = 1 - alfa
Galactic foreground absorption with a hydrogen column density of $N_{\rm H, Gal}=2.49\, 10^{20}$ cm$^{-2}$ \citep{Kalberla2005}
was included and modeled with \textsc{tbabs} \citep{Wilms2000}.
The single-epoch Swift data do not tightly constrain the amount of cold absorption, due to the short exposure times.
We therefore caution against treating absorption as a free parameter in single-epoch spectral fits as this can induce spurious flux variations and absorption is not expected to vary on daily timescales.   
In fact, our spectral fitting of XMM-Newton data of OJ 287 has shown that no excess cold absorption beyond the Galactic value is required to fit the X-ray spectra of OJ 287 \citep{Komossa2020, Komossa2021a}. We have therefore fixed the absorption at the Galactic value.  
Single X-ray spectra of OJ 287 typically contain 200--1000 counts. 
In order to treat all $\sim$700 spectra homogeneously, spectral fits were carried out on the unbinned data and using 
the W-statistics of \textsc{xspec}.
% appropriate for spectra dominated by Poisson statistics.  
% https://heasarc.gsfc.nasa.gov/xanadu/xspec/manual/XSappendixStatistics.html

\begin{figure*}
\includegraphics[width=0.32\linewidth, trim=1.5cm 1.4cm 2.0cm 0.0cm]{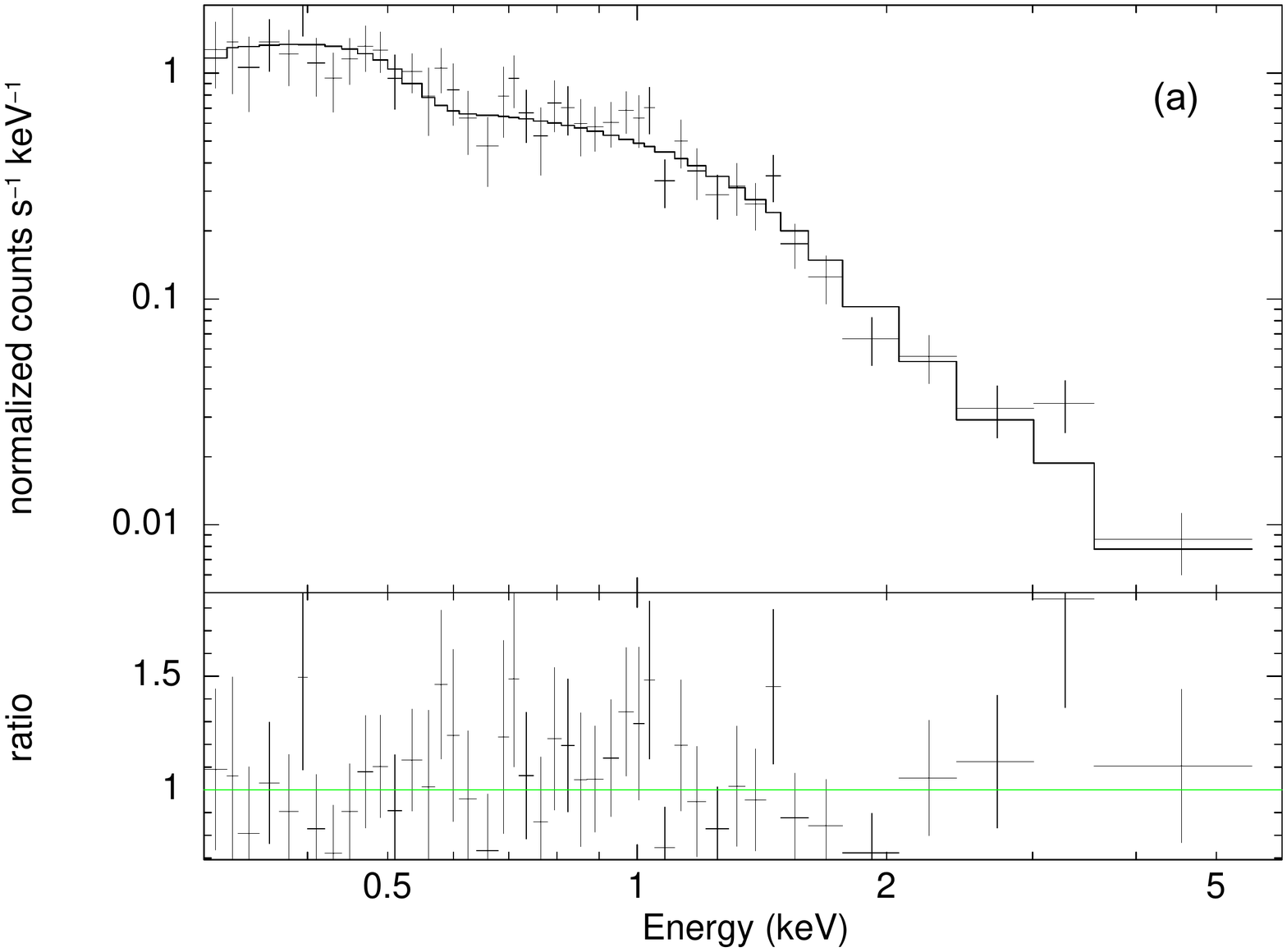}
\includegraphics[width=0.32\linewidth, trim=1.5cm 1.4cm 2.0cm 0.0cm]{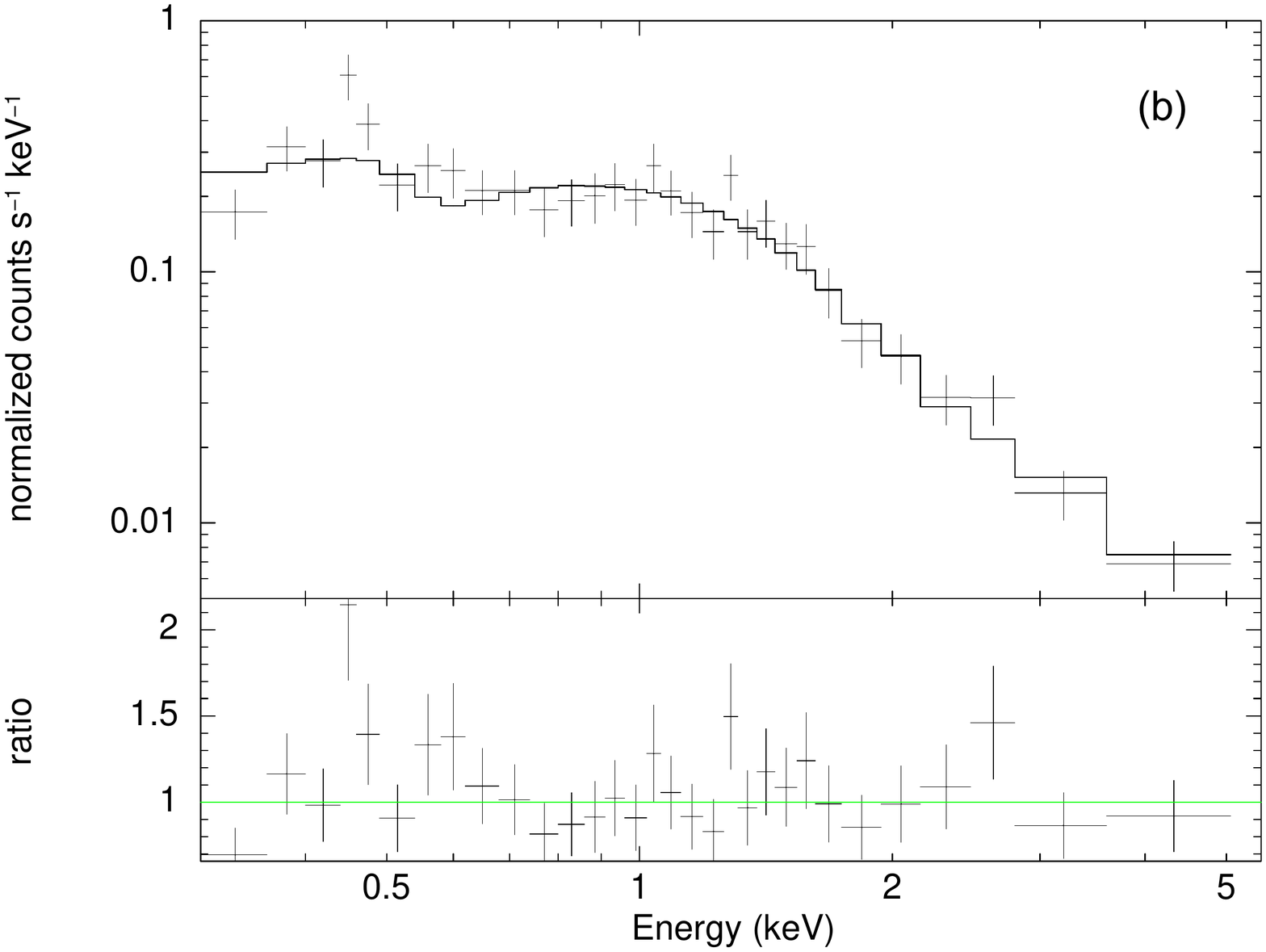}
\includegraphics[width=0.32\linewidth, trim=1.5cm 1.3cm 2.0cm 0.0cm]{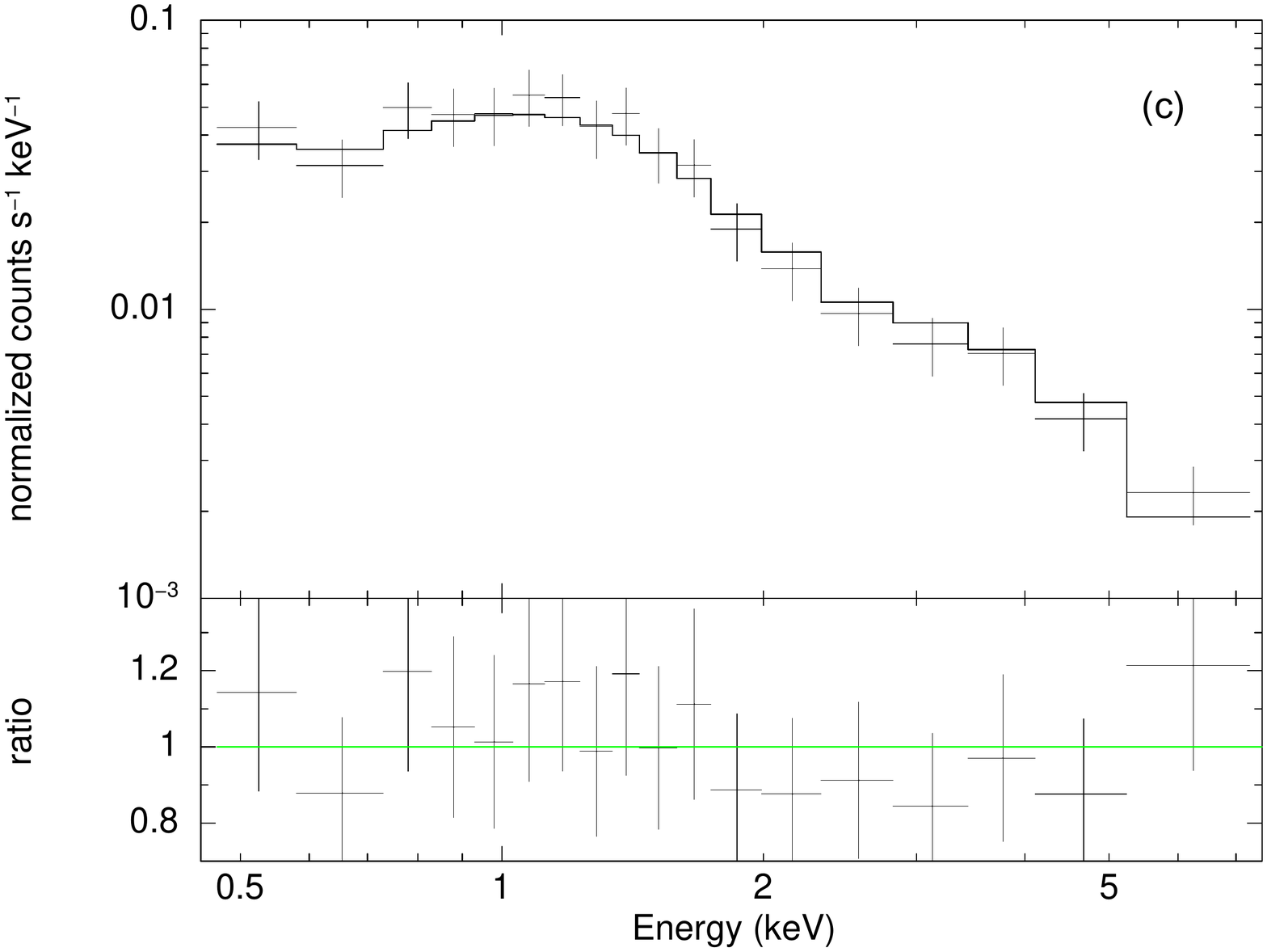}
\includegraphics[width=0.32\linewidth, trim=1.5cm 0.4cm 2.0cm 0.0cm]{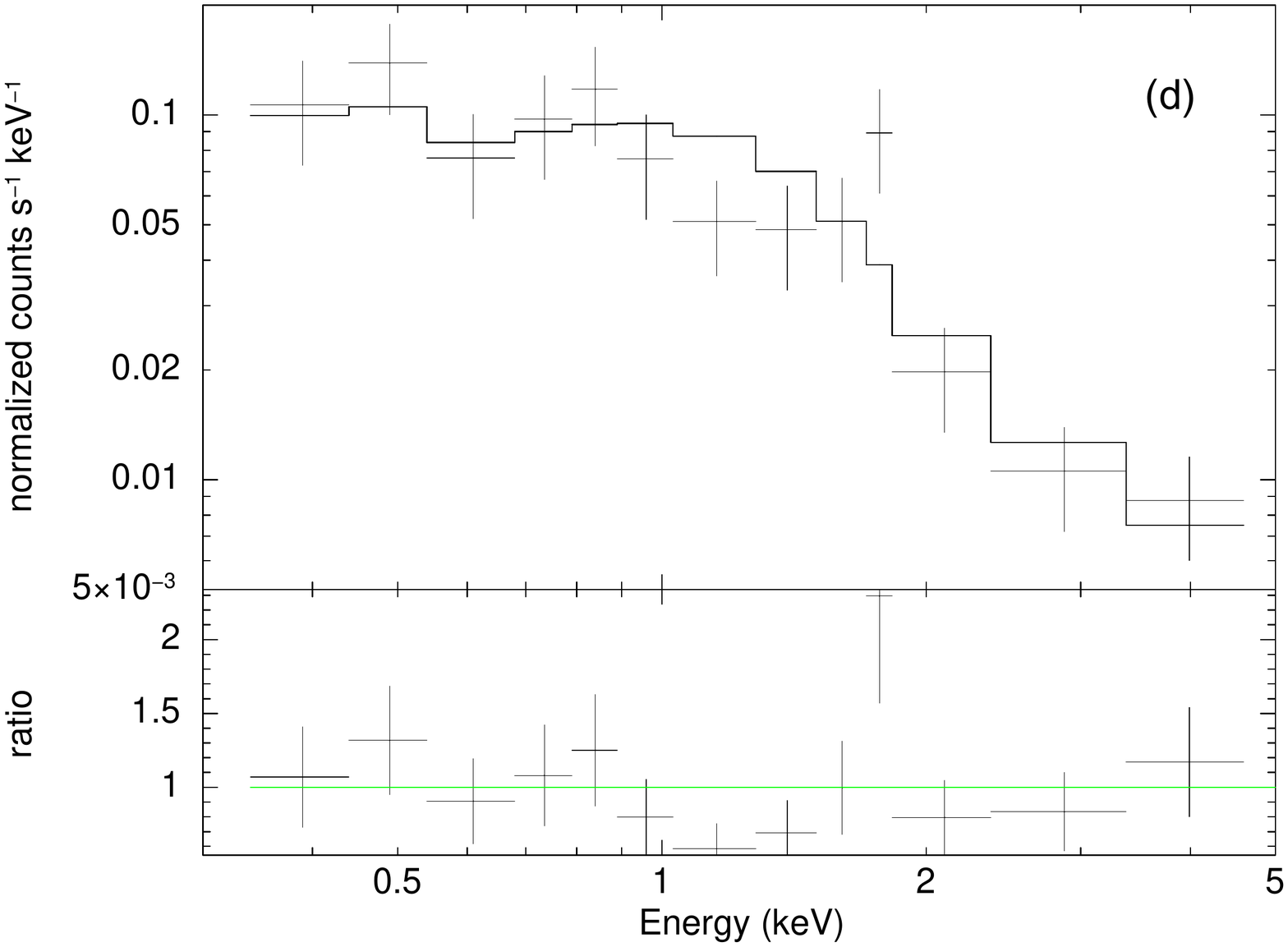}
\includegraphics[width=0.32\linewidth, trim=1.5cm 0.4cm 2.0cm 0.0cm]{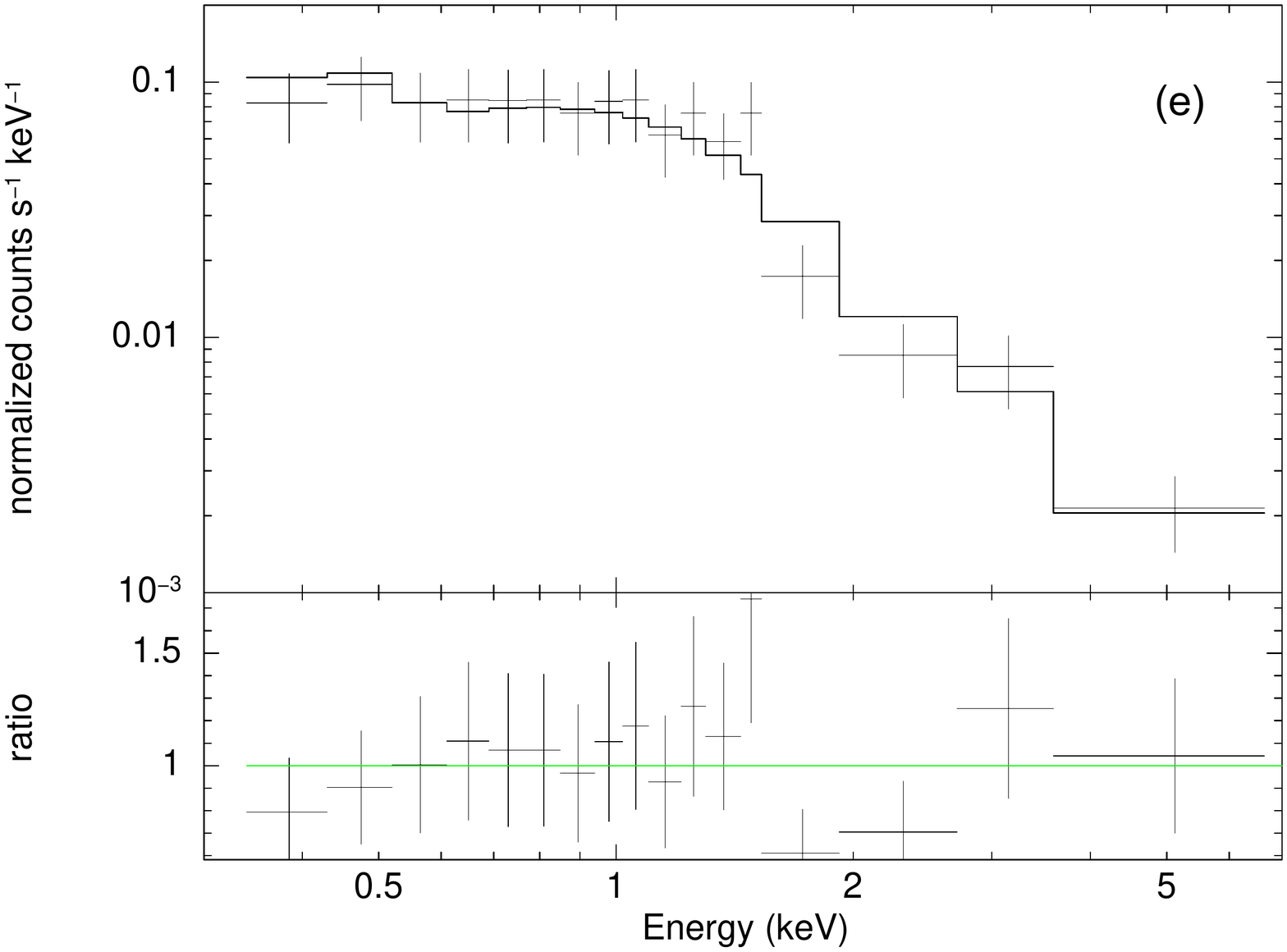}
\includegraphics[width=0.32\linewidth, trim=1.5cm 0.4cm 2.0cm 0.0cm]{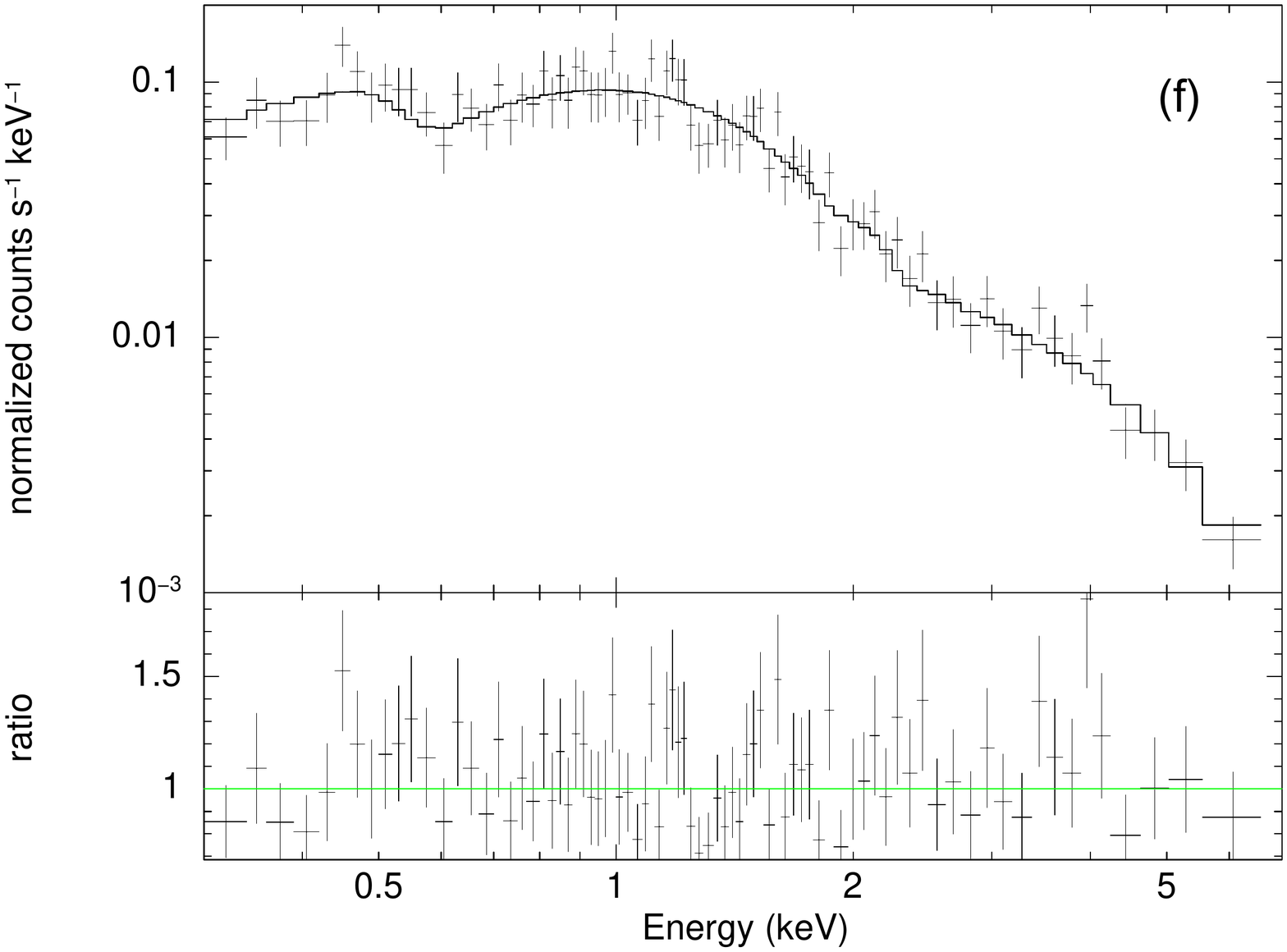}
\caption{Representative single-epoch Swift spectra of OJ 287. The spectra are binned for clarity. 
%% either 20 photons/bin, or 10 photons/bin.
Single power-law fits and their residuals are shown in the six panels, which represent different brightness and spectral states of OJ 287. Upper row, from left to right: {\bf{(a)}} outburst state (1 cts/s; MJD 57800; 2017 February 16), {\bf{(b)}} post-outburst state (0.4 cts/s; MJD 57852; 2017 April 9), {\bf{(c)}} low state (0.1 cts/s; MJD 54396; 2007 October 23).
Lower row, from left to right: {\bf{(d)}}  intermediate--low state (0.2 cts/s; MJD 58589; 2019 April 16) and {\bf{(e)}} intermediate--low state (0.2 cts/s; MJD 59172; 2020 November 19). The last figure {\bf{(f)}} shows the X-ray spectrum during the UV--optical deep fade (MJD 58033 -- MJD 58090; 2017 October 7 -- December 3). The single-epoch spectra were merged into a single spectrum. 
See also Tab. \ref{tab:spectral-fits}. }
    \label{fig:Swift-spectra}
\end{figure*}

The photon index varies between $\Gamma_{\rm X}$=1.5--3.0 (Fig. \ref{fig:lc-Swift-fluxes2005} and \ref{fig:lc-Swift-fluxes2016}). Some representative spectral fits of selected single-epoch spectra are shown in Tab. \ref{tab:spectral-fits} and Fig. \ref{fig:Swift-spectra}. 
In addition, we have extracted and merged all X-ray spectra during the 2017 UV--optical deep fade (MJDs 58033 – 58090) in order to obtain a single spectrum of better quality. It is well fit by a power law of photon index $\Gamma_{\rm X}=2.0\pm{0.1}$ (Tab. \ref{tab:spectral-fits}), further discussed below (Sect. 6.5). 

\begin{table*}
%\scriptsize
	\centering
	\caption{Representative Swift X-ray power-law fit results for the single-epoch spectra shown in Fig. \ref{fig:Swift-spectra}. 
	Absorption was fixed at the Galactic value, $N_{\rm H, Gal}$.  
}
\label{tab:spectral-fits}
\begin{tabular}{llccccl}
	\hline
	MJD & Date & CR [cts/s] & $\Gamma_{\rm x}$ &  $\chi^2$/d.o.f & W-stat/d.o.f & comments \\
	(1) & (2) & (3) & (4)  & (5) & (6) \\
	\hline
	54396 & 2007-10-23 & 0.10 & 1.5$\pm{0.1}$ & - & 207.8/234 & low state 
	\\
	57800 & 2017-02-16 & 1.07 & 2.8$\pm{0.1}$ & 28.8/42 & - & outburst \\
	57852 & 2017-04-09 & 0.40 & 2.4$\pm{0.1}$ & 27.8/29 & - & decline \\  %% near EHT
	58589 & 2019-04-16 & 0.20 & 1.9$\pm{0.1}$ & - & 105.4/107 & % near GMVA
	\\	
	58801 & 2019-11-14 & 0.21 &     1.7$\pm{0.2}$ & - & 61.8/68 &  flattest recent\\
	59172 & 2020-11-19 & 0.22 & 2.2$\pm{0.1}$ & - & 94.9/123 & %random recent 
	\\
	 \hline
	58033 -- 58090 & 2017-10-07 -- 2017-12-03 & 0.20 & 2.0$\pm{0.1}$ & 67.9/79 & - & UV--optical deep fade, merged \\
		\hline
	\end{tabular}
\end{table*}

\subsection{Swift UVOT data analysis}

We have also observed OJ 287 with the UV--optical telescope \citep[UVOT;][]{Roming2005} in all three optical and all three UV photometric bands
[with filters V (5468\AA), B(4392\AA), U(3465\AA),  UVW1(2600\AA), UVM2(2246\AA), and UVW2(1928\AA), where values in brackets are the filter central wavelengths \citep{Poole2008}] since the end of 2015 in order to obtain SED information of this rapidly varying blazar and measure interband time lags. All public archival data since 2005 were added to the analysis.

In each UVOT filter, the observations were first coadded using the tool \textsc{uvotimsum}. Source counts in all six filters were then extracted in a circular region of 5\arcsec~radius centered on OJ 287.
The background was selected in a nearby region of 20\arcsec~radius.
% The tool {\em{ uvotsource}} was used to measure the magnitudes. 
The background-corrected counts were then converted into VEGA magnitudes and 
fluxes based on the latest calibration as described by \citet{Poole2008} and \citet{Breeveld2010}.  
All fluxes are reported as flux density multiplied by the central frequency of the corresponding filter.
For data since 2017, the recently released {\sc{caldb}} update, version 20200925, was employed.{\footnote{\url{https://www.swift.ac.uk/analysis/uvot/index.php}}}

Correction of the UVOT data for Galactic reddening was carried out assuming $E_{\rm{(B-V)}}$=0.0248 \citep{Schlegel1998} and using a correction factor in each filter according to Equ. (2) of \citet{Roming2009}. The reddening curves of \citet{Cardelli1989} were adopted. 

\begin{figure}
\includegraphics[clip, trim=1.8cm 1.6cm 2.2cm 0.3cm, angle=-90, width=\columnwidth]{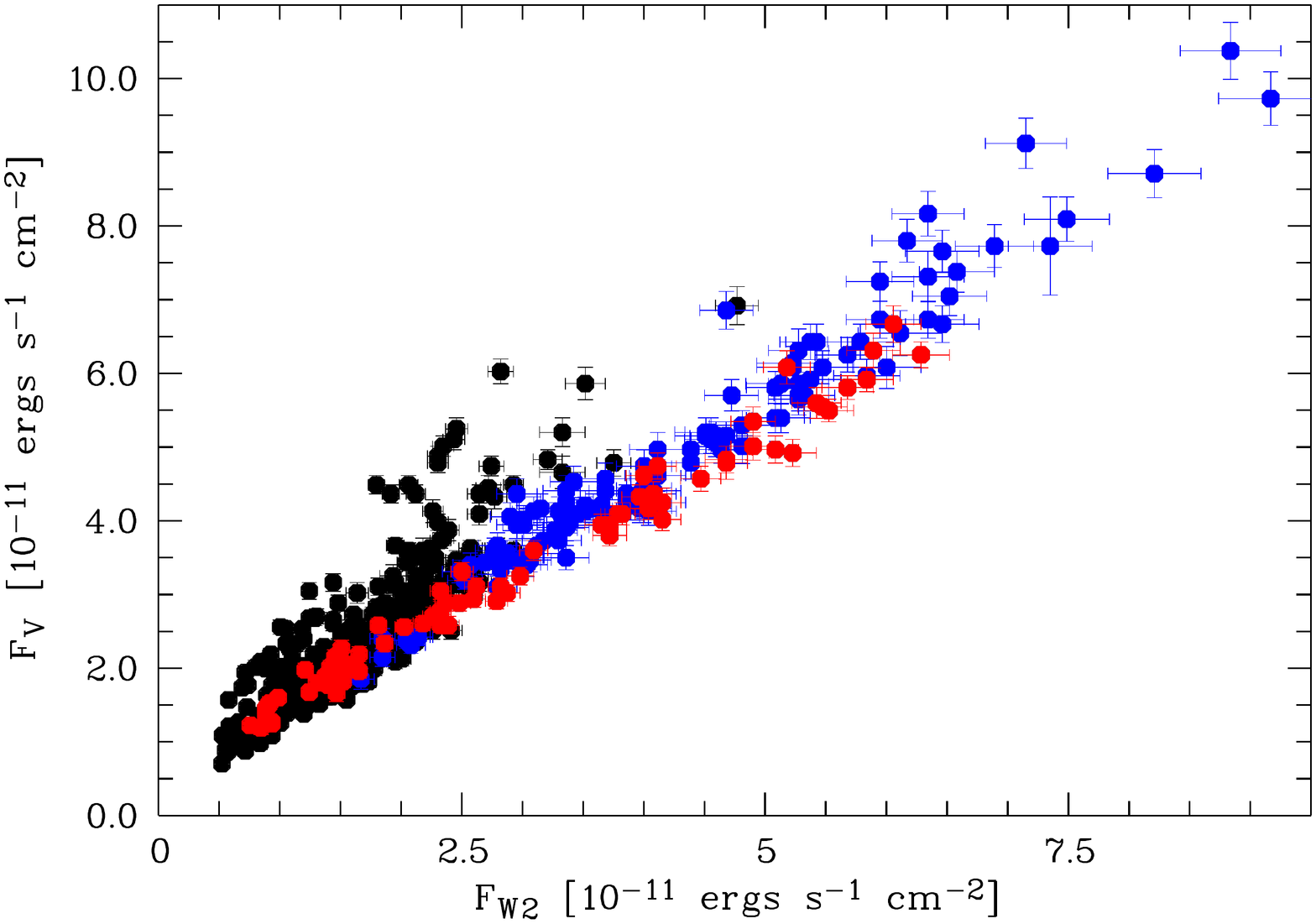}
	\caption{Optical--UV flux correlation based on the Swift data of OJ 287. Optical and UV bands are tightly correlated. Outburst epochs are marked in color (2016/17: blue, 2020: red). 
     }
    \label{fig:fluxcorr}
\includegraphics[clip, trim=1.7cm 2.0cm 1.9cm 0.3cm, angle=-90, width=\columnwidth]{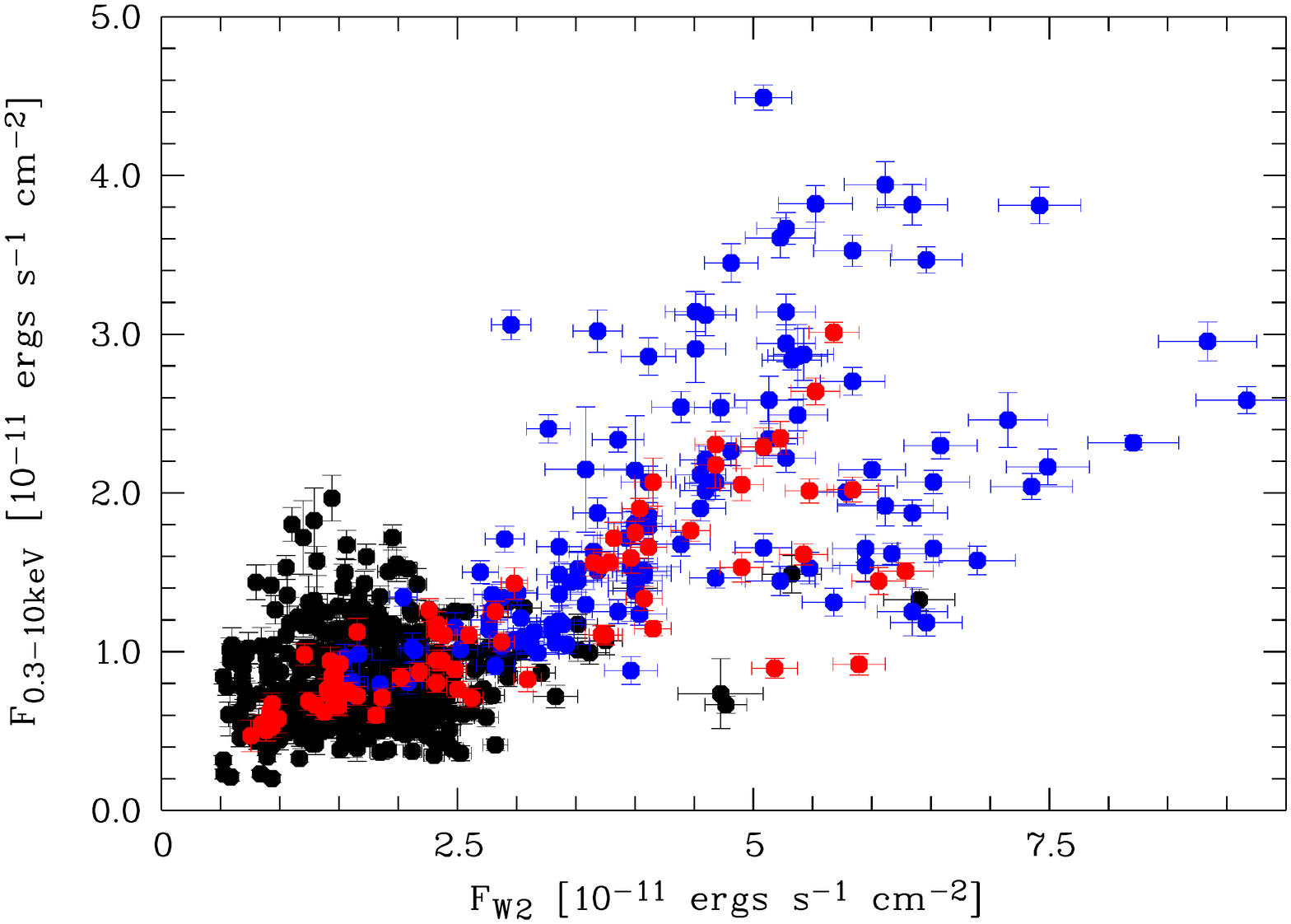}
\caption{UV--X-ray flux correlation based on the Swift data of OJ 287. Outburst epochs are marked in color (2016/17: blue, 2020: red).}
\label{fig:fluxcorr2}
\end{figure}

\subsection{2005--2021 Swift light curve} 

The long-term Swift UVOT and XRT light curve between 2005 and 2021 is shown in Figs \ref{fig:lc-Swift-CR}, \ref{fig:lc-Swift-fluxes2005}, and \ref{fig:lc-Swift-fluxes2016}. 
In Fig. \ref{fig:lc-Swift-CR} we provide the directly observed quantities; the X-ray count rate  and the UVOT magnitudes in the VEGA system uncorrected for reddening. This figure displays the long-term trends of variability and visualizes the  brightness of OJ 287 in different instrumental bands. 
Figs \ref{fig:lc-Swift-fluxes2016} and \ref{fig:lc-Swift-fluxes2005} present an overview of the long-term evolution of the observed (absorption and extinction-corrected) fluxes in each band.  Epochs of special interest are marked in color. The most densely covered epochs are not resolved in these plots.
Annual light curves are displayed in Fig. \ref{fig:lc-Swift-2015}--\ref{fig:lc-Swift-2019} in Appendix A. These serve at resolving finely the epochs of dense coverage, and the dynamic range of each light curve is adjusted to the annual flux range.   

The Swift light curve of OJ 287 between 2005 and 2021 is highly variable. The X-ray flux varies by a factor of 28 in total, the optical--UV flux by a factor 18. 
Several long- and short-term features stand out in the Swift light curve:
In addition to two major X-ray--UV--optical outbursts in 2016-17 and 2020 \citep{Komossa2017,Komossa2020}, OJ 287 exhibits many low-amplitude flares that occur on 
biweekly to monthly time intervals. 
The amplitude of variability of these ``mini flares" is remarkably constant between 2009 and 2021, with CR=0.1 cts/s at minima and CR=0.5 cts/s in X-rays at maxima, implying a very constant underlying emission mechanism.  
The X-ray count rate  rarely drops below 0.1 cts/s. 
An exception is the epoch in 2005-2007, where OJ 287 was very sparsely covered with Swift \citep{Massaro2008}, but was always found in a deep low state
with CR=0.05--0.1 cts/s, implying an epoch of particular inactivity. 
At the end of 2017, a remarkable, deep symmetric UV--optical minimum is observed 
(marked blue in Figs. \ref{fig:lc-Swift-CR} and \ref{fig:lc-Swift-fluxes2016}), not accompanied by any similar drop in X-rays. 
Another optical--UV low state is seen in 2020 September when OJ 287 was observable with Swift again after Swift Sun constraint.{\footnote{In line with the wording in this paragraph, throughout this article we continue to refer to different flux states as follows: the two outbursts (e.g., Fig. \ref{fig:Swift_Xlight-MOMO}) are events lasting for months with a flux increase, from pre-outburst to maximum, by a factor $>5$. The deep low state in X-rays refers to the long-lasting epoch with CR$<$0.1 cts s$^{-1}$ when X-rays were faintest. The deep low state (deep fade) in the UV--optical refers to the epoch of the symmetric dip when the UV--optical was faintest. If the flux is persistently low in subsequent epochs, but not as low as during the deep fade within the errors, we loosely refer to such a state as low state (without any further interpretation). Finally, when we make comparisons with theoretical predictions of the binary model, we adopt the terminology commonly used in the context of that model; specifically, we use the terms impact flare and after-flare (see Sects 1 and 6.1).}} 

\subsection{UVOT--XRT flux correlations}

The X-ray and UV fluxes of OJ 287 are closely correlated, especially during the two outbursts in 2016/17 and 2020 (Fig. \ref{fig:fluxcorr2}).
An even closer correlation is detected between the UV and optical fluxes of OJ 287 at all epochs (Fig. \ref{fig:fluxcorr}). 
A Spearman rank-order correlation analysis for UV-W2 and V flux data gives a correlation 
coefficient $r_{\rm s} = 0.93$ and a Student's T-test of $T_{\rm S} = 63.7$. 
For $N$= 605 data points, this corresponds to a probability of a random result of $P<10^{-8}$. 
For the UV-W2 and X-ray fluxes we find $r_{\rm s} = 0.56 , T_{\rm S} = 17.6$, and $P<10^{-8}$ ($N$=655). 
This result implies closely correlated emission mechanism(s). The characteristic variability properties of OJ 287 are analyzed in greater detail in Sects. 3--5.  

\begin{table*}
%\scriptsize
	\centering
	\caption{Fractional variability amplitude $F_{\rm var}$ of the Swift UVOT and XRT fluxes of OJ 287. }
	\label{tab:Fvar}
	\begin{tabular}{lccccc} 
		\hline
		 & Epoch 1 & Epoch 2 & Epoch 3 & Epoch 4 & Epoch 5  \\
		 & 2015 Dense Coverage & 2016/17 Outburst & 2017/18 Quiescence & 2018/19 Quiescence & 2019/20 Outburst \\
		\hline
$f_{\mathrm{x}}$ & $0.232\pm0.012$ & $0.418\pm0.009$ & $0.244\pm0.011$ & $0.199\pm0.015$ & $0.517\pm0.007$ \\
$f_{\mathrm{W2}}$ & $0.183\pm0.005$ & $0.317\pm0.005$ & $0.289\pm0.005$ & $0.180\pm0.005$ & $0.605\pm0.005$ \\
$f_{\mathrm{M2}}$ & $0.190\pm0.005$ & $0.322\pm0.006$ & $0.283\pm0.006$ & $0.175\pm0.006$ & $0.593\pm0.005$ \\
$f_{\mathrm{W1}}$ & $0.189\pm0.004$ & $0.320\pm0.006$ & $0.280\pm0.005$ & $0.175\pm0.005$ & $0.578\pm0.005$ \\
$f_{\mathrm{U}}$ & $0.192\pm0.004$ & $0.312\pm0.005$ & $0.279\pm0.005$ & $0.174\pm0.005$ & $0.542\pm0.005$ \\
$f_{\mathrm{B}}$ & $0.196\pm0.004$ & $0.302\pm0.005$ & $0.278\pm0.005$ & $0.180\pm0.005$ & $0.500\pm0.005$ \\
$f_{\mathrm{V}}$ & $0.206\pm0.004$ & $0.303\pm0.004$ & $0.271\pm0.006$ & $0.176\pm0.006$ & $0.479\pm0.005$ \\
		\hline
%\begin{tablenotes}
%{Notes: }
%\end{tablenotes}
	\end{tabular}
\end{table*}

\section{Fractional variability amplitude}

\subsection{Swift}

We have calculated the fractional rms variability amplitude $F_{\rm var}$ \citep{Vaughan2003} separately for each 
UVOT filter and in X-rays (Tab. \ref{tab:Fvar}). To do so, we have split the total data set since 2015 into five subsets 
that represent different activity states of OJ 287, and different coverage: the epoch of dense monitoring in 2015 
(epoch 1 hereafter); two epochs including the 2016/17 and 2020 outbursts, respectively (epoch 2 and 5); and two epochs 
of relative quiescence in 2017--2018 and 2018--2019 (epochs 3 and 4). Epochs 2--5 of 9 month duration each are separated by $\sim$3 months 
each where OJ 287 is unobservable by Swift due to the Sun constraint. These same epochs are also used for the 
analyses in Sects 4 and 5. 
$F_{\rm var}$ was calculated as
\begin{equation}
F_{\rm var} = \sqrt{S^2 - \overline{\sigma^{2}_{\rm err}} \over \overline{x}^2},
\end{equation}
where $S^{2}$ is the variance of the light curve, $\overline{\sigma^{2}_{\rm err}}$ is the mean square of the
measurement errors, and $\overline{x}$ is the mean flux. The error of $F_{\rm var}$ was calculated according 
to the prescription outlined in Appendix B of \citet{Vaughan2003} as
\begin{equation}
\sigma_{F_{\rm var}} = \sqrt{ \left(\sqrt{1 \over 2N} {\sigma^{2}_{\rm err} \over \overline{x}^2 F_{\rm var}} \right)^2 + \left( \sqrt{\sigma^{2}_{\rm err} \over N} {1 \over \overline{x}^2} \right)^2 },
\end{equation}
where $N$ is the number of data points used in the computation of $F_{\rm var}$.

Results (Tab. \ref{tab:Fvar}) show that the optical, UV, and X-rays show overall similar values of $F_{\rm var}$ and therefore arise from 
closely correlated processes in cospatial regions.
During the epochs of outbursts, $F_{\rm var}$ is higher than during the epochs of low-level activity.  In X-rays, the average value during epochs of outburst and quiescence is $F_{\rm var,x}$ = 0.52 $\pm$ 0.01 and 0.27 $\pm$ 0.01, respectively. 

\subsection{XMM-Newton}

\begin{figure}
\includegraphics[width=\columnwidth]{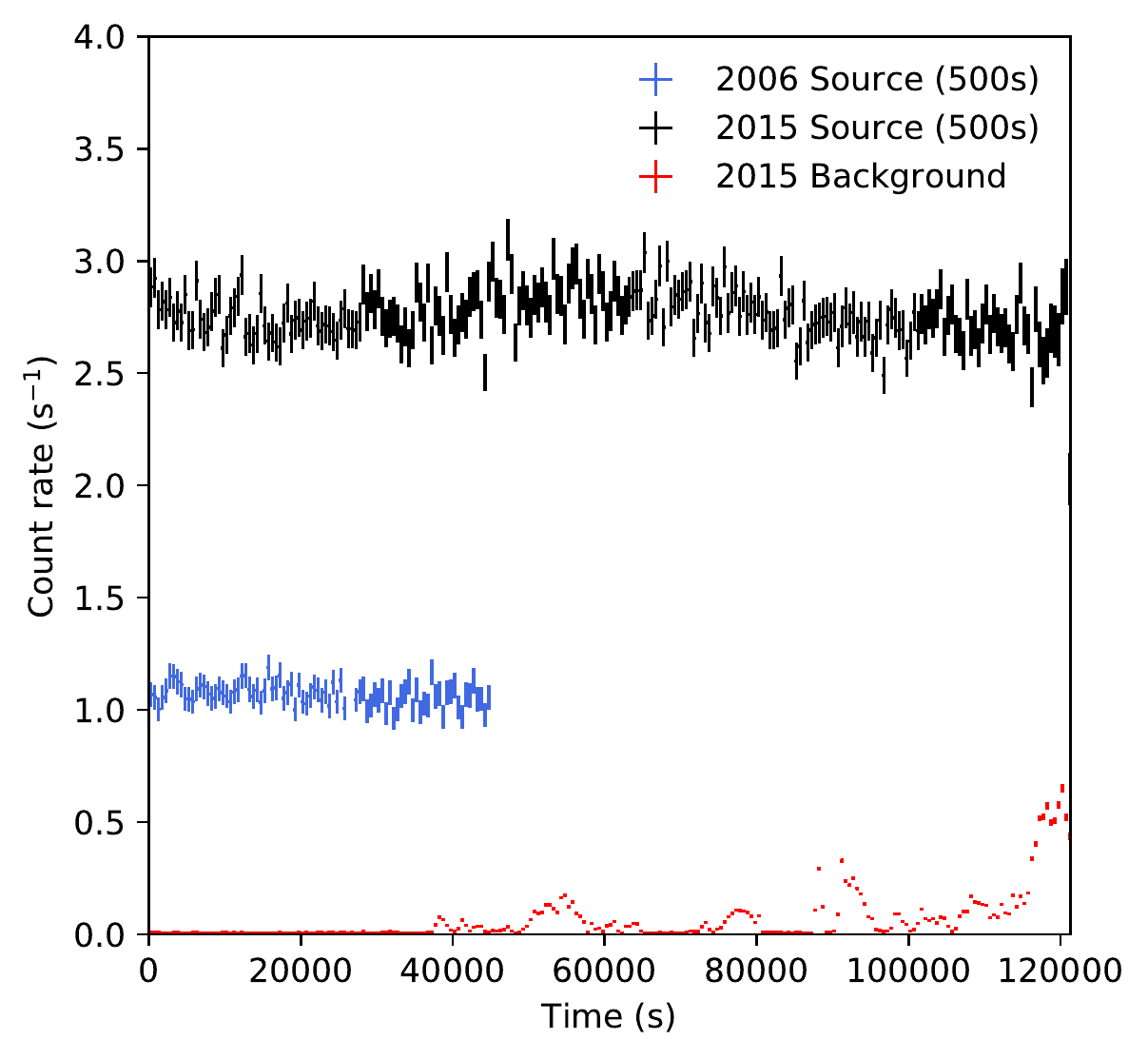}
	\caption{XMM-Newton light curves of OJ 287 at 500s binning during the two longest observations in 2006 November (green) and 2015 May (black). The background during 2015 is shown in red. During 2006 it was constant and completely negligible and is not plotted.}
    \label{fig:XMM-lightcurve}
\end{figure}

During the 2020 outburst, we used our XMM-Newton observation to search for short-time variability within the observation and did not detect any \citep{Komossa2020}. This is in contrast with the high variability detected with Swift on longer timescales. 
We have therefore reanalyzed the two longest XMM-Newton observations.  
We compare the 2015 May 7 long-look XMM-Newton light curve (120 ks duration) with the 2006 November 17 XMM-Newton light curve (46 ks duration) of OJ 287 in Fig. \ref{fig:XMM-lightcurve}. The 2015 observation  was affected by a flaring particle background but at low levels (Fig. \ref{fig:XMM-lightcurve}), which only affect the hard band. 
The 2006 observation was unaffected by background flaring. 
During 2015, $F_\mathrm{var}$=$0.033\pm0.002$.  
During 2006, no significant variability is detected. 
$F_\mathrm{var}$=$0.01\pm0.01$ (0.3-10 keV band).  These values are consistently much smaller than those measured with Swift at timescales of days to weeks. 

\begin{figure*}
\includegraphics[clip=, trim=1.5cm 1.5cm 2.55cm 2.7cm, width=0.33\linewidth]{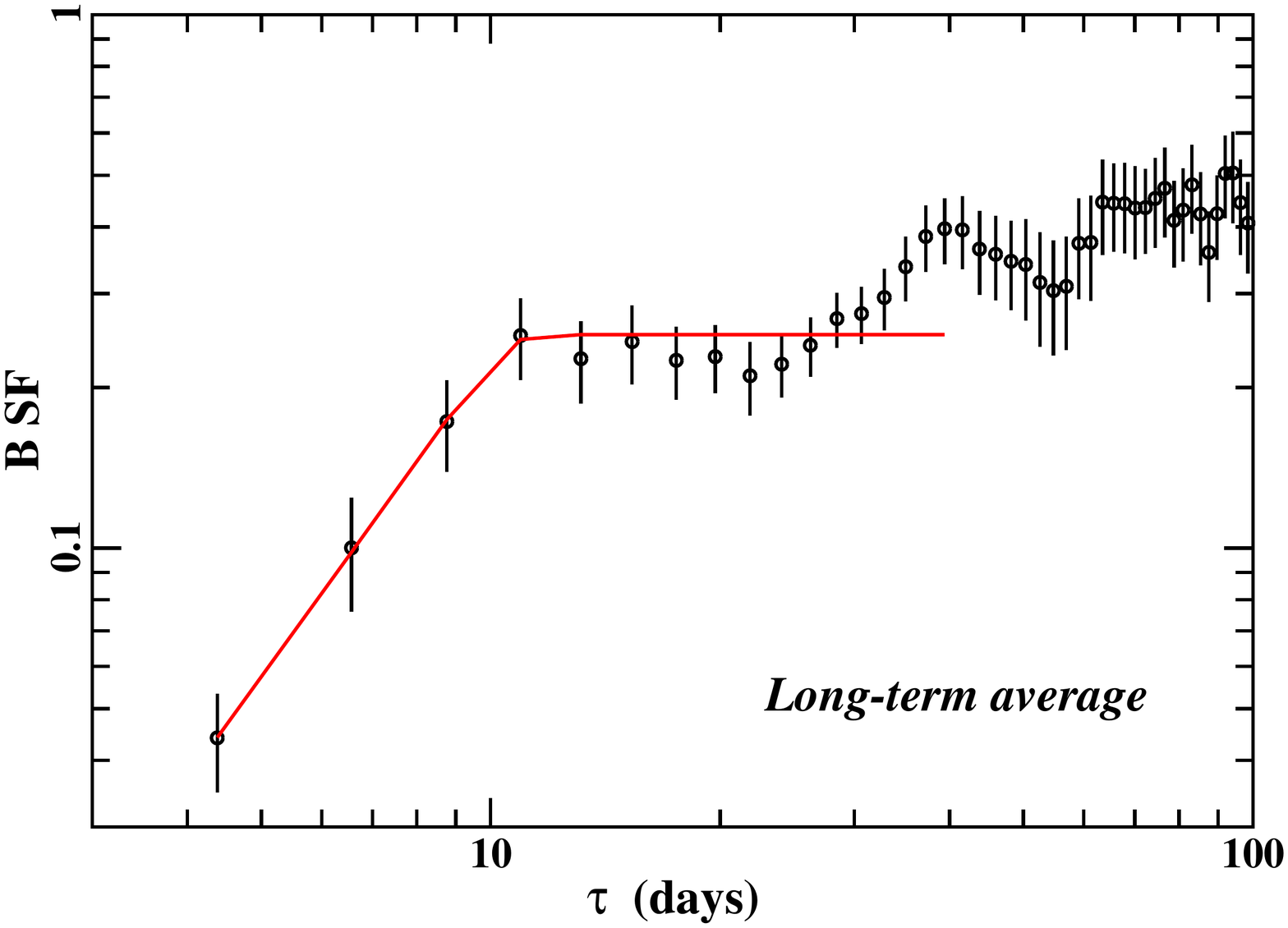}
\includegraphics[clip=, trim=1.5cm 1.5cm 2.55cm 2.7cm, width=0.33\linewidth]{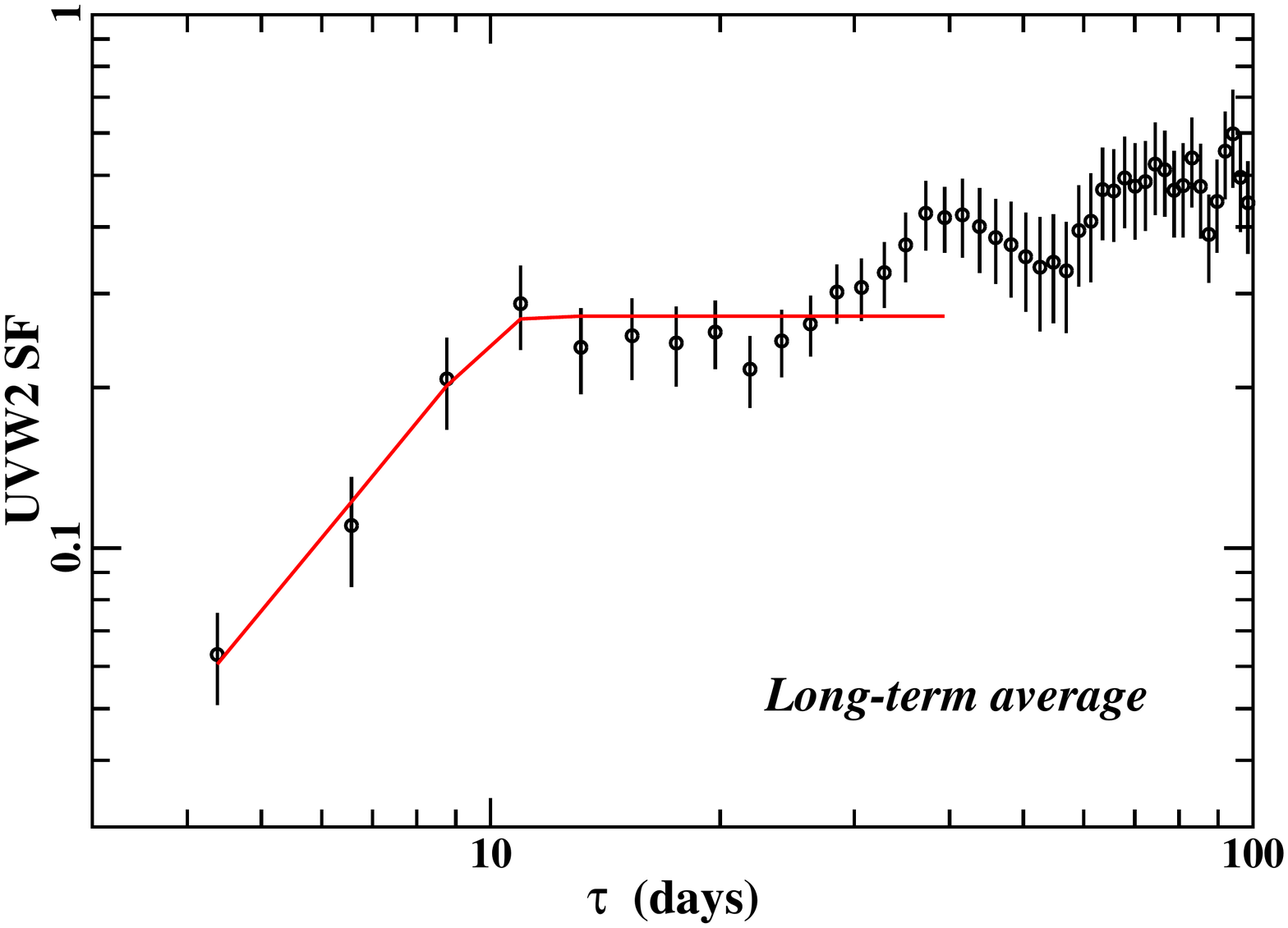}
\includegraphics[clip=, trim=1.5cm 1.5cm 2.55cm 2.7cm,width=0.33\linewidth]{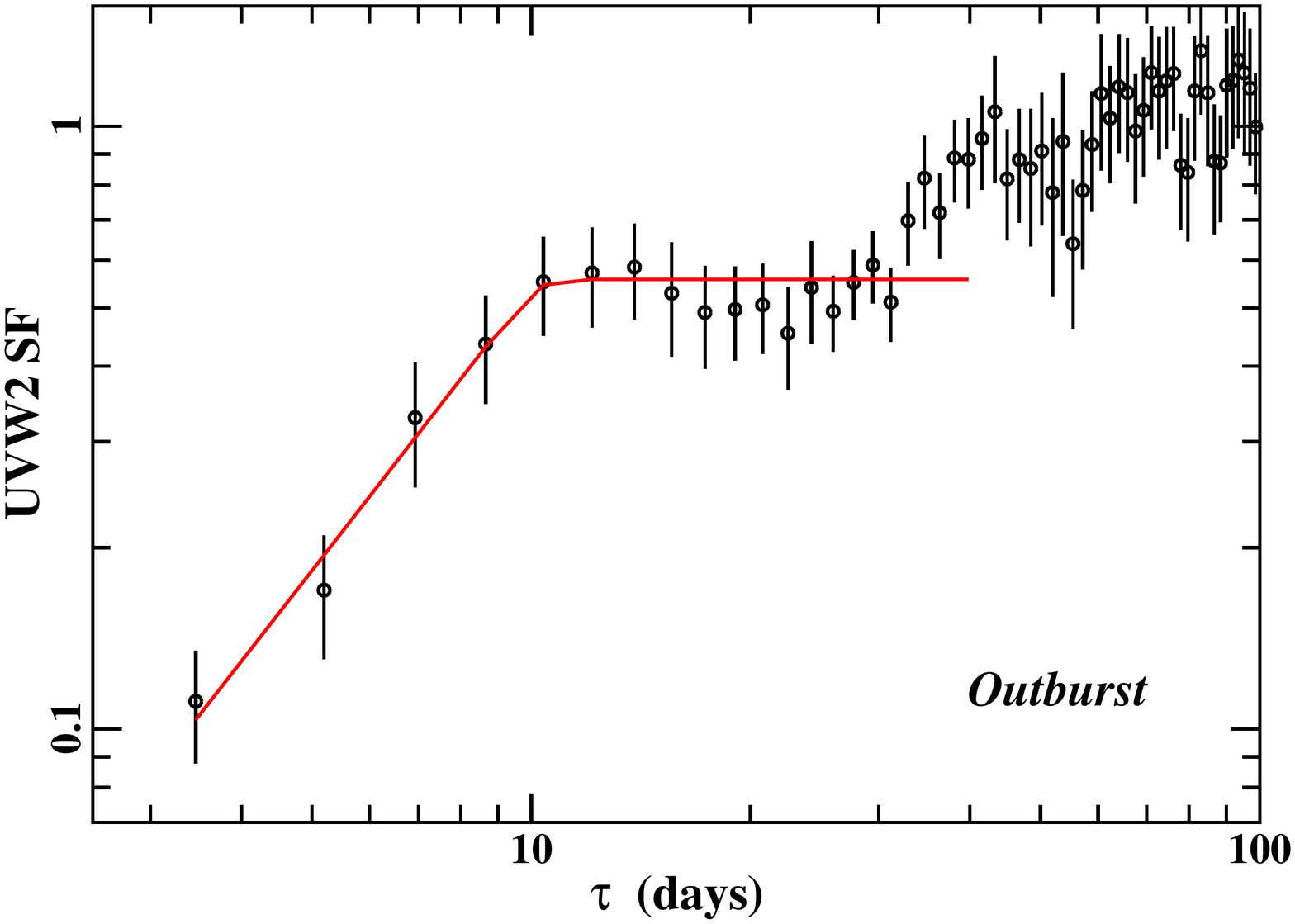}
\includegraphics[clip=,trim=1.5cm 1.5cm 2.55cm 2.7cm, width=0.33\linewidth]{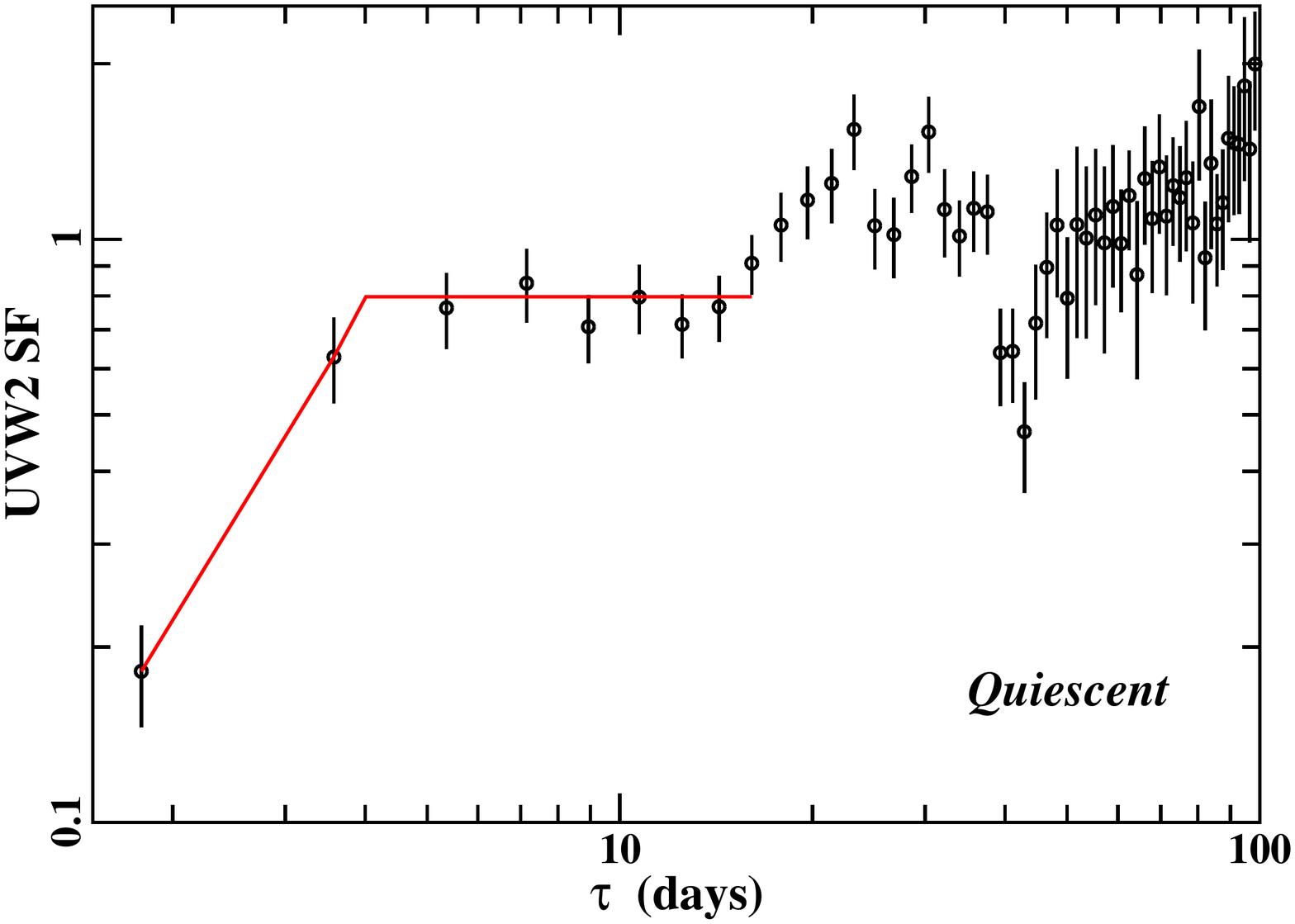}
\includegraphics[clip=,trim=1.5cm 1.5cm 2.55cm 2.7cm, width=0.33\linewidth]{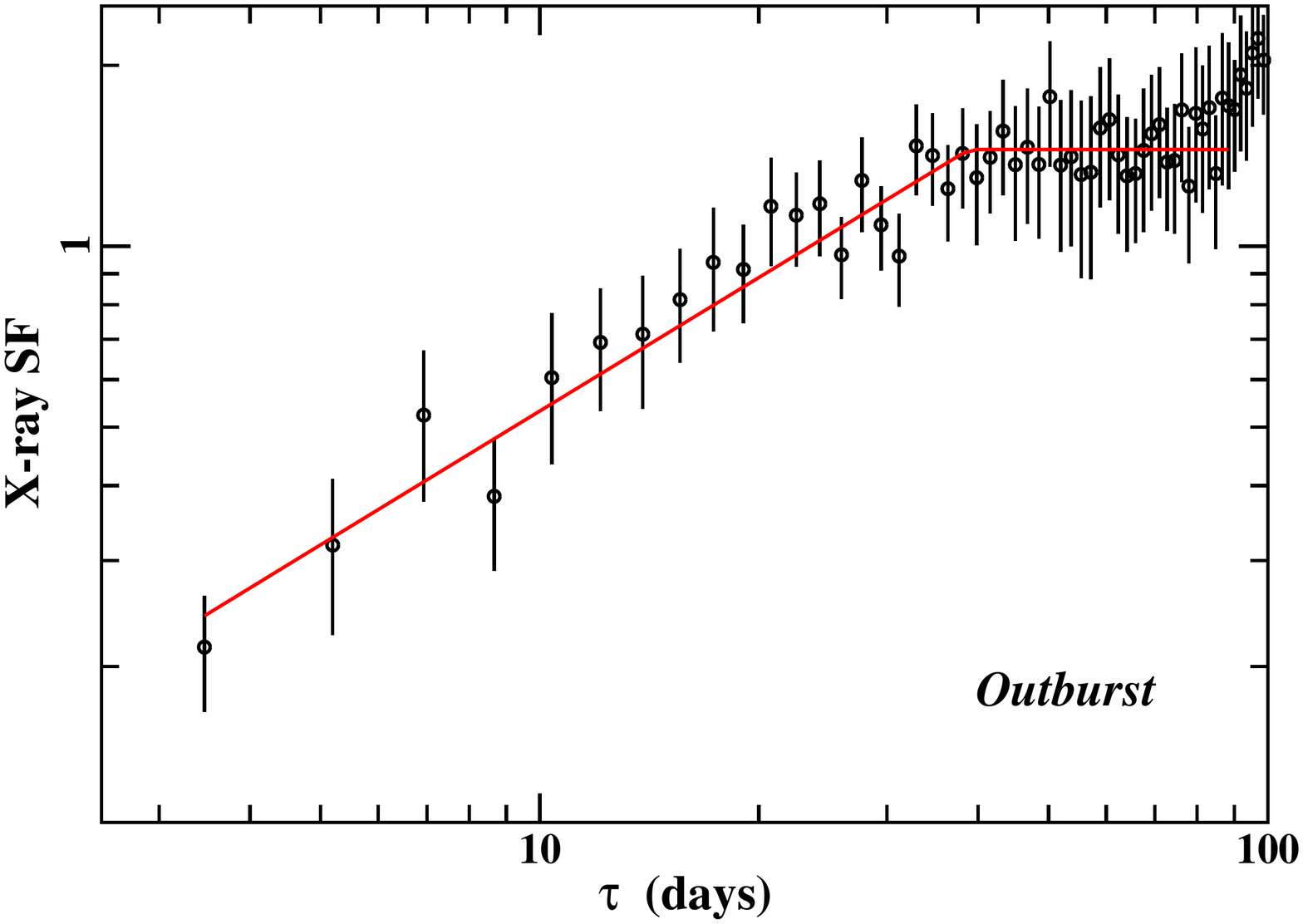}
\includegraphics[clip=,trim=1.5cm 1.5cm 2.55cm 2.7cm, width=0.33\linewidth]{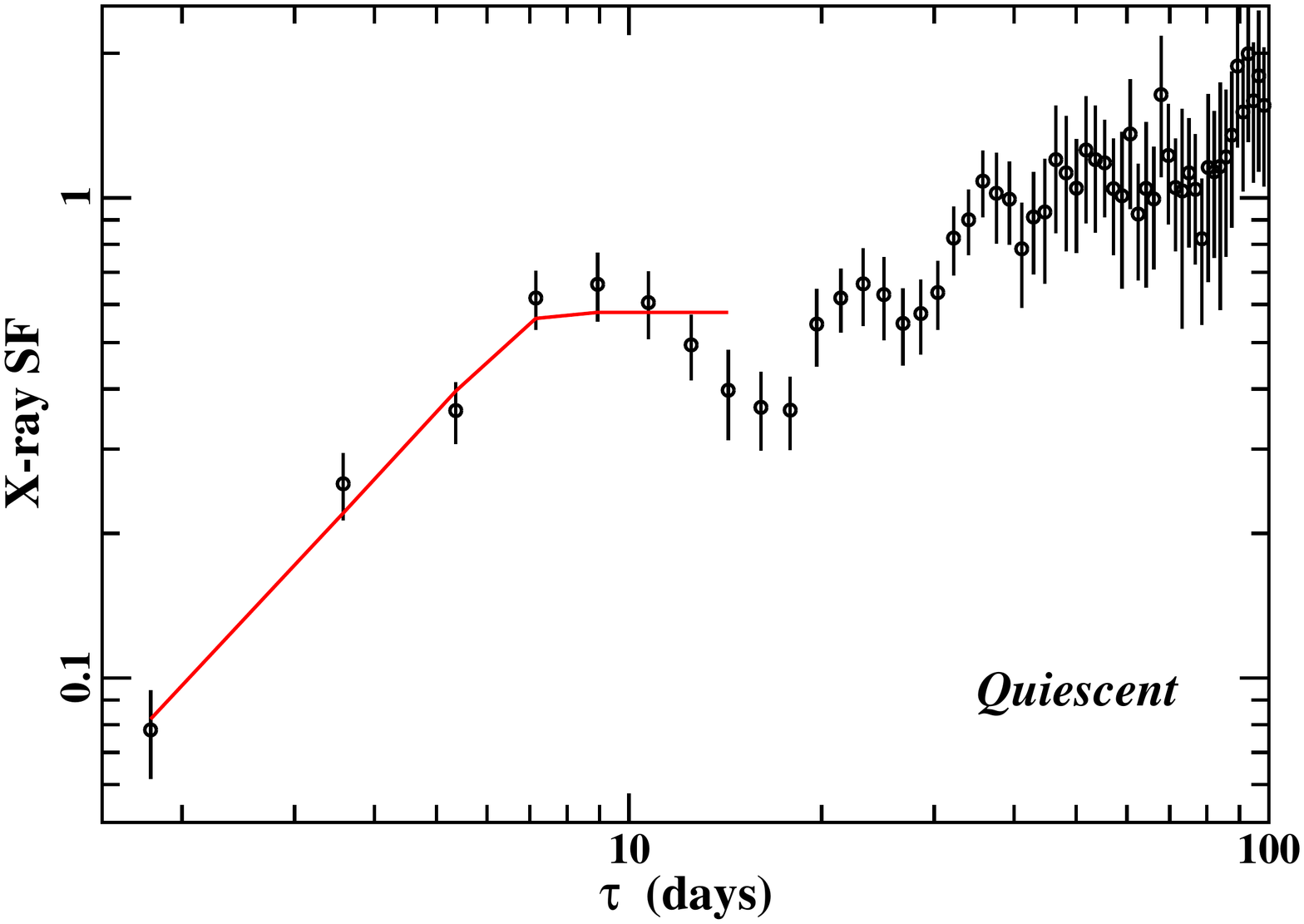}
\includegraphics[clip=,trim=1.5cm 1.5cm 2.55cm 2.7cm, width=0.33\linewidth]{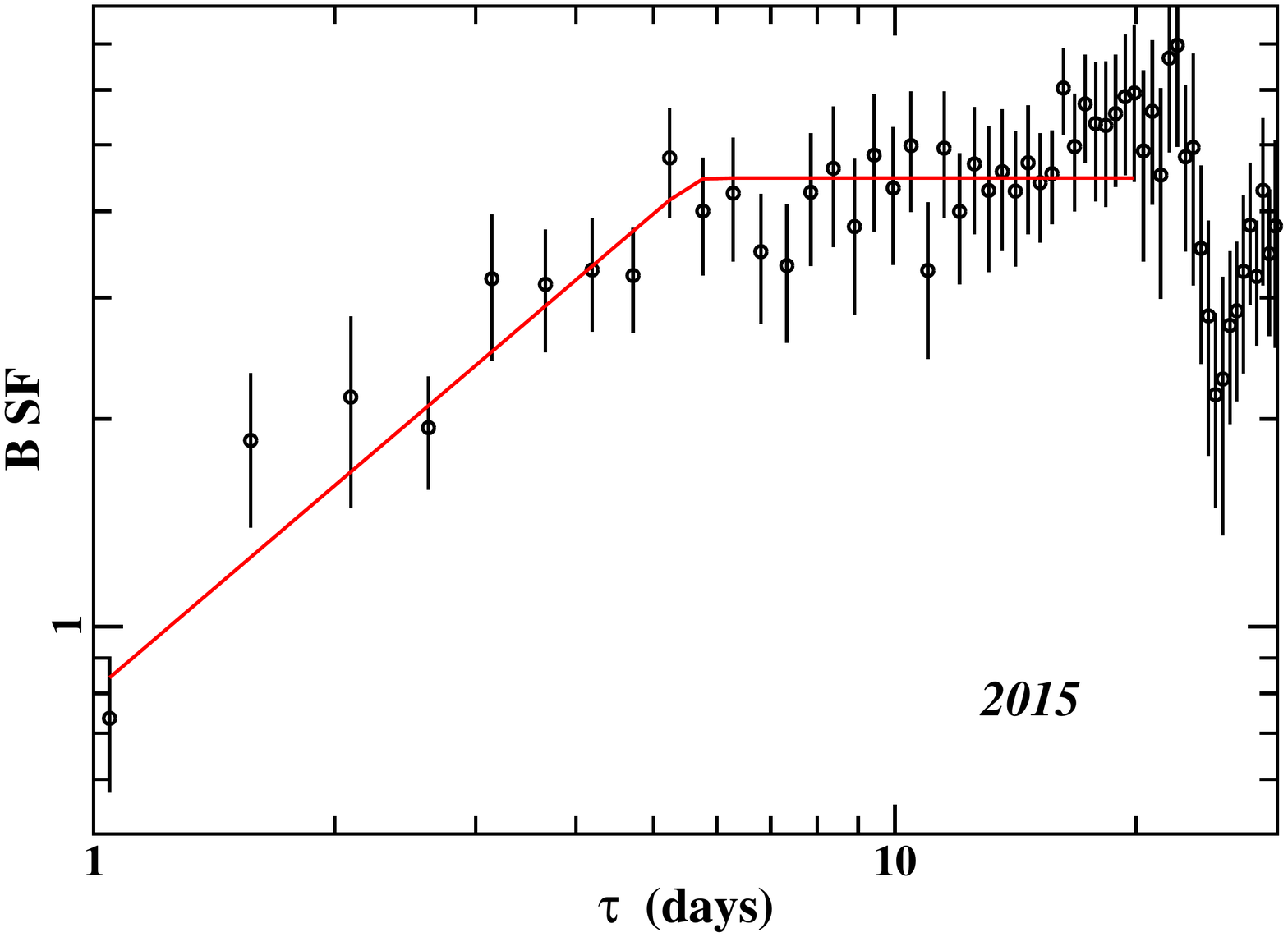}
\includegraphics[clip=,trim=1.5cm 1.5cm 2.55cm 2.7cm, width=0.33\linewidth]{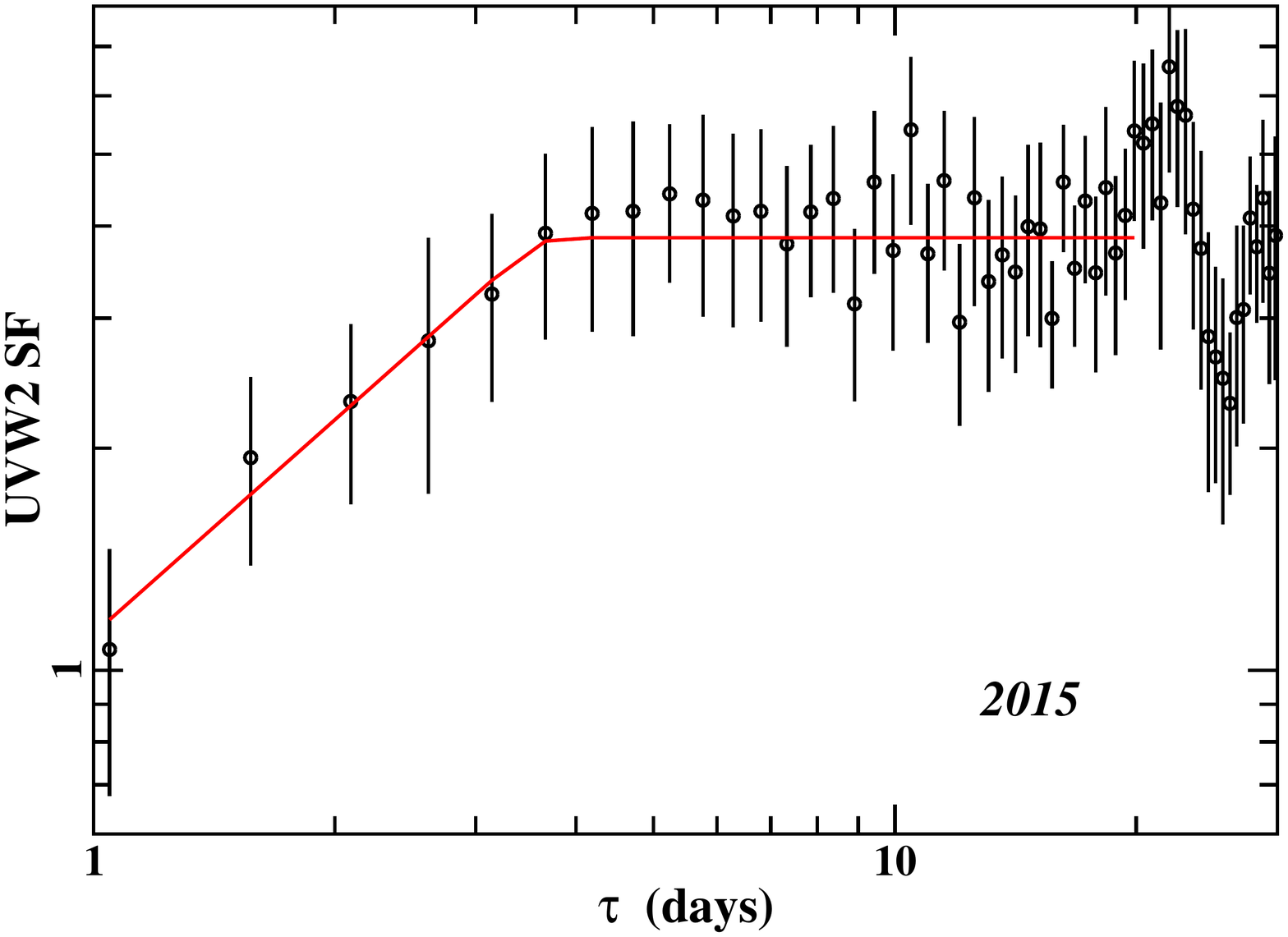}
\includegraphics[clip=,trim=1.5cm 1.5cm 2.55cm 2.7cm, width=0.33\linewidth]{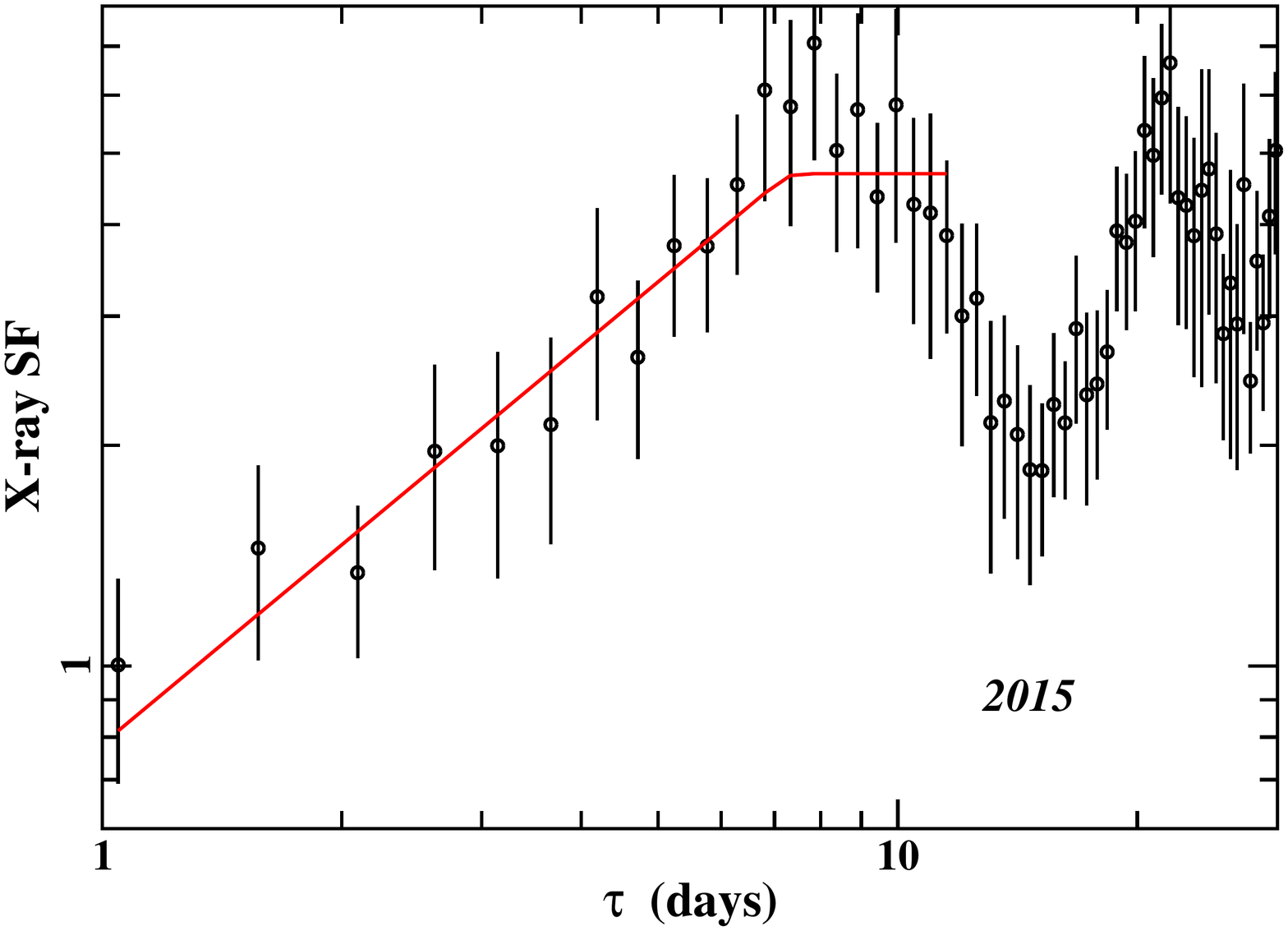}
\caption{Broken power laws fitted to the structure functions of the Swift B, W2, and X-ray flux of OJ 287. The panels show the SF over different flux states. Upper panel: B band long-term average (left), UV-W2 band long-term average (middle), and UVW2 outburst epochs (right). Middle panel: UV-W2 quiescent epochs (left), X-ray outburst epochs (middle), and X-ray quiescent epochs (right). 
Lower panel: B (left), UV-W2 (middle), and X-rays (right) during the densely covered quiescence epoch 1 (note the different scale in $\tau$ in this panel). 
}   
    \label{fig:SF}
\end{figure*}

\section{Structure function}
\label{sec:SF}

Like the power spectral density (PSD), the structure function (SF) measures the distribution of power in time series.  
SFs are commonly applied to analyze the variability of blazars \citep{Simonetti1985, Tanihata2001, Emmanoulopoulos2010} and radio-quiet active galactic nuclei
\citep{DiClemente1996, Collier2001, Gallo2018} in unevenly sampled data sets, where the application of Fourier-transform techniques is problematic. 
The definition used in this work follows \citet{Collier2001} and \citet{Gallo2018}\footnote{https://github.com/Starkiller4011/SFA}:
\begin{equation}
SF\left(\tau\right)=\frac{1}{N\left(\tau\right)}\sum_{i<j}\left[f\left(t_i\right)-f\left(t_i+\tau\right)\right]^{2}
\label{eq:sf}
\end{equation}
where $\tau = t_j - t_i$ is the time differences between pairs of points, $i$ and $j$ (where $j > i$), in a series with $N(\tau)$ pairs. 

Between some minimum and maximum timescale where the variations are correlated, the SF is of power-law shape with index $\beta$. At some break timescale, the power law flattens to approximately two times the time-series variance ($2\sigma^{2}$) (Hughes1992).  At timescales where the SF is no longer well defined, the function will start to exhibit oscillations.

The median temporal sampling of the light curve is $\delta=1.8$~days and is used to define the binning of the SFs. The statistical uncertainties in the SFs are defined as $\frac{\sigma_i}{\sqrt{N_i/2}}$, where $N_i$ is the number of pairs in bin $i$ and $\sigma_i$ is the root mean square deviation about the mean SF value in that bin.

We have fitted all the SFs with a once-broken power law with index $\beta$ and break timescale $t_{\rm break}$.  The index above the break is fixed to zero.  The uncertainties on the measured parameters correspond to the 90 per cent confidence region.  In the figures, all the SFs (except epoch 1) are shown up to 100 days though they are fitted over a smaller range.  For the average and outburst epochs, the X-ray and optical--UV SFs are fitted up to 90 and 40 days, respectively.  In the quiescent phase, the SFs are fitted up to 15 days.

The optical and UV SFs over the entire observing campaign (2015-2020, average of epochs 1--5 defined above) are very similar, with a slope $\beta\approx -2$ and characteristic timescale $t_{\rm break}\approx 10.5$~days,
and we only show results for the B and W2 band (Fig. \ref{fig:SF}). 

Because different processes prevail at epochs of outbursts and quiescence at least in the X-ray band, the SF analysis was carried out separately on epochs 1, 3, and 4 (low-level activity) and epochs 2 and 5 (outbursts). 
In the UV-W2, a slope similar to that of the long-term SF is found.  The break time is $t_{\rm break}$ = 10$^{+3}_{-2}$ days at outbursts
and $t_{\rm break} = 4^{+2}_{-1}$ days in quiescence.
In X-rays, at outbursts, the break time is larger than in the optical-UV; $t_{\rm break}$ = $39^{+10}_{-7}$ days. At epochs of quiescence, the X-ray SF is more similar to the optical-UV SF. It breaks at $t_{\rm break}=7^{+2}_{-1}$ days, but measurements of SFs at quiescent epochs are more uncertain, as noise starts to dominate beyond the break. 

Next, we have obtained the SF for epoch 1. During this epoch, Swift observations of OJ 287 were taken at a very high cadence of $\sim$0.5 days between 2015 April 28 and 2015 June 14 (Fig. \ref{fig:lc-Swift-2015}) with a gap during May 22-25 when OJ 287 was unobservable due to the Swift moon constraint.  We find $t_{\rm break,B} = 6^{+3}_{-1}$ days, $t_{\rm break,W2} = 4^{+2}_{-1}$ days, $t_{\rm break,X} = 7^{+2}_{-2}$ days in the B, UV-W2 and X-rays, respectively. Numbers are very consistent within the errors with those obtained for the other epochs of quiescence that span longer timescales of 9 months at a lower cadence.

To estimate the significance of the break detections in the X-ray light curves we simulated 1000 light curves with similar power spectra as the measured light curves, but without a break.  We then calculated the structure function for each of these light curves and determined the improvement of fitting a broken power law to the SF (in the same manner we did for the data) over fitting a single power law. 
For the quiescent period, only 2 of the 1000 SFs exhibited an improvement of $\Delta \chi^2 \ge 6$, as we found in the measured SF.  This implies a detection at $\sim99.8$ per cent confidence.  Similarly, for the high-cadence 2015 epoch and the outburst period, the breaks were detected at $96.6$ and $99$ per cent confidence, respectively.

Finally, we would like to note the following: Our data since 2016 used for the SF analysis represent the densest coverage ever obtained for OJ 287 in the UV and X-ray bands, spanning multiple years of outbursts and more quiescent phases at cadences as dense as $\sim$1 day. These are therefore the best data we have available to obtain SFs in these bands (an optical PSD was derived by \citet{Wehrle2019} based on optical Kepler data at a quiescent epoch in 2015, discussed further below). However, even these multiyear data we present here have uneven gaps, and especially, they do not cover timescales less than $\sim$0.5--1 day.
We present the SFs with these cautionary comments in mind, as they are the best available and as they will also be used to guide future monitoring campaigns of OJ 287 to improve the coverage even further.

\begin{figure*}
\includegraphics[width=0.49\linewidth]{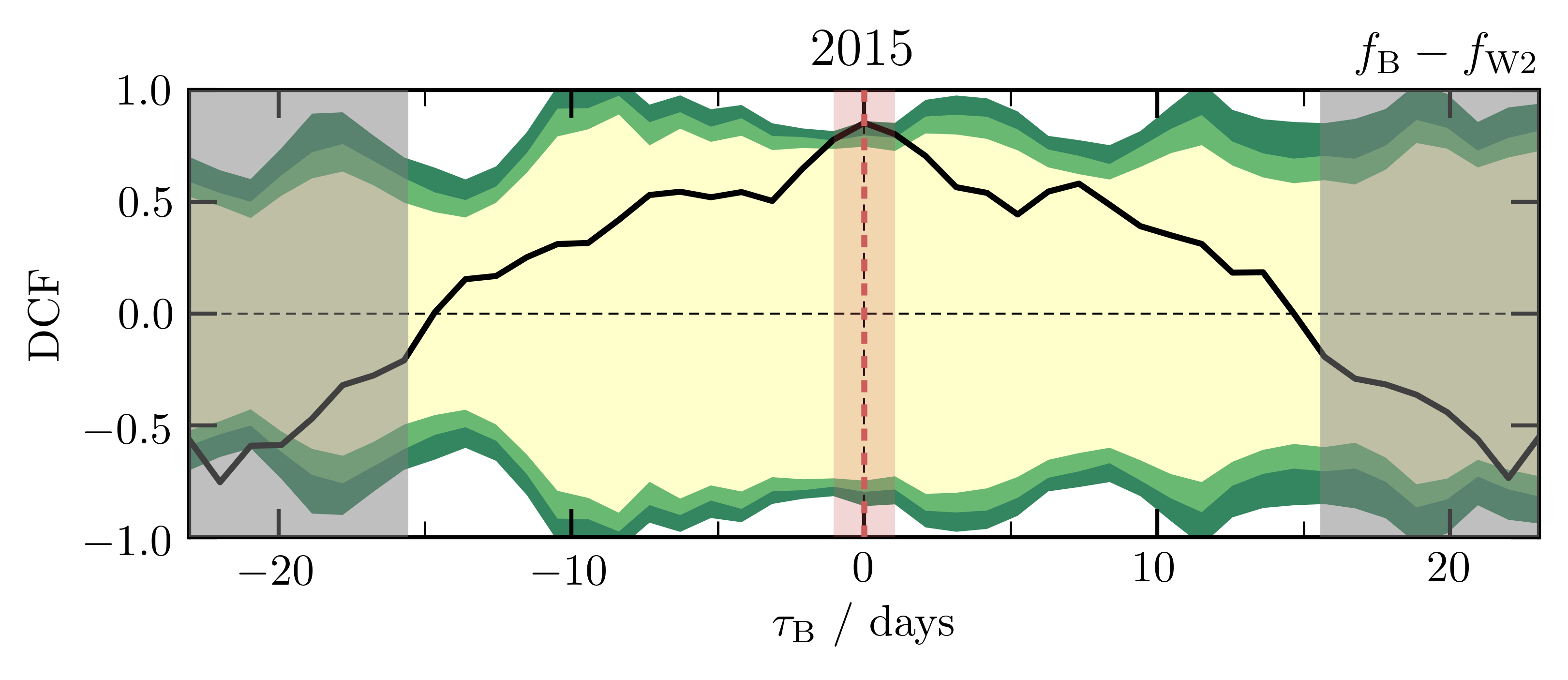}
\includegraphics[width=0.49\linewidth]{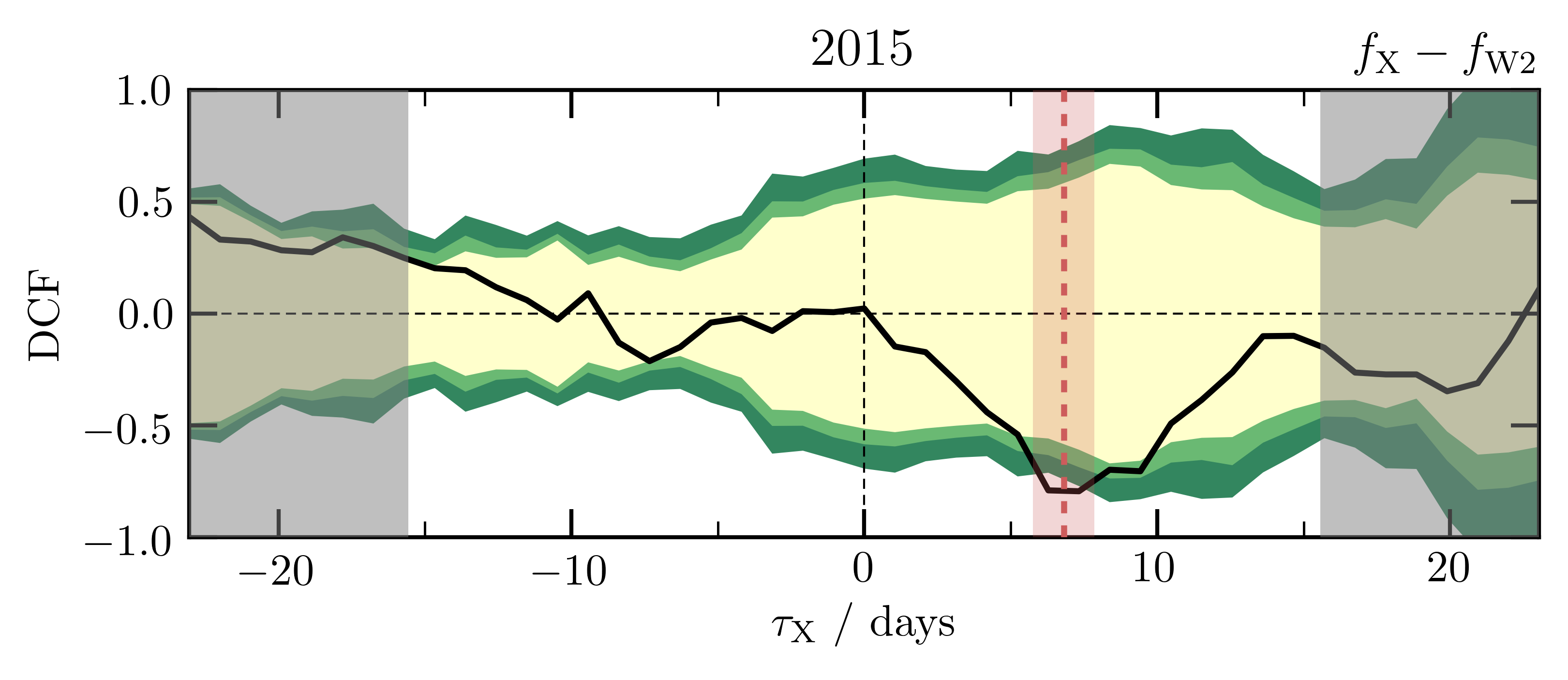}
\includegraphics[width=0.49\linewidth]{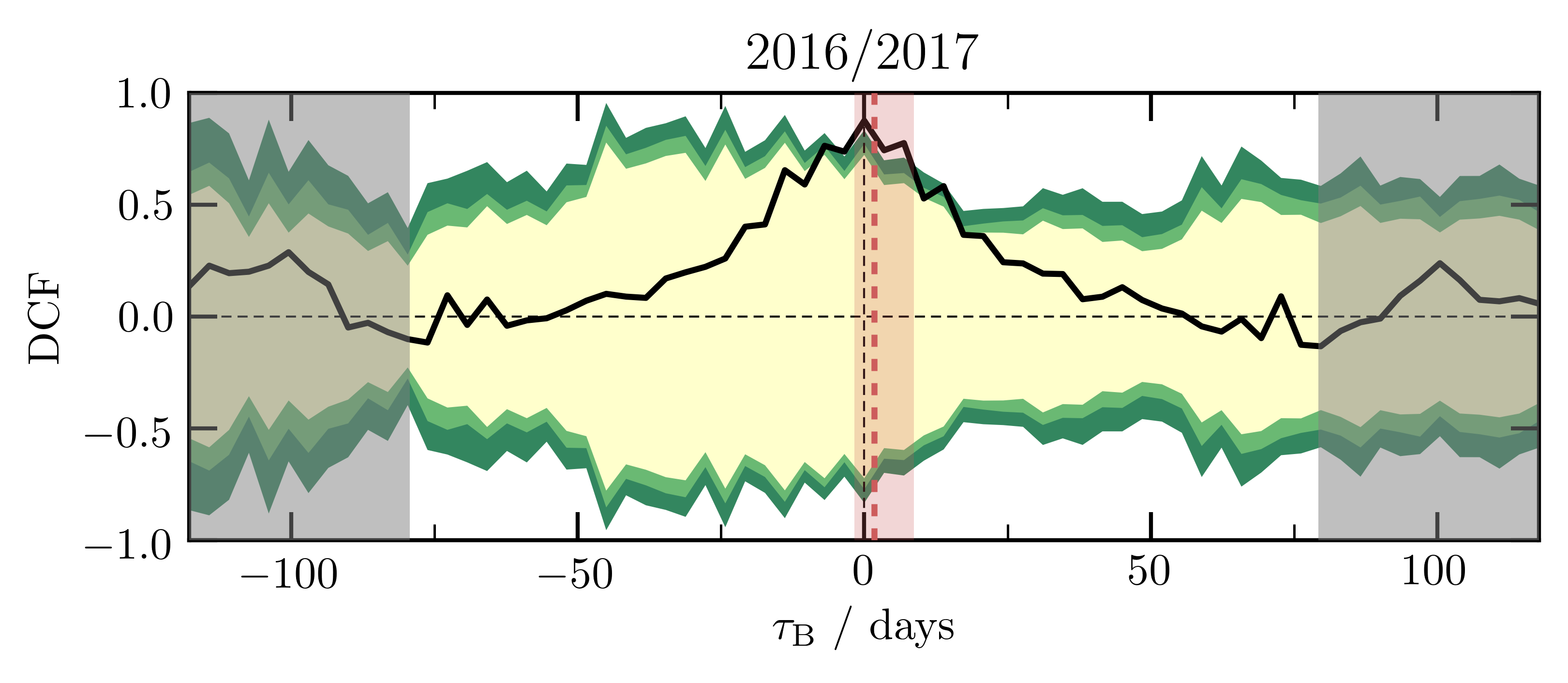}
\includegraphics[width=0.49\linewidth]{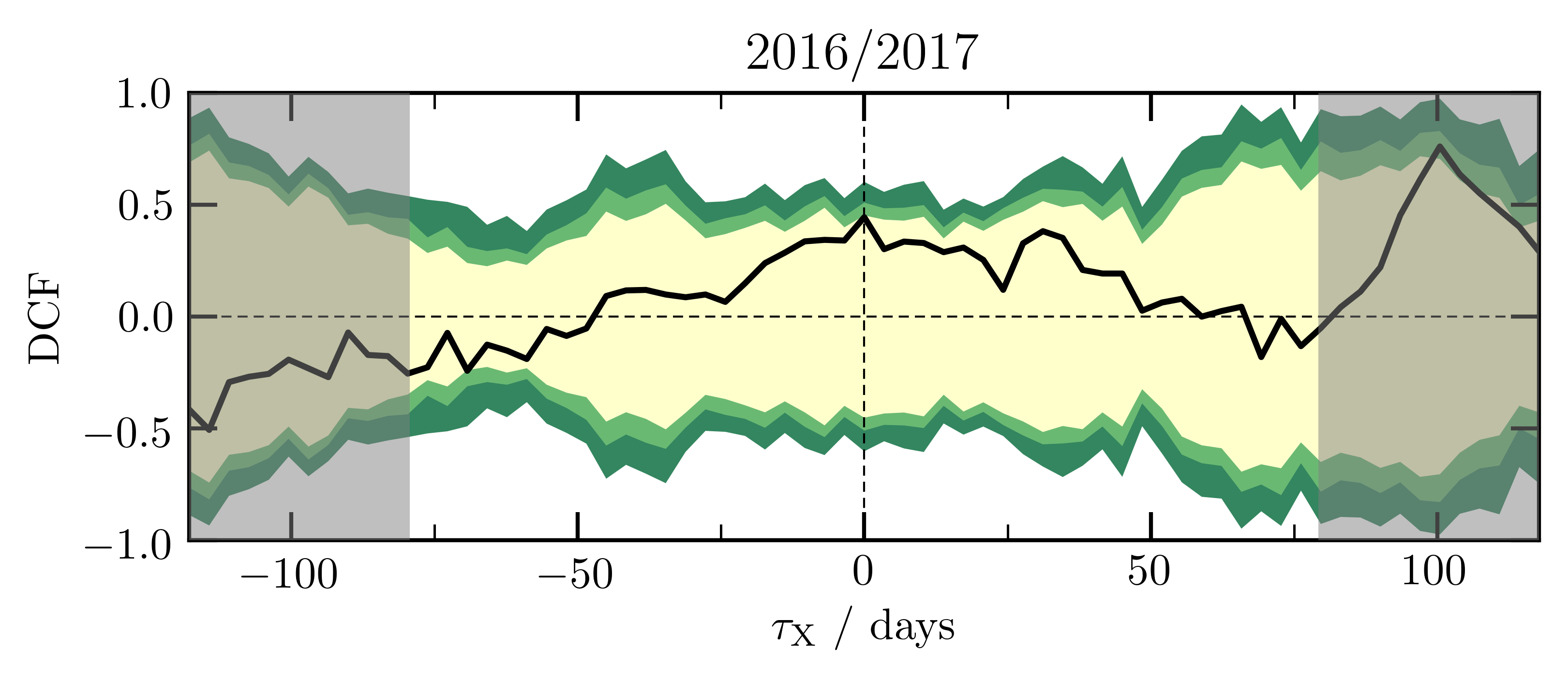}
\includegraphics[width=0.49\linewidth]{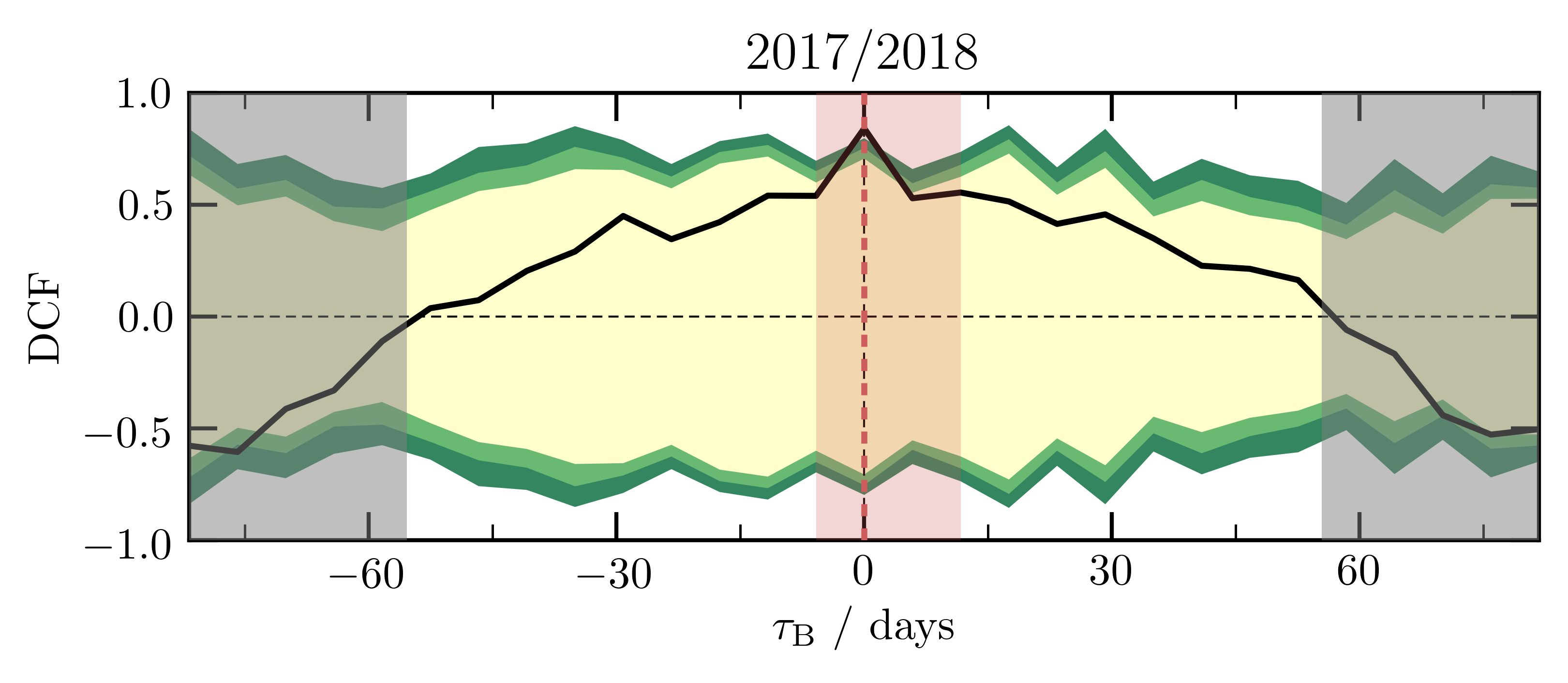}
\includegraphics[width=0.49\linewidth]{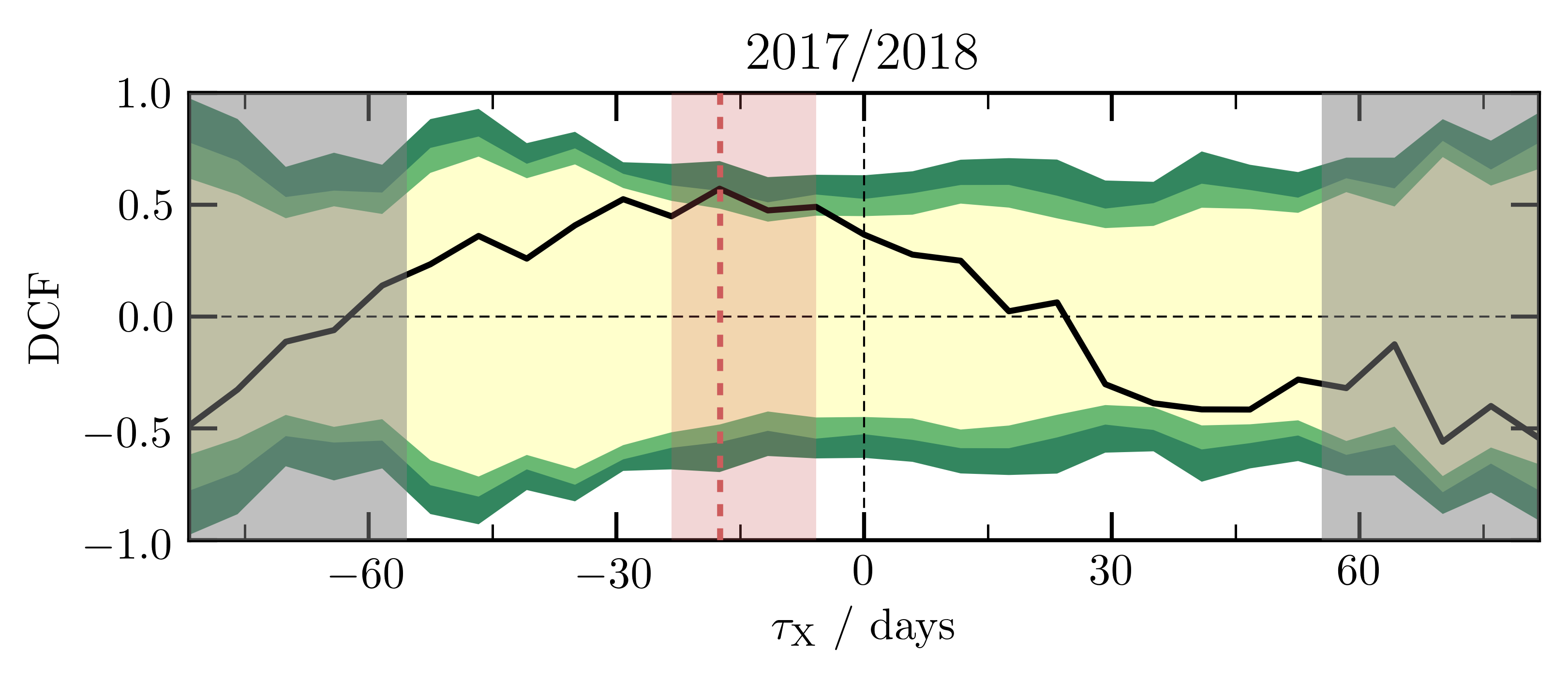}
\includegraphics[width=0.49\linewidth]{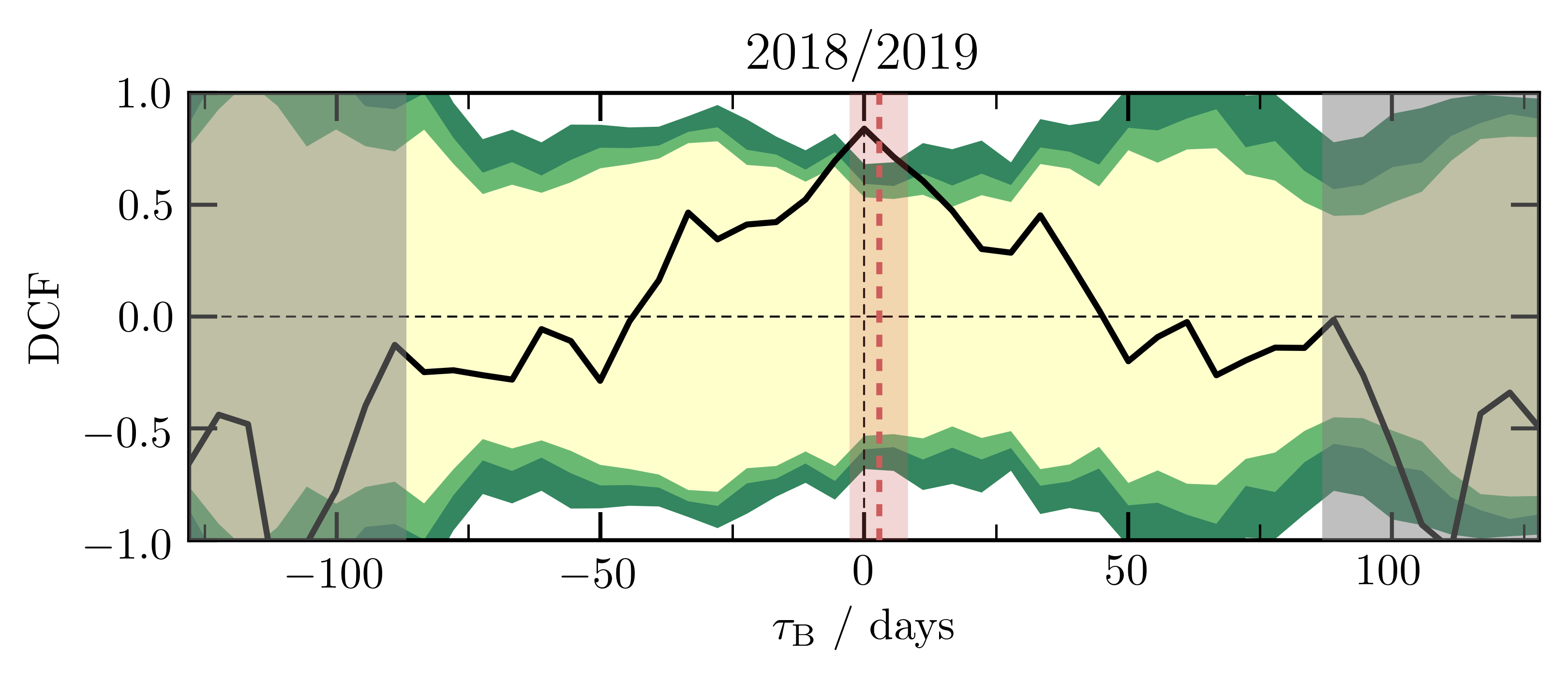}
\includegraphics[width=0.49\linewidth]{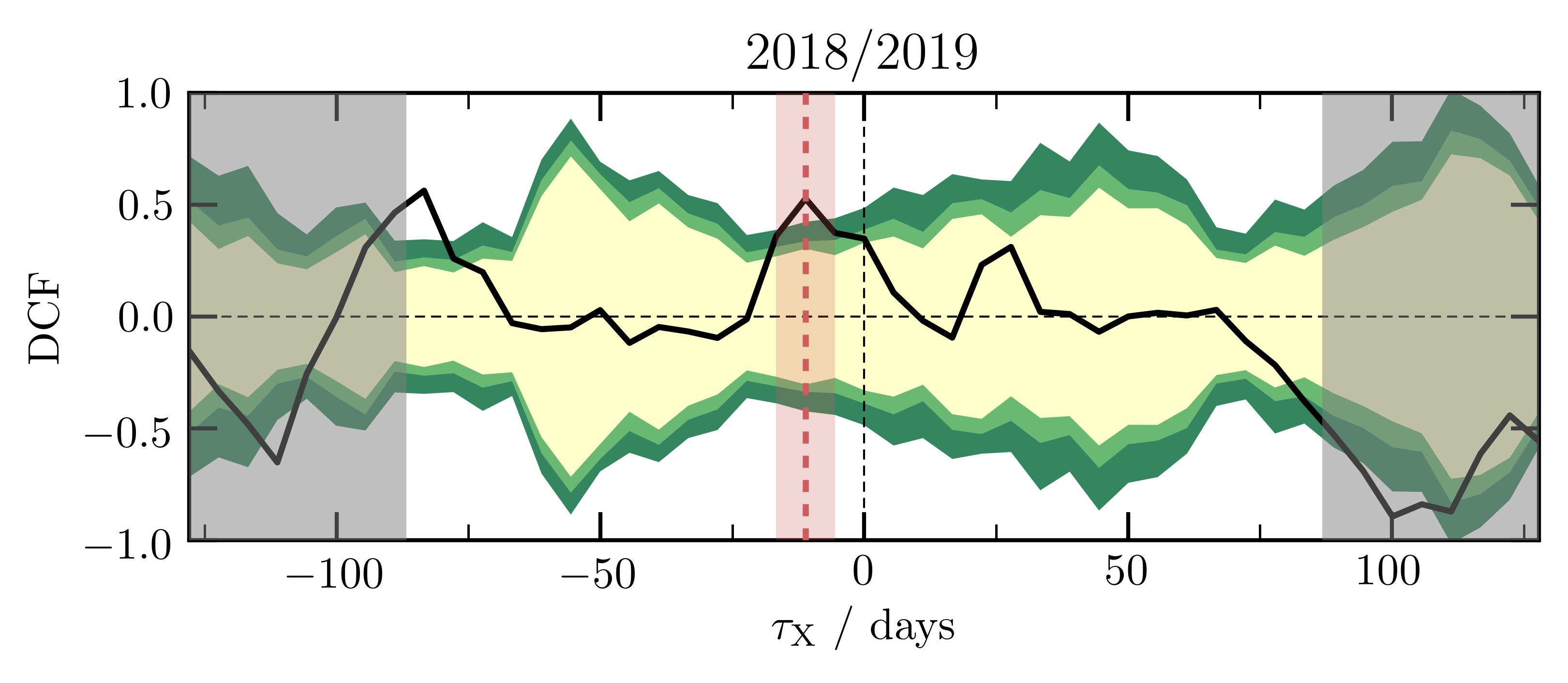}
\includegraphics[width=0.49\linewidth]{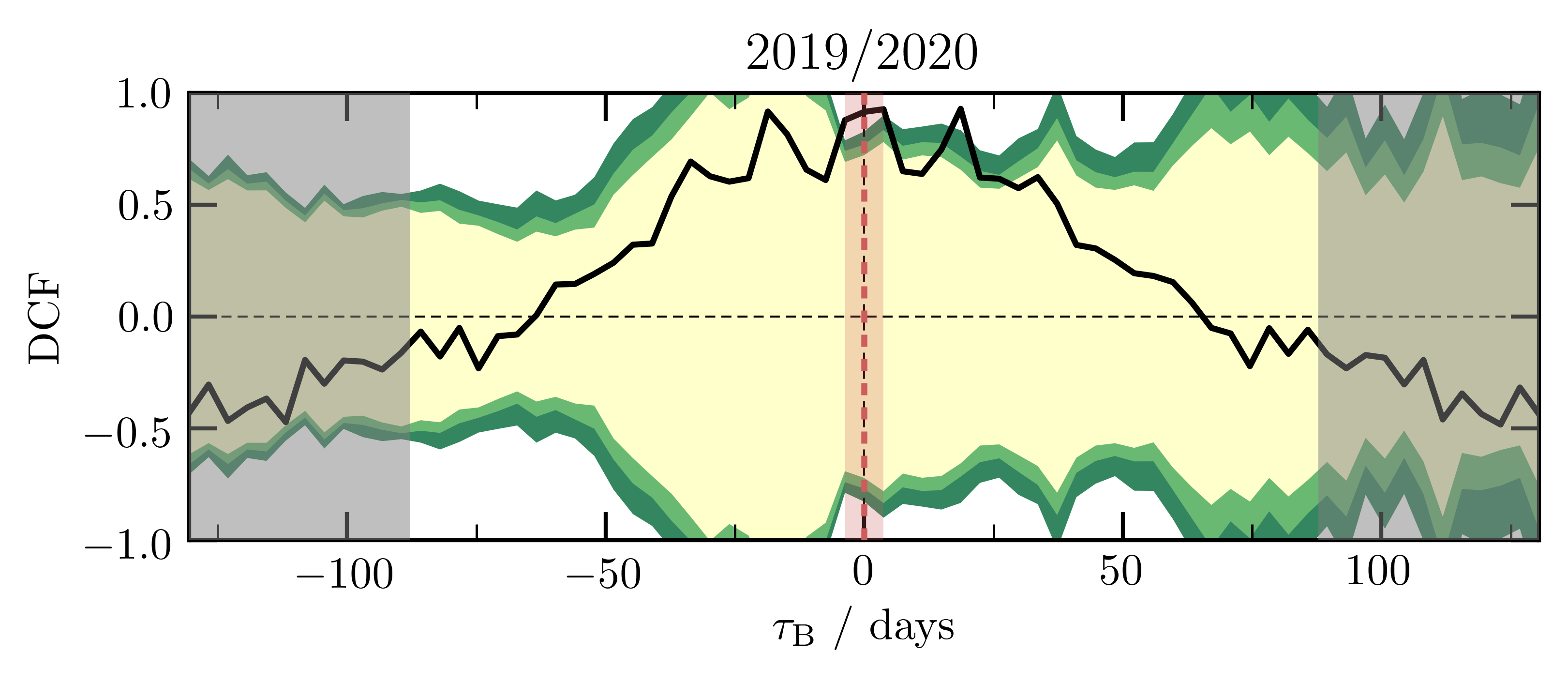}
\includegraphics[width=0.49\linewidth]{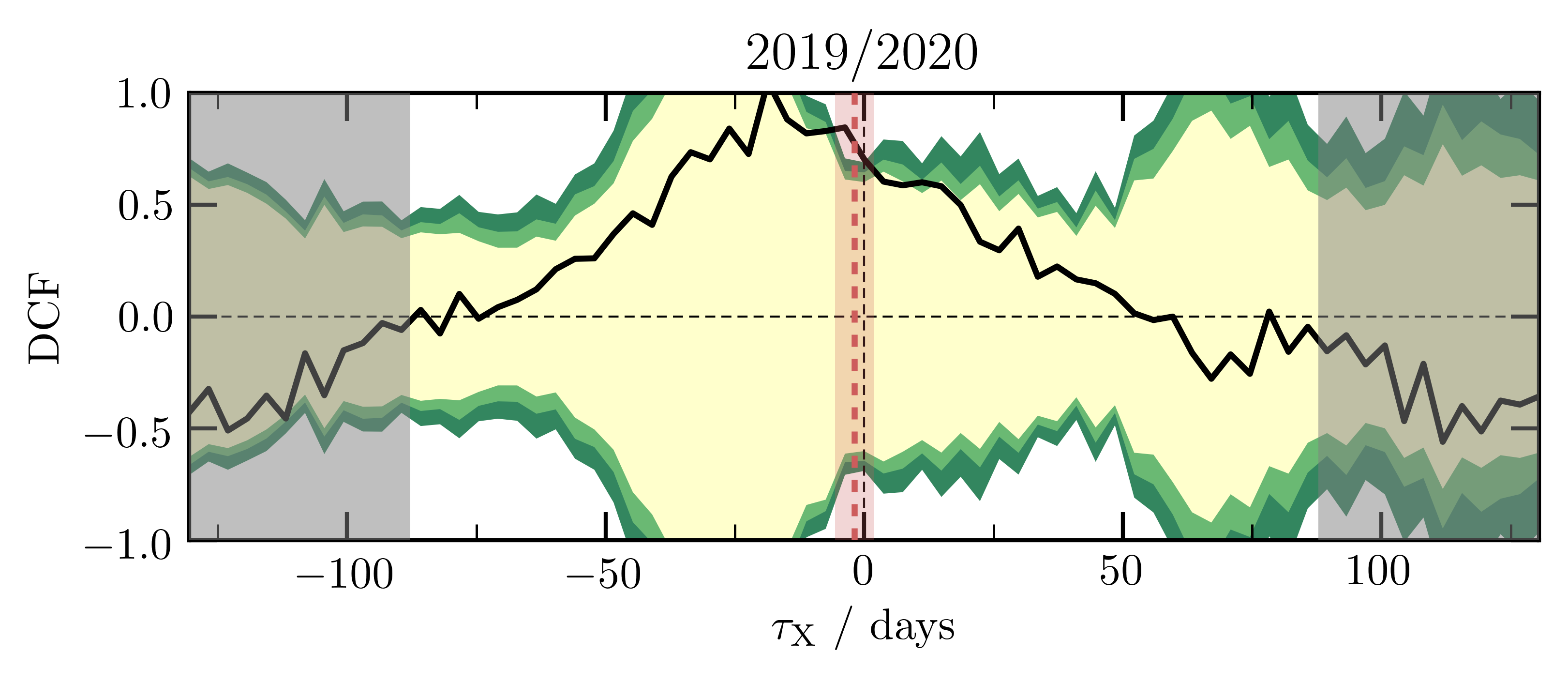}
\caption{{\bf{Left panel}}: the B$-$W2 DCF computed for each of the five epochs 1 to 5 (defined in Sect. 3.1). Epoch 3 was truncated before MJD=58100 to exclude the UV--optical deep fade. 
Filled regions indicate the $\pm90^{\mathrm{th}}$ 
(yellow), $\pm95^{\mathrm{th}}$ (medium green), and $\pm99^{\mathrm{th}}$ (dark green) percentiles from the $N=10^3$ 
light-curve simulations. 
Horizontal dashed lines indicate $\mathrm{DCF}=0$,
% mark zero correlation
and vertical dashed lines indicate $\tau_{\mathrm{B}} = 0~\mathrm{days}$. Negative $\tau_{\mathrm{B}}$ values indicate B lagging W2, positive values indicate leading.
Gray regions are between one-third to one-half the total light-curve length where results become more unreliable. 
The vertical red line indicates the measured time lag and its error. 
At all epochs, the UV and optical fluxes are very closely correlated with a lag consistent with zero days.   
{\bf{Right panel}}: same as the left panel, but for X-ray and W2 fluxes. Negative $\tau_{\mathrm{X}}$ values indicate X-rays leading W2, positive values indicate lagging. While the X-rays show time lags near zero during outburst epochs (2016/2017 and 2019/2020), they are lagging (2015) or leading (2017/2018 and 2018/2019) during quiescent epochs.
Note the different abscissa scale in the uppermost panel (epoch 1, 2015). } 
\label{fig:DCF-opt-UV}
\end{figure*}

\section{Discrete correlation functions} 

The discrete correlation function (DCF) technique was designed to analyze unevenly sampled data sets 
\citep{Edelson1988}. We employ the DCF to search for correlations between the Swift optical, 
UV, and X-ray wave bands. The Swift data were split into five epochs, as described above (Sect. 3.1); two
epochs of relatively low-level activity, two epochs of outbursts, and one shorter epoch of low-level activity in 2015 of densest coverage.

We computed the DCFs as prescribed in \citet{Edelson1988} using the 
\textsc{R} package \texttt{sour}\footnote{Available at https://github.com/svdataman/sour.}
\citep{Edelson2017}. The time step, $\tau$, over which the DCFs were computed corresponds 
to twice the median time step across the entire light curve in each epoch. 
% This binning scheme corresponds to the following $\Delta\tau$ for each epoch: (1) $1.05~\mathrm{days}$, (2) $3.46~\mathrm{days}$, (3) $5.98~\mathrm{days}$, (4) $5.56~\mathrm{days}$, and (5) $3.73~\mathrm{days}$. 
% 
To evaluate the significance level of measured lags we produced confidence contours for each DCF by simulating 
$N=10^3$ artificial UV-W2 light curves following the prescription of \citet{Timmer1995}
assuming a power spectral density (PSD) of $P\left(f\right) \propto f^{-\alpha} = f^{-3}$ based on the results of 
the SF analysis for UV-W2, taking $\alpha = \beta + 1$. The artificial light curves were then 
used to compute artificial DCFs with the band-of-interest light curve, allowing for the computation of 
$90^{\mathrm{th}}$, $95^{\mathrm{th}}$, and $99^{\mathrm{th}}$ percentiles based on the distribution of 
artificial DCFs at each time step. 

To evaluate the error on the measured lags, we computed the autocorrelation function (ACF) of the UV-W2 light curve, 
following the same procedure outlined above to produce confidence contours for the ACF. Because the ACF peaks at 
$\tau = 0~\mathrm{days}$ we estimated the error on the lag measurement as all ACF values in excess of 
the $99^{\mathrm{th}}$ percentile contour around $\tau = 0~\mathrm{days}$. We also use the UV-W2 ACF to 
compare with each DCF to evaluate whether a measured lag corresponds to a true detection or if it in fact
corresponds to a feature in the ACF, rendering it unphysical in origin (see Appendix C). 

Measured lags are determined as those times where the DCF exceeds the $99^{\mathrm{th}}$ percentile contour for 
either a correlation or an \emph{anti}correlation. Furthermore, we restrict the times for which we report 
measured lags to those times corresponding to $\leq 1/3$ the light-curve length in each epoch
such that any  periodicity is detected at least three times.

\subsection{Optical--UV lags}

DCFs for B and UV-W2 were computed for all five epochs (Fig. \ref{fig:DCF-opt-UV}, left panel).
At all times, the optical and UV fluxes are very closely correlated, and  
lags during all five epochs are consistent with zero days. Epoch 1 (2015) was used to derive the best lag measurement, as cadence is exceptionally high during this epoch with a measurement approximately every half-day for about 6 weeks. 
This then results in a time lag of $\tau = 0 \pm 1$ day at the 95--99\% significance level.

\begin{table*}
\footnotesize
	\centering
	\caption{Summary of interesting spectral and flux states of OJ 287 observed with Swift during the last two decades (upper table entries) and noteworthy events related to the SMBBH model taken from the literature (lower table entries). Columns are: (I) type of event, (II) energy band it was observed in, (III) epoch it was observed, (IV) No. of the event as marked in Figs \ref{fig:lc-Swift-CR} and \ref{fig:lc-Swift-fluxes2016}, (V) comments, and (VI) journal reference that first published the event (excluding {\sl{Astronomer's Telegrams}}).}
	\label{tab:states-summary}
	\begin{tabular}{lcccll} 
		\hline
		Event &  Wave Band & Obs. Date & No. & Comments & 1st reference \\
		(I) & (II) & (III) & (IV) & (V) & (VI) \\
		\hline
		low/hard state &  X-rays & 2005-2007 &  & flattest $\Gamma_{\rm x}$, sparse coverage & \citet{Massaro2008} \\
		outburst & all &  2016 Sept - 2017 April & (2) & strong soft X-ray excess & \citet{Komossa2017} \\
		deep fade &  UV-opt  &  2017 Oct-Dec & (3) & symmetric; X-rays do not follow & \citet{Komossa2020};  \\
		 &    &   &  &  & this work \\
		outburst &  all  & 2020 Apr-Jun & (4) & strong soft X-ray excess & \citet{Komossa2020}   \\ 
		low state & UV-opt & 2020 Sept & & & this work  \\
		precursor flare epoch &  opt& 2020 Dec - 2021 Jan & & & this work \\
		\hline 
		impact flare (centennial flare) & opt & 2015 Dec & (1) &  & \citet{Valtonen2016}   \\
		impact flare (Eddington flare) & IR & 2019 Jul & & & \citet{Laine2020}   \\
		\hline
	\end{tabular}
\end{table*}

\subsection{UV--X-ray lags}
DCFs for UV-W2 and X-rays were obtained for all five epochs (Fig. \ref{fig:DCF-opt-UV}, right panel). 
During the two epochs of outbursts, lags between UV and X-rays are consistent with zero days. 
We measure a correlation, with X-rays found to lead the UV by $\tau = -2 \pm 4$ days at the $>$99\% significance level during the 2020 outburst while no statistically significant lag is measured during the epoch of the 2016/17 outburst.
At epoch 3 (2017/18) and epoch 4 (2018/19) of low-level activity, the X-rays and UV are correlated and the X-rays are found to lead the UV. During epoch 4, $\tau = -11 \pm 6$ days at the $>$99\% significance level.  During epoch 3, the UV--optical deep fade (Sect. 6.5) dominates the first 2 months of the data set. Therefore, epoch 3 was only analyzed beyond MJD 58100.  
The X-rays and UV are correlated and the X-rays lead the UV with $\tau = -18^{+12}_{-6}$ days at the 95-99\% significance level.

Different behavior is observed during epoch 1 (2015). During this epoch, the UV and X-rays are anti-correlated, and the X-rays are lagging with $\tau = +7 \pm 1$ days at the $>$99\% significance level.
The analysis was rerun after applying a Gaussian smoothing kernel to the data.
This produces a similar overall result with a time lag of $\tau = +7.5$ days  with $>$99\% confidence.
Possible reasons for the discrepant behavior in epoch 1 are discussed below. Because epoch 1 is of shorter duration than the other epochs, it is possible that it is more sensitive to the influence of competing processes like synchrotron and IC emission with different time lags. 

\section{Discussion} 

\subsection{Binary SMBH model of OJ 287} 

In the following sections, we will refer to aspects of the SMBBH model of OJ 287 and we therefore briefly describe some key features of that model that we will come back to below. 
In the SMBBH model of \citet{Valtonen2016}, impact flares arise when the secondary SMBH crosses the accretion disk twice during its orbit. The impact drives two supersonic bubbles of hot, optically thick gas from the disk. The bubbles then cool as they expand. 
Once they become optically thin, they start emitting and only then does the flare become observable
\citep[see][for hydrodynamic simulations]{Ivanov1998}.
The most recent impact flares were reported in 2015 and 2019 \citep{Valtonen2016, Laine2020}. At such epochs, there is an additional optical-IR emission component, and the total optical flux is less polarized \citep{Valtonen2016}.   
In addition to the impact flares, the model predicts after-flares when the impact disturbance reaches the inner accretion disk \citep{Sundelius1997} and triggers new jet activity. 
The latest two impact flares (Tab. \ref{tab:states-summary}) received special designations: That of 2015 December is dubbed the  ``centenary flare" \citep{Valtonen2016} because it coincided with the centenary of GR. That of July 2019 is called the ``Eddington flare" \citep{Laine2020} because it coincided with the centenary of Eddington's measurement of light deflection during a solar eclipse. 

\subsection{Length scales in (the host galaxy of) OJ 287}

%Obs timescales (1+z) longer than resframe.
We perform some estimates of typical length scales in the nucleus of OJ 287. 
These will also be used later in the context of evaluating and interpreting some of the observed timescales of variability of OJ 287. 
Our estimates are repeated for two SMBH masses;
$M_{\rm BH, primary} = 1.8 \times 10^{10}$ M$_{\odot}$, or $M_{\rm BH} = 10^{8}$. The second estimate accounts for the fact that the primary SMBH may be overmassive w.r.t. the host galaxy of OJ 287 \citep{Nilsson2020}. In that case, a typical blazar SMBH mass of $M_{\rm BH} = 10^{8}$ is adopted. 

% BLR: 
First, it is useful to know the size of the BLR, as that may be the site that drives the EC emission, and as its length scale could then drive some characteristic time delays. The scaling between BLR radius and SMBH mass 
is given by 
\begin{equation}
M_{\rm BH} = f_{\rm BLR}~~ \frac{R_{\rm BLR} ~~\Delta v^2} {G}, 
\end{equation}
where $R_{\rm BLR}$ is the radius of the BLR, 
$f_{\rm BLR}$ is an inclination-dependent factor that parameterizes the BLR geometry \citep{Peterson2014, Pancoast2014}, 
and the velocity dispersion $\Delta v$ is determined from the width of broad emission lines.
With FWHM(H$\alpha$)=4200 km/s \citep{Nilsson2010} and the average value $f_{\rm BLR}$=1.9 of \citet{Pancoast2014}, Eqn (2) implies  
$R_{\rm BLR}$ = 7.5 lt-yr for $M_{\rm BH, primary} = 1.8 \times 10^{10}$ M$_{\odot}$. (This value can be regarded as an upper limit, if we assume that broad wings in H$\alpha$ escaped detection w.r.t the bright blazar continuum emission, and/or if a flattened BLR with $f_{\rm BLR}$=6.9 \citep{Decarli2011} is assumed, but $R_{\rm BLR}$ remains large in any case.) 
Alternatively, we obtain $R_{\rm BLR}$ = 15.3 light days for $M_{\rm BH} = 10^{8}$ M$_{\odot}$.

Second, the size of any dusty molecular torus in OJ 287 is estimated. Its photon field could provide an alternative source of EC seed photons.  
The torus is located beyond the BLR. Its inner edge{\footnote{Given geometrical considerations, a classical toroidal structure is an unlikely source of EC seed photons in BL Lacs. However, our dust-survival estimate holds more generally for any reservoir of dusty gas along our line of sight.}} is defined by the dust sublimation radius which scales as  
\begin{equation}
R_{\rm sub} = 1.3 \, ( \frac{L_{\rm UV}}{10^{46} \,  \rm{ erg/s}} )^{1/2} \, ( \frac{T_{\rm sub}}{1500 \,  \rm{ K}} )^{-2.8} \, ( \frac{a}{0.05 \,  \mu \rm{m}} )^{-1/2}
\end{equation}
\citep{Barvainis1987},
where $T_{\rm sub}$ is the dust sublimation temperature and $a$ is the average radial grain size. $R_{\rm sub}$ agrees well within a factor of 2-3 with direct imaging observations \citep[e.g.,][]{Kishimoto2007}. 
%% can be easily adjusted by changing grain-size distribution and subl temp)
We assume graphite grains with $T_{\rm sub}$=1500 K and $a$=0.05 $\mu$m,  
% (Barvainis 1987), 
and $L_{\rm UV,disk} \simeq L_{\rm bol,disk}$, where
$L_{\rm bol,disk} = 0.1 L_{\rm Edd}$ \citep{Valtonen2019}.
Then, $R_{\rm torus}=R_{\rm sub}= 20.6$ ly for $M_{\rm BH, primary} = 1.8 \times 10^{10}$ M$_{\odot}$. Alternatively, $R_{\rm torus}=R_{\rm sub}= 1.5$ ly for $M_{\rm BH} = 10^{8}$ M$_{\odot}$. 
We will use the results of this section, when discussing DCF lags below. 

\subsection{X-ray--UV--optical variability and time lags}

% {\em{Note: All timescales from data done in observer frame. When moving to rest frame: devide (1+z).}} 

\subsubsection{Fractional variability amplitude} 

Results of the fractional variability analysis over a timescale of several years (Tab. \ref{tab:Fvar}) show that the optical, UV, and X-rays show similar values of $F_{\rm var}$ and therefore arise from closely related processes. During the epochs of outbursts, $F_{\rm var}$ is higher in all bands than during the epochs of low-level activity.

Based on an analysis of our \citep{Komossa2017} and archival 2016 XRT observations of OJ 287, \citet{Siejkowski2017} reported $F_{\rm var}$ = 0.357 $\pm$ 0.016 in X-rays, which agrees well with the value reported here (Tab. \ref{tab:Fvar}), given that the 2017 evolution of the outburst was not included in that other analysis. 
Interestingly, the fractional variability amplitude we measured at much shorter
time intervals during XMM-Newton observations
is {\em{much smaller}}.
During the 2020 outburst, no variability was observed at all during our 10 ks XMM-Newton observation \citep{Komossa2020}. 
During a 120 ks XMM-Newton long look in 2015 at intermediate
flux levels binned to 500s intervals, $F_{\rm var}$ = 0.033 $\pm$ 0.002, 
%%\citep{Komossa2021a}. 
and during a 46 ks observation in 2006, $F_{\rm var}$ = 0.01 $\pm$ 0.01.
These findings suggest a significant change in the processes that drive the X-ray variability on these different timescales between $t\simeq 500$ s -- 0.5 day (XMM-Newton) on the one hand and flux-doubling times as fast as a few days (Swift) on the other hand. A similar conclusion is also evident in the principal-component analysis (PCA) of OJ 287 by \citet{Gallant2018}. Using all XMM-Newton observations between 2005 abd 2015 they showed that power-law variations dominated the long-term changes (on the order of years), but on short time scales (e.g. 3 hr), the PCA showed a hard component above $\sim2$ keV that dominated the (very small-amplitude) variability.

\subsubsection{Structure Function} 

Based on our SF analysis, we have measured a  break timescale of 4--11 days in the UV--optical bands during epochs of quiescence and outburst. This compares to a characteristic PSD timescale of 5.8 days reported by \citet{Wehrle2019} based on optical Kepler data of OJ 287 obtained during a quiescent epoch in 2015. That result agrees well with our B-band break timescale of 6$^{+3}_{-1}$ days measured during the same year. 
The SF break time we measure is not a disk timescale \citep{Wehrle2019} as outbursts and epochs of quiescence are dominated by nonthermal jet emission but is consistent with the timescale (10-20 days) expected for a shock front swiping through a helical magnetic field as it moves forwards and backwards in the jet \citep{Villforth2010}.
% as deduced from optical polarimetry observations.

\subsubsection{Discrete correlation function} 

DCFs at very different epochs and flux states (outbursts, low-level activity) allow us to identify the crucial interband timescales on days to months and provide us with valuable constraints on emission processes and particle distributions \citep[e.g.][]{
Marscher2014, Sokolov2004, 
Weaver2020}.  

Using the optical--UV DCF, we have measured a time lag  of $\tau = 0\pm1 $ day, best constrained during the 2015 quiescence epoch, but similar during outbursts. These wave bands are always dominated by synchrotron emission and the small delays are consistent with synchrotron theory \citep{Kirk1998}.  

The UV--X-ray DCF during epochs of outbursts implies time lags  near zero as well, consistent with the previous interpretation that X-ray outbursts are driven by a strong soft X-ray synchrotron component, which is closely linked to the optical-UV emitting component \citep{Komossa2020}.     

The UV--X-ray DCF during most epochs of quiescence has revealed a time lag of $\sim$2 weeks (X-rays leading UV). This result implies different emission physics than during outbursts.   
During quiescence, the X-ray spectra are much flatter and can be explained by an inverse Compton contribution ($\Gamma_{\rm x} \approx 1.5$) in addition to residual soft synchrotron emission \citep{Komossa2021a}. 

The (shorter) epoch 1 in 2015 stands out in that the UV--X-ray DCF reveals a different result, with the X-rays {\em lagging} by 7 days.  The 2015 epoch precedes by several months the observation of an impact flare interpreted in the context of the binary SMBH model \citep{Valtonen2016}. 
This raises the question whether processes related to the binary's presence could operate at epoch 1, explaining the different DCF result. According to the binary model, the initial impact of the disk happened in 2013, driving a two-sided expanding bubble of hot gas, which only became optically thin and therefore visible in 2015 December. No particular emission components are predicted for mid-2015
\citep{Pihajoki2013b}. While it is conceivable that the impact caused additional streams of high-velocity gas that interact with their gaseous environment and emit radiation, it is unlikely that they dominate the X-ray emission during epoch 1. The simultaneous deep XMM-Newton spectrum did not reveal any additional or peculiar emission components  beyond a soft synchrotron component and a hard IC component, and the 2015 spectrum is identical in X-ray flux and very similar in spectral shape to a later spectrum taken in 2018 quasi-simultaneous with the EHT \citep{Komossa2021a}. Further, the overall Swift flux variability in epoch 1 is very similar to other epochs of low-level activity.     The only immediate difference between epoch 1 and later epochs 4 and 5 is its much shorter duration of only 6 weeks. It is therefore most likely that we are seeing one of the several emitting components that always contribute to the X-ray spectrum, representing several jet knots or several substructures within the same jet component.
The longer 9 month epochs average over these individual components and events.    

In fact, both lags or leads between optical and high-energy radiation on the timescale of days to weeks are predicted by the SSC and EC models. For EC, the BLR or torus is considered to be the major source of the external seed photon field \citep{Sikora2009}.  
SSC models predict time delays (both lags or leads) that are dominated by energy stratification and geometry of the emitting regions \citep{Sokolov2004}. 
Time delays are generally of lower amplitude in this model; typically lower than the lags we reported here. We therefore favor EC. 
% EC is thought to be more common in FSRQs and less in BL Lacs. However, 
OJ 287 has a detectable BLR at low states \citep{SitkoJunkkarinen1985, Nilsson2010} that provides seed photons. 
BLR (and torus) are found at large spatial separations from the nucleus in OJ 287, if the primary SMBH mass drives scaling relations. 
In Sect. 6.2., we estimated a
BLR radius in the range $R_{\rm BLR}$ = 15.3 lt-days (using $M_{\rm BH} = 10^{8}$ M$_{\odot}$) to $R_{\rm BLR}$ = 7.5 lt-yr (using $M_{\rm BH, primary} = 1.8 \times 10^{10}$ M$_{\odot}$). 
However, travel time delays and geometry will also play a role.
For EC to be efficient, the proximity of the jet region to a major BLR clump is required.
Such a scenario can be tested further with spatially resolved radio observations that (after deprojection) provide the spatial separations of major radio knots from the nucleus. 

Finally, we note that the correlated variability between optical-UV and (non-synchrotron) X-rays we detect at low states favors a leptonic jet model as these two parts of the SED vary independently in hadronic models.

\subsection{Notes on outstanding epochs in the 2005--2021 Swift light curve of OJ 287}

 Several epochs in the 2005--2021 light curve of OJ 287 stand out; including outbursts, low states, mini flares, and particularly flat or ultrasteep spectral states. 
 Several of these epochs have already been reported in previous publications. 
 %% citations of those who first published each data sub-set given below
We briefly rediscuss these here in context and add new information.  

\subsubsection{2005--2007 X-ray inactivity}

OJ 287 was undergoing a phase of inactivity in X-rays in this period (Figs \ref{fig:lc-Swift-CR} and \ref{fig:lc-Swift-fluxes2005}) and reached the lowest levels of X-ray emission ever observed with Swift.  
These early Swift data were first published by
\citet{Massaro2008}. 
The X-ray spectrum is flat with $\Gamma_{\rm X}=1.5$ and consistent with pure IC emission at this epoch. In late 2007, one can recognize an epoch of optical flaring that is not accompanied by a significant rise in X-rays, unlike most other epochs.  

\subsubsection{2015 December ``centennial" impact flare}

Enhanced emission levels were recorded with Swift in 2015 December. While the UV(--optical) was very bright at this epoch, the X-rays were within
their normal range of variability (marked in pink in Fig. \ref{fig:lc-Swift-CR} and Fig. \ref{fig:lc-Swift-fluxes2016}). 
Swift data of this epoch were first published by \citet{Valtonen2016}. 
The variability behavior at this
epoch is very different from later epochs, where outbursts are seen at all wave bands and the optical–UV and X-rays follow each other closely. 
The 2015 December flare  was interpreted as a thermal impact flare \citep{Valtonen2016} within the binary SMBH model.
Ground-based optical observations covered this epoch more densely. 
The Swift X-ray spectrum at this epoch is well represented by a photon index of $\Gamma_{\rm x} \simeq 2$, comparable to subsequent epochs, indicating the contribution of a  moderate synchrotron component in the soft band, similar to the one observed at other epochs \citep{Komossa2021a}.  

\subsubsection{2016/2017 outburst}

In the course of our Swift monitoring of OJ 287, we found a bright outburst in all bands in 2016, extending into the first months of 2017
(marked in dark blue in Fig. \ref{fig:lc-Swift-CR} and Fig. \ref{fig:lc-Swift-fluxes2016}). The outburst, which is the brightest of OJ 287 recorded in X-rays, was first published by \citet{Komossa2017}. Despite early speculations about an accretion flare at this epoch (to explain the soft X-ray spectrum), multiple lines of evidence then clearly established this outburst as nonthermal in nature: 
It was accompanied by VHE emission \citep{OBrien2017} and a radio flare \citep{Myserlis2018, Lee2020}, and the optical band showed high levels of polarization \citep{Valtonen2017} as did the radio \citep{Goddi2021}. Further, the Swift X-ray spectra can be well explained by a soft synchrotron emission component and show the very same softer-when-brighter pattern also seen during the nonthermal 2020 outburst \citep{Komossa2020}. 
Finally, with Swift, we detect X-ray flux-doubling timescales during the outburst as short as 4 days. This is shorter than the light-crossing time at the last stable orbit of the accretion disk around the primary SMBH of OJ 287, again ruling out an accretion disk origin.  

The 2016 outburst, covered independently by optical ground-based observations, was suggested to represent an after-flare predicted by the binary SMBH model \citep{Valtonen2017}. 

\subsubsection{2020 outburst} 

The 2020 April--June outburst (marked in red in Fig. \ref{fig:lc-Swift-CR} and Fig. \ref{fig:lc-Swift-fluxes2016}) was the second-brightest X-ray outburst of OJ 287 we detected with Swift \citep{Komossa2020}. Our accompanying XMM-Newton spectrum confirmed the supersoft synchrotron component seen with Swift and allowed detailed spectral modeling. Our NuSTAR spectrum  revealed a spectral component extending to $\sim$70 keV, remarkably soft for that high-energy band ($\Gamma_{\rm x}=-2.2$), and softer than pure IC emission.
Rapid flux variability detected with Swift is faster than the light-crossing time at the last stable orbit of the accretion disk around the primary SMBH. All these observations, plus the detection of radio flaring \citep{Komossa2021c}, confirm the nonthermal nature of this
bright outburst. \citet{Komossa2020} concluded that the timing of the outburst is consistent with an after-flare predicted by the binary SMBH model \citep{Sundelius1997}.  

Optical and UV fluxes are very tightly correlated at all times (Figs \ref{fig:fluxcorr}, \ref{fig:DCF-opt-UV}), at low states, at epochs of mini flaring, and during the two major outbursts.  This finding leaves very little room for any additional emission contribution to the optical--UV other than synchrotron radiation during the 2015--2021 epoch of dense light-curve coverage. Any additional component, no matter whether a thermal UV--EUV component, for instance, or an IR--optical reprocessing component, would have disturbed the tight correlation.{\footnote{During outbursts, the optical and X-rays are closely correlated, too, but with some larger scatter (Fig. \ref{fig:fluxcorr2}). In particular, during the 2020 outburst, one optical--UV subflare does not have an X-ray counterpart, and during the 2016/17 outburst, the highest optical state is reached in 2016, the highest X-ray state in 2017.}}      

\subsubsection{Frequent mini flaring (2009--2021)} 
Small-amplitude flaring, which we call ``mini flares", is ongoing most of the time, and OJ 287 is rarely in a phase of complete inactivity over weeks.  
The amplitude of variability of the mini flares is remarkably constant between 2009 and 2021, always varying between 0.1--0.5 cts s$^{-1}$ (Fig. \ref{fig:lc-Swift-CR}). This implies a very constant underlying emission mechanism(s).

\subsection{2017 UV--optical deep fade}

At the end of 2017, a remarkable, symmetric UV--optical deep fade is seen, 
marked in light blue in Fig. \ref{fig:lc-Swift-CR} and \ref{fig:lc-Swift-fluxes2016} and displayed in higher resolution in Fig. \ref{fig:lc-Swift-2017}.
The deep fade was already reported by \citet{Komossa2020} but not discussed further. It was independently noticed in optical ground-based monitoring and was used to trigger imaging of the host galaxy of OJ 287, which is difficult to measure at epochs where the blazar glare is bright \citep{Nilsson2020}. At this epoch, the optical--UV magnitudes reached the lowest value ever recorded with Swift, even though not as low as during a previous deep fade of OJ 287 in 1989 where V decreased from $\sim$16 mag to 17.4 mag within a month \citep{Takalo1990}. 
%  V_minimum = 0.5 mJy. 

One possible scenario that might explain deep fades in the light curve of OJ 287 is a short misalignment of the jet due to the perturbation from the secondary SMBH when it passes near the jet \citep{Takalo1990, Ingram2021}. However, according to the model of \citet{Dey2018},  the secondary SMBH is behind the accretion disk in late 2017 and not crossing the jet between observer and the primary. 

The deep fade  is symmetric, reminiscent of an occultation event. This raises the question of whether a dusty cloud could have passed our line of sight, temporarily extincting the optical-UV emission by a factor of 3. 
However, such an event would cause a significant and systematic reddening of the optical-UV color as the flux decreases and then re-rises. This is not observed.{\footnote{While there {\em is} a reddening seen during the deep fade with OJ 287 becoming redder in the NIR versus optical, this is consistently explained by an increasing contribution of the host galaxy in the NIR as the optical fades \citep{Valtonen2020}, the host being an elliptical missing young stars \citep{Nilsson2020}.}}  
The flux ratio $f_{\rm W2}$/$f_{\rm U}$ is constant within the errors throughout the deep fade. 
Even if the absorber was dust free and would only affect the UV--optical spectrum by deep absorption lines, it would still affect different filters differently.
Further, we note such a dusty (or dust-free) cloud would cause strong X-ray absorption. However, the X-rays are remarkably constant during the event (Fig. \ref{fig:lc-Swift-fluxes2016}), with an average X-ray photon index of $\Gamma_{\rm x}=2.0\pm0.1$ (Tab. \ref{tab:spectral-fits}).  
 
The constancy of the {\em{observed}} X-rays, while the optical--UV fades so strongly, is difficult to explain in any synchrotron model in which those X-rays and the UV are coupled. The observed constant X-ray emission at that epoch must therefore arise from a different component that is not causally connected with the optical-UV synchrotron component within the observational time interval of 2 months that the deep fade lasts. The observed X-rays plausibly contain a strong IC contribution and arise in different jet regions at that epoch.   
 
We speculate that the deep fade is caused by a temporary dispersion or swing of that part of the jet that dominates the optical--UV emission at that epoch. Such an event could plausibly happen either in the radio core itself or in the area of a bright quasi-stationary feature, which is prominent in radio images of OJ 287 \citep{Agudo2012, Hodgson2017}; interpreted as recollimation shock or as a region of maximized Doppler factor in a bent jet.
%% model: Alberdi1997. 
High-cadence radio imaging at the epoch of the deep fade during 2017 October--December could further test this scenario. 

\subsection{2020 September low state}
When OJ 287 emerged from Swift Sun constraint and was observable again in 2020 September, it became immediately clear that the bright 2020 April--June outburst had ended. OJ 287 was found in a  UV--optical(--X-ray) low state instead \citep{Komossa2020b}. 
% see our Atel 
The only other such low state seen during the epoch of dense coverage of OJ 287 within the MOMO project (Fig. \ref{fig:lc-Swift-fluxes2016}) was the 2017 UV--optical deep fade. The 2020 September low state lacks the symmetric appearance of the deep fade, but we also have to keep in mind that we did not see OJ 287 enter the low state in 2020 because of its unobservability with Swift.  It is interesting to note in passing that both outbursts, in 2016/2017 and in 2020, were followed by such a  low state months later: six months and three months, respectively. 

\subsection{2020 October -- 2021 January rise and search for precursor flare activity} 
As we near the next impact flare predicted by the binary model in 2022 \citep{Dey2018, Laine2020}, it is interesting to ask whether we start detecting precursor flare activity. 
%% Pihajoki suggestion: maybe these are accretion events on the secondary when it meets gas above the primary disk; or sthg else. 
According to \citet{Pihajoki2013b} such flares have been seen to precede the main impact flares. \citet{Pihajoki2013b} predicted the next precursor event around 2020.96$\pm{0.10}$. We have carefully inspected the recent light curve. During the epoch 2020 October -- 2021 February we detect some of the mini-flaring activity similar in amplitude and duration as during other epochs (Sect. 6.4.5). None of them stands out.  Underlying it is a systematic long-term rise in flux which reaches its maximum in early 2021 January. While the timing agrees, the observed duration is longer than that of the precursor flares discussed by \citet{Pihajoki2013b}.
% who also predict a strong rise by $\sim$2 magnitudes, not detected. 
However, our observing cadence is only 3-4 days at this epoch, and any sharp flare or faint flare would have escaped detection. 

\section{Summary and conclusions}
\label{sec:summary}
We have been carrying out a dedicated long-term  project, MOMO, in order to understand the blazar physics and binary black hole physics of OJ 287 during its recent evolution since late 2015.

In particular, we have used Swift to obtain observations in the UV, optical, and X-ray bands. 
Essentially all Swift data of recent years are from our project.  This is the densest monitoring program of OJ 287 so far carried out in the UV and X-rays. It includes simultaneous observations in the three optical bands of Swift UVOT, too. It is accompanied by dense radio monitoring and other follow-up observations involving other observatories and wave bands.
Results are presented in a sequence of publications. Here, we focus on the characteristic variability of OJ 287 derived from the long-term Swift light curve at very different spectral and flux states. Our results can be summarized as follows.

\begin{itemize}

\item The 2005--2021 Swift light curve of OJ 287, densely covered in our project since 2016, has stretched over more than one orbital period predicted by the binary SMBH model of OJ 287. Overall, the light curve  revealed very different activity states of OJ 287 including low states, a deep fade, epochs of mini flaring, episodes of exceptional spectral variability, and two major nonthermal synchrotron outbursts in 2016/17 \citep{Komossa2017} and 2020 \citep{Komossa2020} with supersoft X-ray spectra (up to $\Gamma_{\rm x} \simeq 3$).  These were interpreted as possible after-flares predicted by the binary model and correspond to episodes where new jet components are launched.

\item An important result is the characterization of the intraband and interband variability properties and time lags based on SF and DCF analyses at very different activity states.
% SFs are used to measure the power spectra in all bands. 
We have divided the long-term light curve into epochs of outbursts and lower-level activity (referred to as quiescence) to carry out the analysis. Characteristic SF break timescales of 4 days -- 39 days have been derived, depending on wave band and activity state.

\item Near-zero DCF lags among all wave bands at outbursts are consistent with synchrotron theory.   
Lags and leads in X-rays w.r.t. the UV ($\tau$ = 7--18 days) at epochs of low-level activity can be attributed to the dominance of an IC component.  EC is favored over SSC.

%\item During quiescence, spectra and X-ray--optical time lags change significantly. A time delay of $\sim$7--18 d (X-rays lagging or leading)  and flatter X-ray spectra imply a significant contribution of inverse Compton emission.

\item Scaling relations are used to derive BLR and torus sizes of the host galaxy of OJ 287: The BLR radius is in the range $R_{\rm BLR}$ = 15.3 lt-days (using $M_{\rm BH} = 10^{8}$ M$_{\odot}$) to $R_{\rm BLR}$ = 7.5 lt-yr (using $M_{\rm BH, primary} = 1.8 \times 10^{10}$ M$_{\odot}$). 
If the primary is overmassive w.r.t. its host, then we derive
$R_{\rm torus} = 1.5$ lt-yr (using $M_{\rm BH} = 10^{8}$ M$_{\odot}$), 
otherwise 
$R_{\rm torus} = 20.6$ lt-yr ($M_{\rm BH, primary} = 1.8 \times 10^{10}$).  

\item In between outbursts or low states, OJ 287 exhibits phases of mini flaring, with remarkably constant amplitude between 2009 and 2021. The epoch 2005--2007 is characterized by a phase of particular inactivity in X-rays with flat X-ray spectra in the inverse-Compton regime.

\item A remarkable, symmetric UV--optical deep fade is detected at the end of 2017, reminiscent of an occultation event. The Swift deep fade was briefly mentioned before \citep{Komossa2020} but is investigated here for the first time. While the UV-optical fluxes drop by a factor 3, they do not reach the faintness of the historical deep fade of OJ 287 from 1989.   We can rule out an extinction/occultation event from the passage of a dusty cloud, because of the lack of UV reddening. Further, we can rule out a temporary jet deflection by a close passage of the secondary SMBH because the binary model predicts the secondary's location to be behind the accretion disk and not between observer and primary at that epoch.  We speculate that the deep fade is linked to processes in a bright radio jet component (either the core or the region of a quasi-stationary feature) temporarily dispersing or deflecting the jet in this region.

\item The UV--optical deep fade reveals an additional, causally disconnected, X-ray component that does not follow the two-month fade. These X-rays must arise in a spatially distinct emission region.

\item We have searched for precursor flare activity at the epoch 2020.96$\pm$0.10 predicted by the binary SMBH model. While we do see an emission peak in early 2021 January, it is broader than predictions and we may have missed any actual sharp precursor flare because of the low (3--4 days) cadence at that epoch. Alternatively, the precursor flare could have been fainter than at previous epochs because of the different secondary's angle of disk approach.   

\end{itemize}

In summary, Swift light curves of OJ 287 not only play an important  role  in our understanding of  the emission mechanisms of this bright blazar and binary SMBH candidate, but our Swift data also enabled us to trigger follow-up observations with other missions at epochs of bright outbursts or low states. This  includes XMM and NuSTAR spectroscopy of the exceptional 2020 outburst. In the course of the MOMO program, we will continue to monitor OJ 287 at multiple frequencies.  OJ 287 is of special interest as a multi-messenger source, as the binary model predicts that the orbit is already measurably shrinking due to the emission of gravitational waves. In the future, the planned mission Einstein  Probe will deliver well-covered X-ray light curves of OJ 287. 

% Authors are encouraged to use an online tool at \url{http://authortools.aas.org/FIGSETS/make-figset.html} to generate their own specific figure set mark up to incorporate into their \latex\ articles.

\begin{acknowledgments}
We would like to thank the Swift team for carrying out our observations,
and Ski Antonucci,
Phil Evans, Jose L. Gomez, and Mauri Valtonen for very useful discussions. 
We would like to thank our anonymous referee for very useful comments. 
This work made use of data supplied by the UK Swift Science Data Centre at the University of Leicester. 
This research has made use of the
 XRT Data Analysis Software (XRTDAS) developed under the responsibility
of the ASI Science Data Center (SSDC), Italy. 
In addition to data obtained by us with the Neil Gehrels Swift mission, we also acknowledge the use of public data from the Swift data archive. 
This research is partly based on observations obtained with XMM-Newton, an ESA science mission with instruments and contributions directly funded by ESA Member States and NASA.
This research has made use of the NASA/IPAC Extragalactic Database (NED) which is operated by the Jet Propulsion Laboratory, California Institute of Technology, under contract with the National Aeronautics and Space Administration.
\end{acknowledgments}


\vspace{5mm}
\facilities{Swift (XRT and UVOT), XMM-Newton}


\software{HEASoft (\url{https://heasarc.gsfc.nasa.gov/docs/software/heasoft/}) with XSPEC \citep{Arnaud1996},  ESO-MIDAS (\url{https://www.eso.org/sci/software/esomidas/}), ZDCF and PLIKE \citep{Alexander2013}, the R programming language (\url{https://www.r-project.org/}) with the `sour' package \citep[][\url{https://github.com/svdataman/sour}]{Edelson2017}, SFA \citep[][\url{https://github.com/Starkiller4011/SFA}]{Gallo2018}, and Python (\url{https://www.python.org/}). 
}
%Cloudy \citep{2013RMxAA..49..137F}, 
% Source Extractor \citep{1996A&AS..117..393B}
%          }


\appendix

\section{Annual light curves} 
\label{sec:appendix-light}

\begin{figure}
\includegraphics[clip=, trim=1.8cm 5.6cm 1.3cm 2.6cm, width=0.49\linewidth]{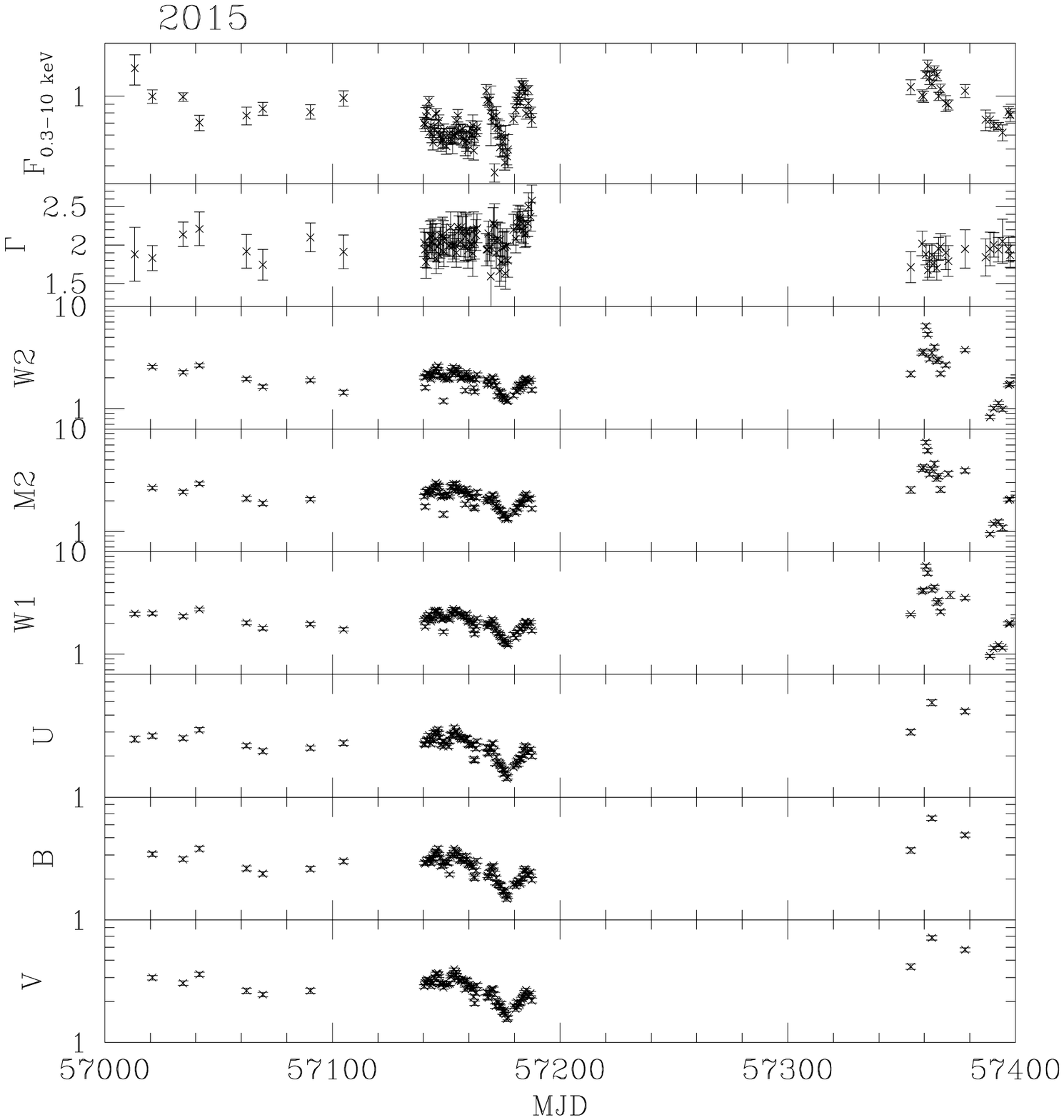}
\includegraphics[clip=, trim=1.8cm 5.6cm 1.3cm 2.6cm, width=0.49\linewidth]{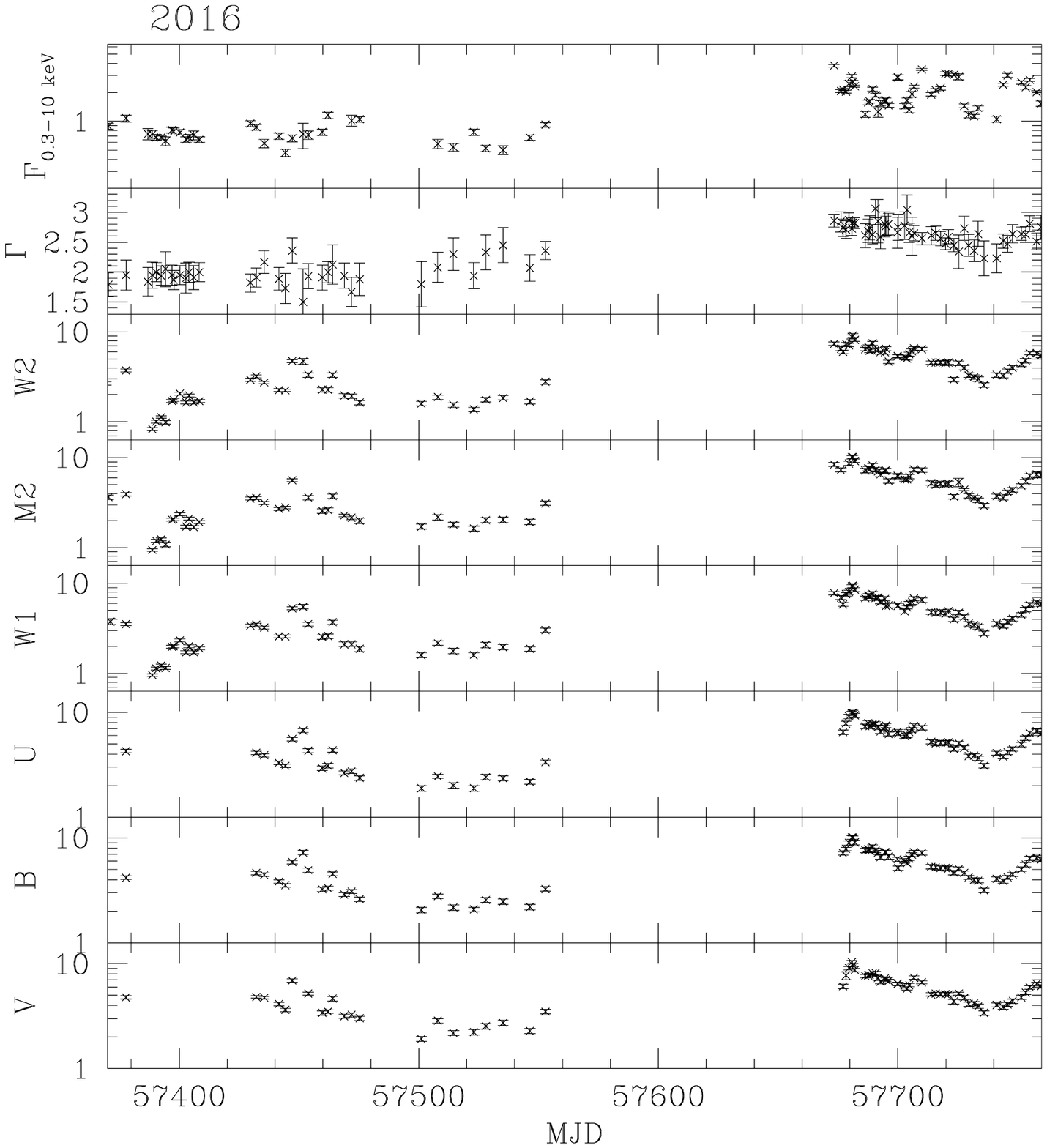}
\caption{Left: 2015 Swift XRT and UVOT light curve of OJ 287. The observed fluxes are absorption and extinction corrected and are given in units of 10$^{-11}$ erg/s/cm$^2$. $\Gamma_{\rm x}$ is the X-ray power-law photon index. Right: 2016 Swift XRT and UVOT light curve. The flux axis scale is adjusted for optimal display of the dynamic range of the data during 2016. This panel includes the beginning of the 2016/17 outburst (marked in dark blue in Figs \ref{fig:lc-Swift-CR} and \ref{fig:lc-Swift-fluxes2016}). 
}
\label{fig:lc-Swift-2015}
\end{figure}

\begin{figure}
\includegraphics[clip=, trim=1.8cm 5.6cm 1.3cm 2.6cm, width=0.49\linewidth]{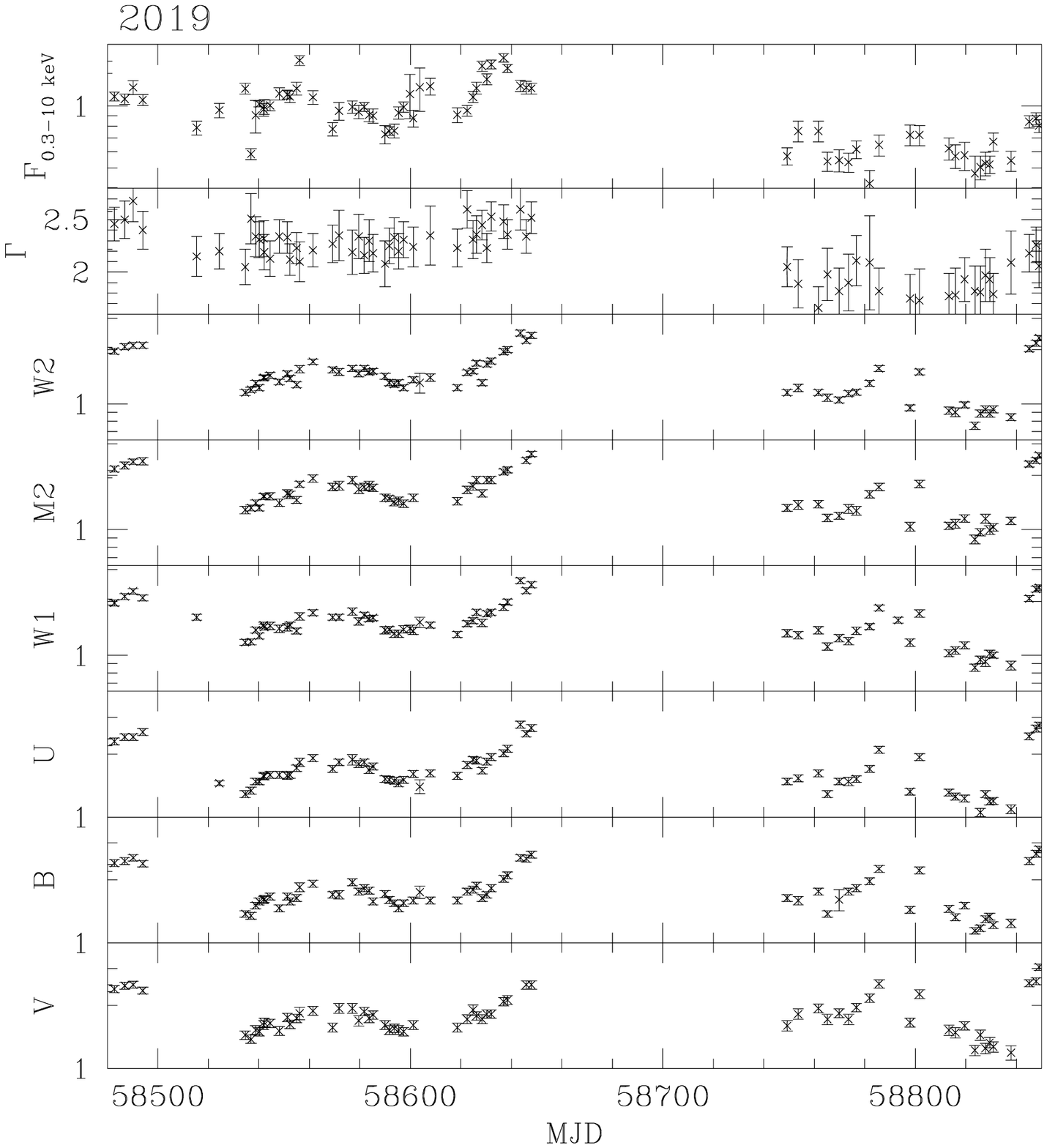}
\includegraphics[clip=, trim=1.8cm 5.6cm 1.3cm 2.6cm, width=0.49\linewidth]{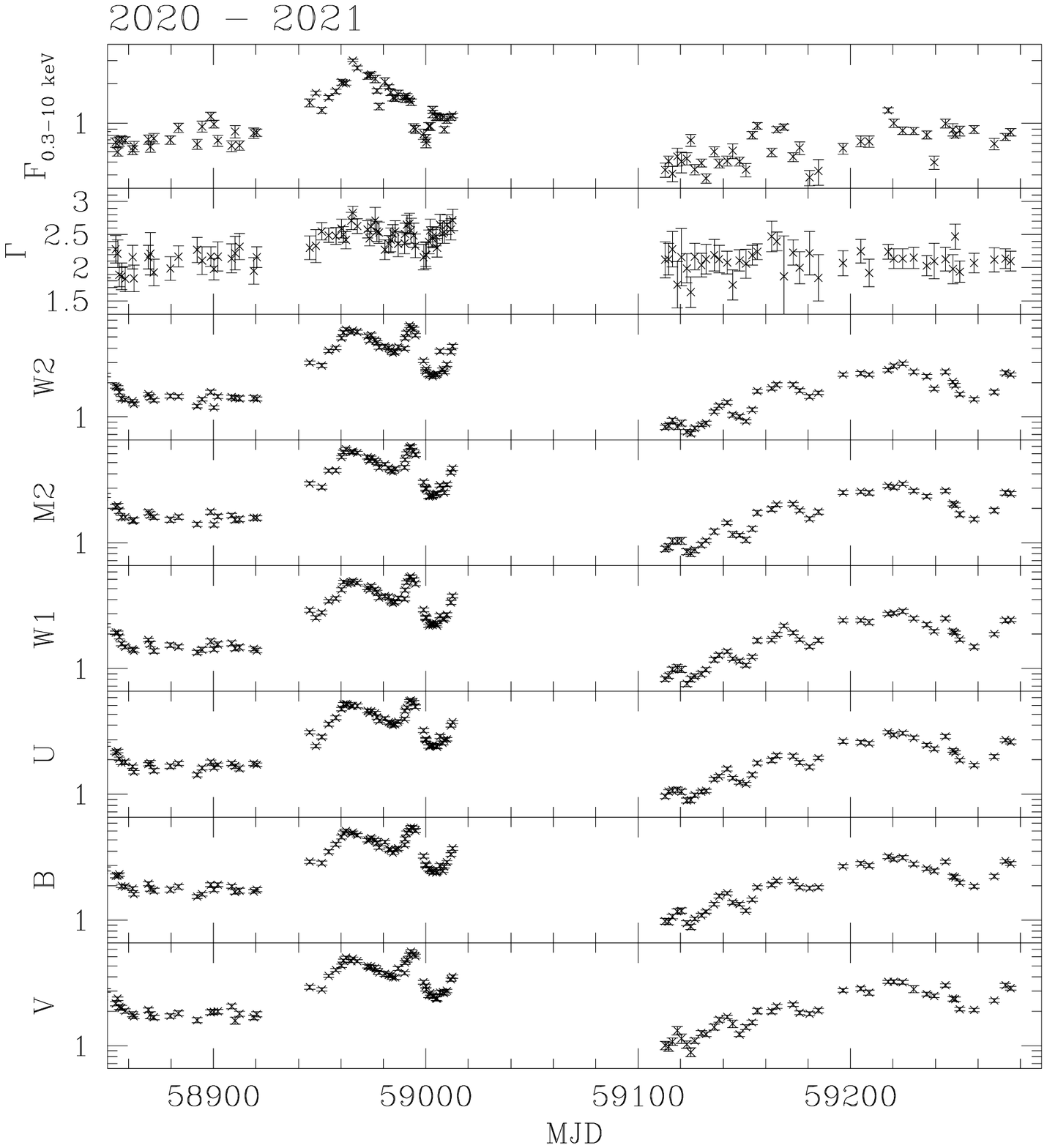}
\caption{Left: 2019 Swift XRT and UVOT light curve of OJ 287. Units as in Fig. \ref{fig:lc-Swift-2015}, but the flux axis scale is adjusted for optimal display of the dynamic range of the data during 2019.  
Right: 2020-2021 Swift XRT and UVOT light curve. This panel includes the 2020 outburst (marked in red in Figs \ref{fig:lc-Swift-CR} and \ref{fig:lc-Swift-fluxes2016}). The last data point is from 2021 March 1. 
}
\label{fig:lc-Swift-2019}
\end{figure}

In this Appendix, we show the annual Swift XRT and UVOT light curves starting with the year 2015 when OJ 287 was covered more densely with Swift for the first time. Figures \ref{fig:lc-Swift-fluxes2005} and \ref{fig:lc-Swift-fluxes2016}  best visualized the long-term evolution of OJ 287 in the optical, UV, and X-rays, and the range of the flux axis was fixed. However, this approach means that lower-amplitude variability  and epochs of very dense data coverage are not displayed and resolved in an optimal way. Therefore, annual light curves are shown in Figs \ref{fig:lc-Swift-2015}...\ref{fig:lc-Swift-2019} between 2015 and 2021 March 1. In each plot, the flux axis is chosen such that the dynamic range of the data is matched within that year.  

\begin{figure}
\includegraphics[clip=, trim=1.8cm 5.6cm 1.3cm 2.6cm, width=0.49\linewidth]{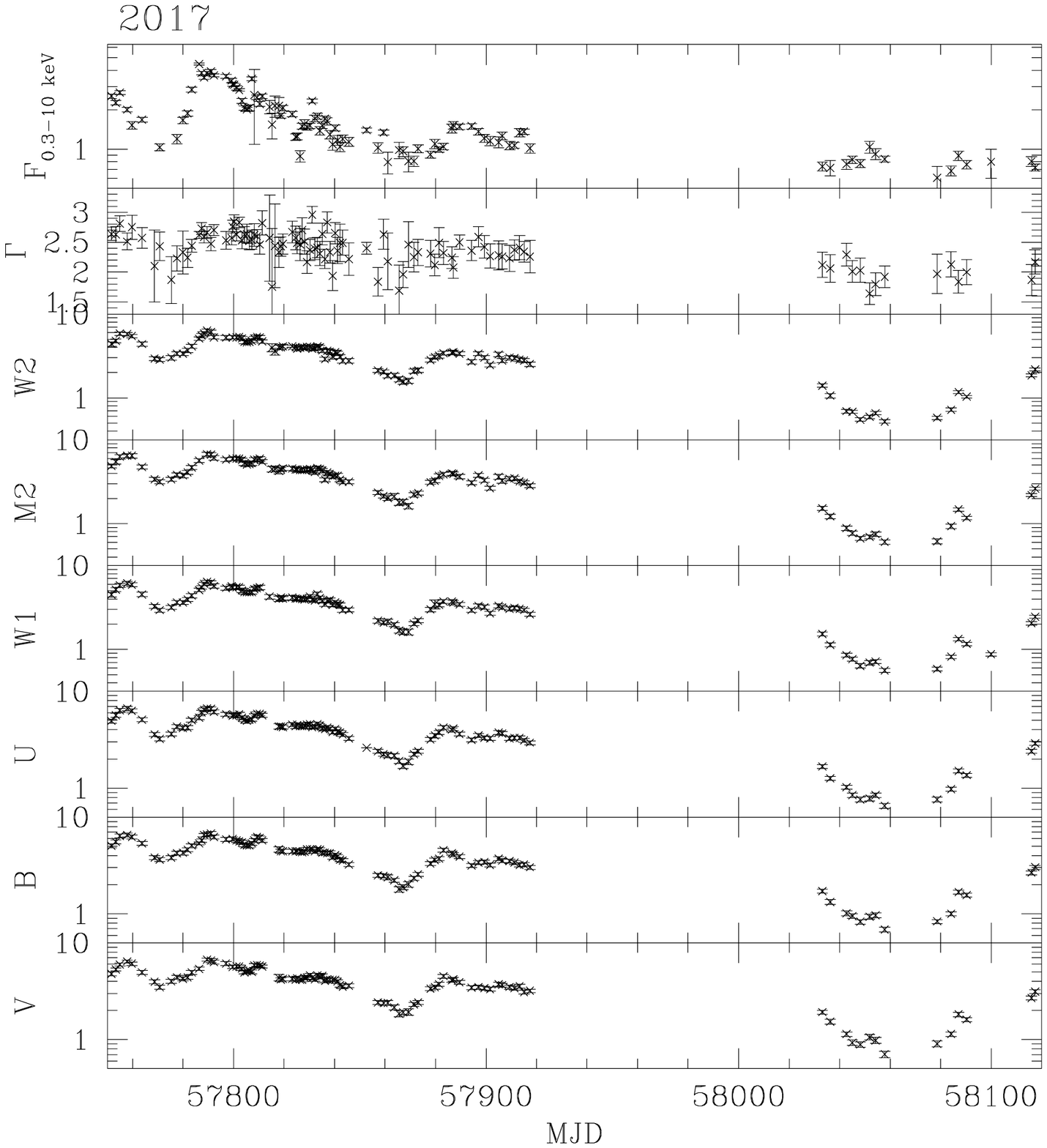}
\includegraphics[clip=, trim=1.8cm 5.6cm 1.3cm 2.6cm, width=0.49\linewidth]{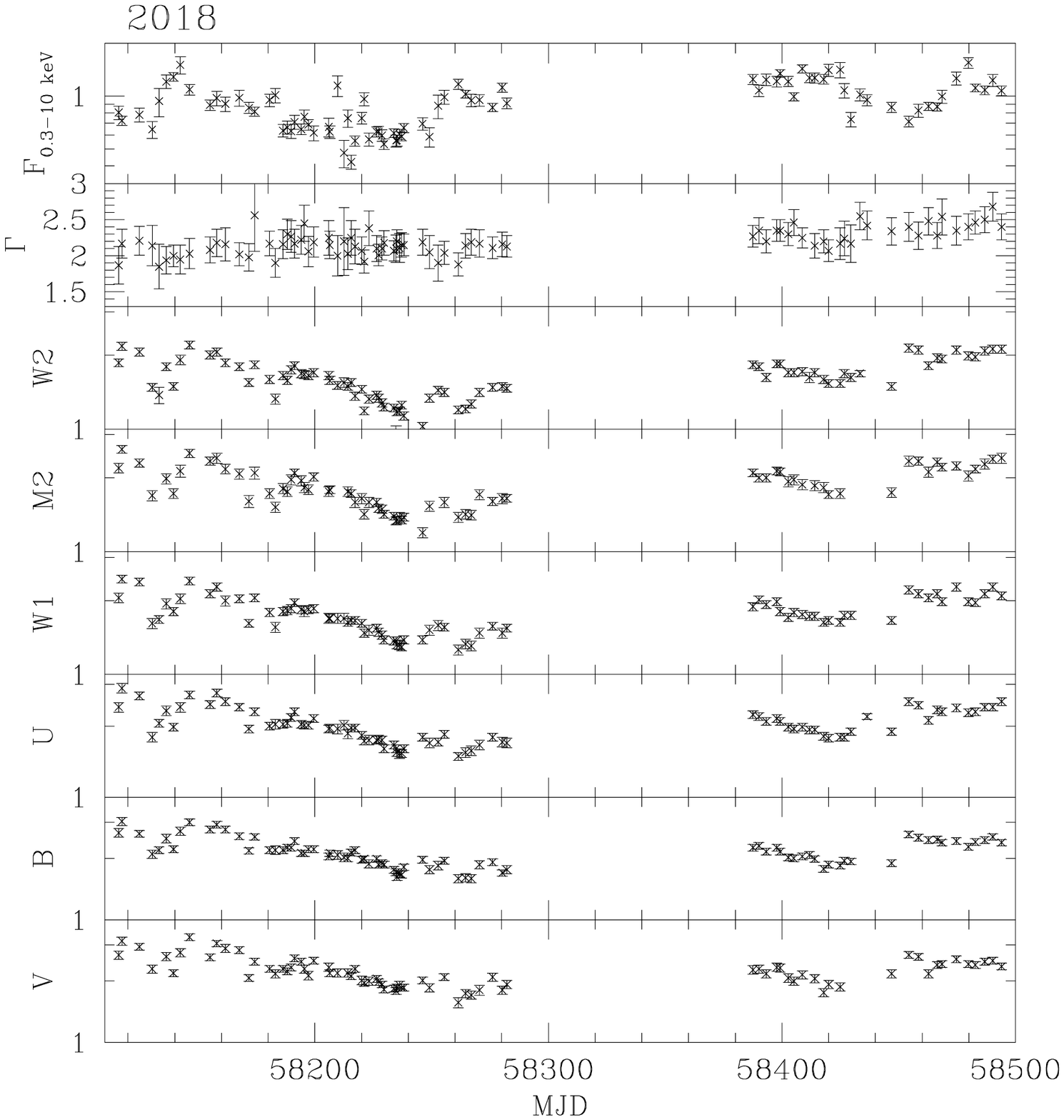}
\caption{Left: 2017 Swift XRT and UVOT light curve of OJ 287. Units as in Fig. \ref{fig:lc-Swift-2015}, but the flux axis scale is adjusted for optimal display of the dynamic range of the data during 2017. This panel includes the end of the 2016/17 outburst and the 2017 UV--optical deep fade (marked in dark blue and light blue, respectively, in Figs \ref{fig:lc-Swift-CR} and \ref{fig:lc-Swift-fluxes2016}). Right: 2018 Swift XRT and UVOT light curve.      
}
\label{fig:lc-Swift-2017}
\end{figure}

\section{X-ray sources in the Swift XRT field of view}

The X-ray field of OJ 287 based on 593 Swift (PC-mode) observations 
%% between 2005 and 2021 January 30 
is shown in Fig. \ref{fig:Xima-Swift-main}. The total on-source exposure time amounts to 670 ks. 
While only two or a few sources are detected in single exposures, the stacked image reveals a large number of X-ray sources in the field of view. 
Here, we provide a list of the brightest (CR$> 0.5 \times 10^{-3}$ cts s$^{-1}$, S/N $>$ 3) serendipitous X-ray sources in the field (Tab. \ref{tab:serendip-sources}). 
These are of special interest 
when assessing possible counterparts of VHE and/or neutrino emission (or other multi-messenger emission recorded with intrinsically large astrometric uncertainty) in the vicinity of OJ 287, and alternative to OJ 287 itself. 
Based on the source detection implementation
of \citet{Evans2007}, 127 X-ray sources above 3$\sigma$ are detected in the total (0.3--10) keV band in the field. 
All X-ray sources in the field are much fainter than OJ 287. The brightest one, at CR=0.0138$\pm{0.0004}$ cts/s, is a factor of 7 weaker than OJ 287 in its faintest state. 

\begin{table}
%\scriptsize
	\centering
	\caption{X-ray sources in the field of view of OJ 287 above a count rate (CR) of 0.5 $\times 10^{-3}$ cts s$^{-1}$. }
	\label{tab:serendip-sources}
	\begin{tabular}{llllc} 
		\hline
		Source & Name & R.A. & Decl. & CR \\
		\hline
OJ 287--X1 & SWXRT J085400.39 +201351.6 &08h 54m 00.39s& +20° 13\arcmin~ 51.6\arcsec & 0.0138$\pm$0.0004 \\
OJ 287--X2 & SWXRT J085511.19 +200032.1 &08h 55m 11.19s & +20° 00\arcmin~ 32.1\arcsec & 0.0023$\pm$0.0001 \\
OJ 287--X3 & SWXRT J085424.87 +201122.3 &08h 54m 24.87s & +20° 11\arcmin~ 22.3\arcsec & 0.0020$\pm$0.0001 \\
OJ 287--X4 & SWXRT J085409.35 +201340.5  &08h 54m 09.35s& +20° 13\arcmin~ 40.5\arcsec & 0.0045$\pm$0.0002 \\
OJ 287--X5 & SWXRT J085408.05 +200304.9  & 08h 54m 08.05s& +20° 03\arcmin~ 04.9\arcsec & 0.0015$\pm$0.0001 \\	
OJ 287--X6 & SWXRT J085403.26 +200748.3  &08h 54m 03.26s& +20° 07\arcmin~ 48.3\arcsec & 0.0016$\pm$0.0001 \\
OJ 287--X7 & SWXRT J085512.04 +200439.4  &08h 55m 12.04s& +20° 04\arcmin~ 39.4\arcsec & 0.0012$\pm$0.0001 \\
OJ 287--X8 & SWXRT J085413.97 +200355.8  &08h 54m 13.97s& +20° 03\arcmin~ 55.8\arcsec & 0.0009$\pm$0.0001 \\
OJ 287--X9 & SWXRT J085355.18 +195803.5  &08h 53m 55.18s& +19° 58\arcmin~ 03.5\arcsec & 0.0075$\pm$0.0004 \\
OJ 287--X10 & SWXRT J085458.85 +201248.4  &08h 54m 58.85s& +20° 12\arcmin~ 48.4\arcsec & 0.0007$\pm$0.0001 \\
OJ 287--X11 & SWXRT J085426.01 +201631.5  &08h 54m 26.01s& +20° 16\arcmin~ 31.5\arcsec & 0.0011$\pm$0.0001 \\	
OJ 287--X12 & SWXRT J085400.48 +200247.1 &08h 54m 00.48s& +20° 02\arcmin~ 47.1\arcsec & 0.0047$\pm$0.0004 \\
OJ 287--X13 & SWXRT J085437.63 +201643.3 &08h 54m 37.63s& +20° 16\arcmin~ 43.3\arcsec & 0.0009$\pm$0.0001 \\
OJ 287--X14 & SWXRT J085530.58 +200709.0 &08h 55m 30.58s& +20° 07\arcmin~ 09.0\arcsec & 0.0009$\pm$0.0001 \\
OJ 287--X15& SWXRT J085503.95 +195526.0  &08h 55m 03.95s& +19° 55\arcmin~ 26.0\arcsec & 0.0013$\pm$0.0001 \\
OJ 287--X16 & SWXRT J085409.98 +195901.4 &08h 54m 09.98s& +19° 59\arcmin~ 01.4\arcsec & 0.0006$\pm$0.0001 \\
OJ 287--X17 & SWXRT J085539.68 +200019.3 &08h 55m 39.68s& +20° 00\arcmin~ 19.3\arcsec & 0.0016$\pm$0.0002 \\	
OJ 287--X18 & SWXRT J085346.13 +201004.4 &08h 53m 46.13s& +20° 10\arcmin~ 04.4\arcsec & 0.0018$\pm$0.0002 \\
OJ 287--X19 & SWXRT J085533.55 +201433.6 &08h 55m 33.55s& +20° 14\arcmin~ 33.6\arcsec & 0.0010$\pm$0.0001 \\
OJ 287--X20 & SWXRT J085406.94 +201751.3 &08h 54m 06.94s& +20° 17\arcmin~ 51.3\arcsec & 0.0012$\pm$0.0001 \\
OJ 287--X21 & SWXRT J085551.40 +200340.5 &08h 55m 51.40s& +20° 03\arcmin~ 40.5\arcsec & 0.0023$\pm$0.0006 \\
		\hline
%\begin{tablenotes}
%{Notes: $^{1}$ ..}.
%\end{tablenotes}
	\end{tabular}
\end{table}

\section{ACF and ZDCF}

\subsection{ACF during epoch 5}

\begin{figure*}
\scalebox{1.0}{\includegraphics[width=\linewidth]{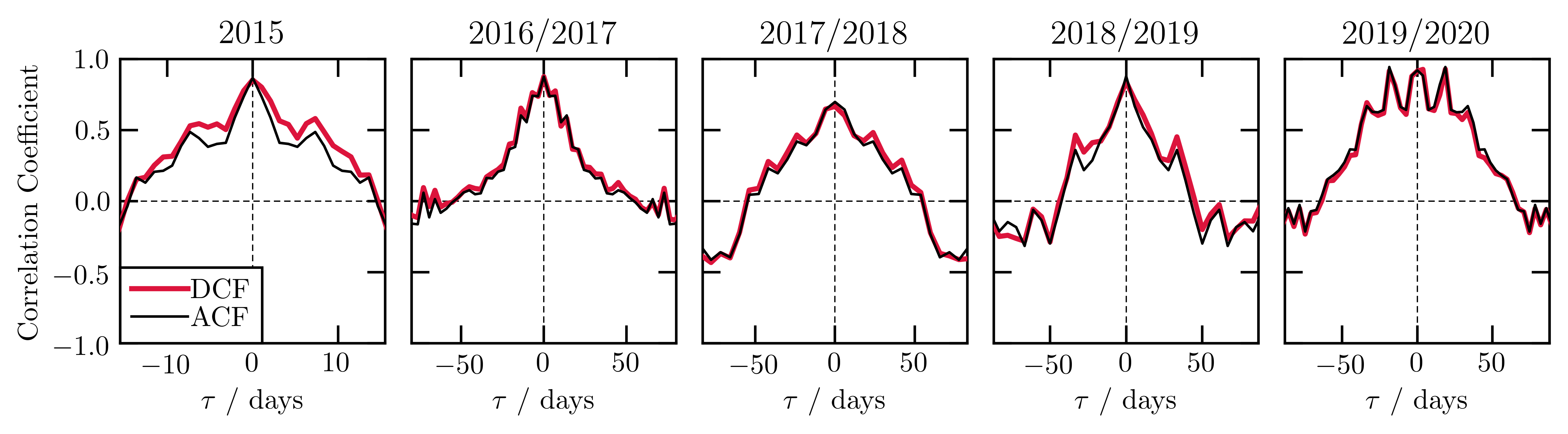}}
\caption{The W2 ACF (black) and B$-$W2 DCF (red) in each of the five epochs. %Horizontal dashed lines indicate a correlation coefficient of $0$, and 
Vertical dashed lines indicate $\tau = 0~\mathrm{days}$.  }
\label{fig:ACFvsDCF}
\end{figure*}

An additional $f_{\mathrm{B}}$ lag of $\tau_{\mathrm{B}}=19\pm6~\mathrm{days}$ over $f_{\mathrm{W2}}$ at the $>99\%$ significance level was found in epoch 5 (2020 outburst) but is not reported in the main body or findings of the paper because it suggests itself as an artifact of the nature of the light curve at that epoch, with multiple similar flares that are present during the 2020 outburst. 
We investigate the nature and robustness of the 19 day lag result by computing the $f_{\mathrm{W2}}$ auto-correlation function (ACF) in each epoch and compare it to the corresponding $f_{\mathrm{B}}-f_{\mathrm{W2}}$ DCF. We find that in epochs 2$-$5 the $f_{\mathrm{B}}-f_{\mathrm{W2}}$ DCF resembles nearly identically the $f_{\mathrm{W2}}$ ACF, only differing by $\sim10\%$ where the correlation coefficients are highest (see Fig. \ref{fig:ACFvsDCF}). Only in epoch 1 are moderate differences between the two curves evident, though overall shape and level are still quite similar. For this reason, we conclude that the $\tau_{\mathrm{B}}=19\pm6~\mathrm{day}$ lag behind the W2 flux in epoch 5 is not a physically relevant result, as it is also present at the $>99\%$ significance level in the $f_{\mathrm{W2}}$ ACF. The significant correlation in these optical and UV light curves of epoch 5 is likely due to the multiple flares observed near the end of the light curves, wherein spurious flare alignments produce highly correlated lag results. Indeed, by truncating the epoch 5 light curves at $\mathrm{MJD}=58985$, effectively removing the last two flares, the significance of the $\tau_{\mathrm{B}}\sim19~\mathrm{day}$ lead is reduced to the $<90\%$ significance level.

\subsection{ZDCF}

For selected epochs, we have also run a $z$-transformed discrete correlation function (ZDCF) in addition to the DCF. The ZDCF calculates the 
correlation function for unevenly sampled data \citep{Alexander1997}. 
The code provided by \citet{Alexander2013} was used to obtain the ZDCF{\footnote{\url{https://www.weizmann.ac.il/particle/tal/research-activities/software}}}. 
The error of the correlation function is estimated by 10.000 Monte Carlo runs, adding randomly drawn errors to the light curves based on measurement errors \citep{Alexander2013}. 
A maximum likelihood function for the ZDCF peak location, also
introduced by \citet{Alexander2013}, is used to estimate the time lag and its 1 $\sigma$ error. 

For epoch 1, we have inspected the lags between all optical bands and UV-W2. The lags and their 1$\sigma$ error, estimated using the maximum likelihood method in \citet{Alexander2013}, are
$\tau_{\rm{V-W2}}$ = +0.52$^{+0.68}_{-0.92}$ days, 
$\tau_{\rm{B-W2}}$ = +0.50$^{+0.48}_{-0.99}$ days, and
$\tau_{\rm{U-W2}}$ = +0.54$^{+0.45}_{-0.87}$ days.
A positive sign indicates that W2 is leading. Results are consistent with the DCF within their errors (Sect. 6). We have also run
the ZDCF on the X-ray -- UV-W2 fluxes of the quiescent epochs 1, 3, and 4, excluding data before MJD\,58100 in epoch 3 as done before. Again the results are consistent within the errors with the DCF. 
For epochs 1, 3, and 4, $\tau_{\rm{X-W2}}$ = $+7.4^{+2.7}_{-1.7}$ days, $\tau_{\rm{X-W2}}$ = $-21.4^{+6.2}_{-9.6}$ days, and  $\tau_{\rm{X-W2}}$ = $-6.9^{+2.2}_{-4.7}$ days, respectively. A negative sign indicates that X-rays are leading.

%\section{Author publication charges} \label{sec:pubcharge}

%Finally some information about the AAS Journal's publication charges.
%In April 2011 the traditional way of calculating author charges based on the number of printed pages was changed.  The reason for the change was due to a recognition of the growing number of article items that could not be represented in print. Now author charges are determined by a number of digital ``quanta''.  A single quantum is 350 words, one figure, one table, and one enhanced digital item.  For the latter this includes machine readable tables, figure sets, animations, and interactive figures.  The current cost for the different quanta types is available at 
%\url{https://journals.aas.org/article-charges-and-copyright/#author_publication_charges}. 
%Authors may use the ApJL length calculator to get a {\tt rough} estimate of the number of word and float quanta in their manuscript. The calculator is located at \url{https://authortools.aas.org/ApJL/betacountwords.html}.


%\bibliography{sample631}{}
%\bibliographystyle{aasjournal}



\end{document}